\documentstyle[12pt,worldsci]{article}

\input{prepictex}
\input{pictex}
\input{postpictex}
\renewcommand\theequation{\thesection.\arabic{equation}}
\def\beq{\begin{equation}}
\def\eeq{\end{equation}}
\def\eeqlb#1{\label{#1}\eeq}
\def\bea#1#2{\renewcommand{\arraystretch}{#1}\begin{array}{#2}}
\def\eea{\end{array}}
\def\beqa#1#2{\beq \renewcommand{\arraystretch}{#1} \begin{array}{#2} }
\def\eeqa{\end{array} \eeq}
\def\eeqalb#1{\end{array} \label{#1} \eeq}
\def\be{\begin{eqnarray}}
\def\ee{\end{eqnarray}}
\def\seqno#1#2{\eqno(\mbox{\ref{#1}#2})}
\def\and{\hskip 5mm \mbox{and} \hskip 5mm}
\def\dps{\displaystyle}
\def\ts{\textstyle}

\def\ptr#1#2{\put(#1){\makebox(0,0)[r]{#2}}}
\def\ptl#1#2{\put(#1){\makebox(0,0)[l]{#2}}}
\def\ptc#1#2{\put(#1){\makebox(0,0)[c]{#2}}}
\def\pl#1{\put(#1){\line(1,0){10}}}
\def\Tr{{\,\mbox{Tr}}}
\def\tr{{\,\mbox{tr}}}
\def\muck{\mbox{muck}}

\def\ve{\varepsilon}

\def\tp#1{\tilde\phi_{ #1 }}

\def\rr{\rangle\!\rangle}

\def\Puu{P_{\uparrow\uparrow}}
\def\Pud{P_{\uparrow\downarrow}}
\def\Pdu{P_{\downarrow\uparrow}}
\def\Pdd{P_{\downarrow\downarrow}}
\def\n{{\{n\}}}

\def\D{{\cal D}}

\def\I{{\cal I}}
\def\K{{\cal K}}
\def\L{{\cal L}}
  
\def\M{{\cal M}}

\def\T{{\cal T}}
\def\Y{{\cal Y}}

\def\hr{{\hat r}}
\def\vr{{\vec r}}
\def\vt{{\vec\tau}}
\def\vs{{\vec\sigma}}

\def\vo{{\vec\omega}}
\def\tdr{\vt\cdot\hr}
\def\tdo{\vt\!\cdot\!\vo}
\def\sdr{\vs\!\cdot\!\hr}

\def\sdt{\vs\!\cdot\!\vt}

\def\vA{{\vec{A}}}

\def\vI{{\vec{I}}}

\def\vJ{{\vec{J}}}

\def\vT{{\vec{\Theta}}}
\def\tp{\tilde{\phi}}

\def\m{\overline{m}}

\def\eff{{\mbox{\scriptsize eff}}}
\def\bff{{\mbox{\scriptsize bf}}}
\def\rot{{\mbox{\scriptsize rot}}}

\def\gQ{g_{_Q}}
\def\fQ{f_{_Q}}
\def\gF{gF^\prime(0)}

\def\sv{v\hskip-1.7mm/}
\def\sp{p\hskip-1.7mm/}

\def\ie{{\it i.e.}}
\def\eg{{\it e.g.}}
\def\viz{{\it viz.}}
\def\etal{{\it et al.}}

\def\ap#1#2#3{{\it Ann. Phys. (N.Y.) }{\bf #1}{, #3}{ (#2)}}
\def\appb#1#2#3{{\it Act. Phys. Pol. }{\bf B#1}{, #3}{ (#2)}}
\def\arnps#1#2#3{{\it Annu. Rev. Nucl. Part. Sci. }{\bf #1}{, #3}{ (#2)}}

\def\pzetf#1#2#3{{\it Pis'ma Zh. Eksp. Teor. Fiz. }{\bf #1}{, #3}{ (#2)}}
\def\hap#1#2#3{{\it Halv. Acta. Phys. }{\bf #1}{, #3}{ (#2)}}
\def\ijmpa#1#2#3{{\it Int. Jour. Mod. Phys. }{\bf A#1}{, #3}{ (#2)}}

\def\jetpl#1#2#3{{\it JETP Lett. }{\bf #1}{, #3}{ (#2)}}

\def\jpg#1#2#3{{\it J. Phys. }{\bf G#1}{, #3}{ (#2)}}
\def\mpla#1#2#3{{\it Mod. Phys. Lett. }{\bf A#1}{, #3}{ (#2)}}

\def\np#1#2#3{{\it Nucl. Phys. }{\bf #1}{, #3}{ (#2)}}
\def\npa#1#2#3{{\it Nucl. Phys. }{\bf A#1}{, #3}{ (#2)}}
\def\npb#1#2#3{{\it Nucl. Phys. }{\bf B#1}{, #3}{ (#2)}}
\def\npc#1#2#3{{\it Nucl. Phys. {\bf B} (Suppl.) }{\bf #1C}{, #3}{ (#2)}}
\def\phyrep#1#2#3{{\it Phys. Rep. }{\bf #1}{, #3}{ (#2)}}
\def\plb#1#2#3{{\it Phys. Lett. }{\bf B#1}{, #3}{ (#2)}}
\def\prl#1#2#3{{\it Phys. Rev. Lett. }{\bf #1}{, #3}{ (#2)}}
\def\pr#1#2#3{{\it Phys. Rev. }{\bf #1}{, #3}{ (#2)}}

\def\prd#1#2#3{{\it Phys. Rev. }{\bf D#1}{, #3}{ (#2)}}
\def\prsa#1#2#3{{\it Proc. Roy. Soc. (London) }{\bf A#1}{, #3}{ (#2)}}
\def\ptp#1#2#3{{\it Prog. Theor. Phys. }{\bf #1}{, #3}{ (#2)}}
\def\rpp#1#2#3{{\it Rep. Prog. Phys. }{\bf #1}{, #3}{ (#2)}}
\def\sjnp#1#2#3{{\it Sov. J. Nucl. Phys. }{\bf #1}{, #3}{ (#2)}}

\def\rmp#1#2#3{{\it Rev. Mod. Phys. }{\bf #1}{, #3}{ (#2)}}
\def\yf#1#2#3{{\it Yad. Fiz. }{\bf #1}{, #3}{ (#2)}}
\def\zpa#1#2#3{{\it Z. Phys. }{\bf A#1}{, #3}{ (#2)}}

\def\err#1#2#3{{\rm (E) }{\bf #1}{, #3}{ (#2)}}
\def\same#1#2#3{{}{\bf #1}{, #3}{ (#2)}}
\def\ibid#1#2#3{{\it ibid. \/}{\bf #1}{, #3}{ (#2)}}

\begin{document}

\newcounter{subeq}

\pagenumbering{roman}

\def\arraystretch{1.5}

\noindent\noindent
{Version 6.0} \hfill {SNUTP-94/117}         

 \hfill {T94/149}                           

\hfill {November 1994}

\title{HEAVY-QUARK SYMMETRY AND SKYRMIONS}

\author{ \vskip 1.5cm
Dong-Pil Min \\
  {\em Department of Physics and Center for Theoretical Physics,\\
       Seoul National University, Seoul 151-742, Korea}\\
     \vspace{0.3cm}
Yongseok Oh \\
  {\em Department of Physics, National Taiwan University,
       Taipei, Taiwan 10764, R.O.C.}\\
     \vspace{0.3cm}
Byung-Yoon Park  \\
  {\em Department of Physics, Chungnam National University,
       Daejeon 305-764, Korea}\\
     \vspace{0.3cm}
     and \\
     \vspace{0.3cm}
Mannque Rho  \\
  {\em Service de Physique Th\'{e}orique, C. E. Saclay,
       91191 Gif-sur-Yvette Cedex, France}}

     \maketitle
     \setlength{\baselineskip}{2.8ex}

\vskip 1.2cm
\begin{abstract}
We review recent development on combining heavy-quark symmetry and
chiral symmetry in the skyrmion structure of the baryons containing
one or more heavy quarks, c (charmed) and b (bottom). We describe two
approaches: One going from the chiral symmetry regime of light quarks
to the heavy-quark symmetry regime which will be referred to as
``bottom-up" approach and the other going down from the heavy-quark
limit to the realistic finite-mass regime which will be referred to
as ``top-down." A possible hidden connection between the two symmetry
limits is suggested. This review is based largely on the work done --
some unpublished -- by the authors since several years.
\end{abstract}

\vfill

\centerline{({\it To be published in Int. J. Mod. Phys. E})}

\newpage
\setlength{\baselineskip}{3.2ex}

\tableofcontents
\newpage
\setlength{\baselineskip}{2.6ex}

\pagenumbering{arabic}

\section{Introduction}
Hadrons containing a single heavy quark ($Q$) with its mass ($m_Q$)
much greater than a typical scale of strong interaction
($\Lambda_{QCD}$) can be viewed as a freely propagating point-like
color source dressed by light degrees of freedom, namely the chiral
quarks and gluons (sometimes referred to as ``brown muck"). In addition
to the chiral symmetry governing the dynamics of the light quark system,
such a system reveals a new spin and flavor symmetry\cite{VS,HQ1,IW,Ge90},
``heavy quark symmetry," as the heavy quark mass goes to infinity.
(For a review, see Refs.~\citenum{Ge92,hq,Grinstein,Neubert}.) In this
limit, the heavy quark spin $\vec{S}_Q$ decouples from the rest of the
strongly interacting light quark system, since its coupling is a
relativistic effect of order $1/m_Q$. Furthermore, the dynamics of a
heavy quark in QCD depends only on its velocity and is independent of
its mass, \ie, the flavor. Consider two hadrons $A$ and $B$ made of a
single heavy quark of mass $m_Q^A$ and $m_Q^B$, respectively, and
light degrees of freedom of QCD. If the heavy quark masses are much
greater than the scale of the QCD interactions, $m_Q^A, m_Q^B\gg
\Lambda_{QCD}$, then in the rest frame of the heavy quark, how QCD
distributes the light degrees of freedom around the static heavy
quark is independent of the heavy flavor. By boosting, one can extend
the heavy flavor symmetry to any heavy quarks of the same {\em velocity}
(not the same momentum). This symmetry is analogous to the isotope
effect in atomic physics where the electronic structure of an atom is
independent of the number of neutrons in the nucleus and the hyperfine
splitting is of order $\sim1/Am_N$.

The heavy quark symmetry provides an enormous help in reducing the
large number of independent parameters required to describe the low
momentum properties of the strong interaction involving a heavy quark.
One of the well-known examples is that all six form factors for
$B \rightarrow D e \bar{\nu}_e$ and $B \rightarrow D^* e \bar{\nu}_e$
semileptonic decays are described by a universal function of velocity
transfer, the ``Isgur-Wise" function.\cite{IW}

Another consequence of this heavy quark symmetry can be found in the
spectra and strong decay widths of heavy hadrons\cite{IW91}. Due to
the heavy quark spin decoupling, the total angular momentum of the
light degrees of freedom $\vec{J}_\ell(\equiv \vec{J} -\vec{S}_Q$,
with $\vec{J}$ the spin of the hadron) is conserved and the
corresponding quantum number $j_\ell$ can classify the hadrons and
as a consequence, the hadrons come in degenerate doublets with total
spin
\beq
j_\pm = j_\ell \pm \ts\frac12,
\eeq
(unless $j_\ell=0$) which are formed by combining the spin of the
heavy quark with $j_\ell$. Furthermore, the strong transitions
between any two pairs of doubly degenerate states, occurring via
the emission of light hadrons, are related simply by Clebsch-Gordan
coefficients.\cite{IW91} For example, the ratio of the amplitudes
for $\Sigma_Q\rightarrow \Lambda_Q\pi$ and $\Sigma_Q^*\rightarrow
\Lambda_Q \pi$ will be unity. On the other hand, the heavy
quark-flavor symmetry implies that if we line up the ground
states by subtracting the mass of the heavy quark,
then the spectra built on different heavy flavors should look the
same. That is, the splittings are flavor independent.

\setlength{\unitlength}{1.25mm}
\begin{figure}
\begin{center}
{\bf Figure 1.1} : Spectrum of hadrons containing a single heavy
                   quark.  \\
         ($^{a)}$ Ref.~\citenum{PDG}, $^{b)}$ Ref.~\citenum{SKAT})
\vskip 2mm
\hskip -06.cm
\begin{picture}(100,45)   
\put(38,10){\line(1,0){74}}
\put(36,14.75){\line(4,3){4}}
\put(36,18.125){\line(4,3){4}}
\put(36,22){\line(1,0){4}}
\put(36,25.125){\line(4,-3){4}}
\put(36,29.875){\line(2,-3){4}}
\put(75,8){\makebox(0,0)[c]{$j_\ell$=$\frac12$}}
\put(70,8.15){\vector(-4,-1){4}}
\put(70,8.15){\vector(-4,3){4}}
\put(80,8.15){\vector(4,1){4}}
\put(80,8.275){\vector(2,1){4}}
\put(75,16.75){\makebox(0,0)[c]{$j_\ell$=0}}
\put(70,16.75){\vector(-4,1){4}}
\put(80,17.2){\vector(4,1){4}}
\put(75,21.125){\makebox(0,0)[c]{$j_\ell$=$\frac32$}}
\put(70,26){\vector(-1,-1){4}}
\put(70,26){\vector(-2,-1){4}}
\put(75,26){\makebox(0,0)[c]{$j_\ell$=1}}
\put(70,21.125){\vector(-4,1){4}}
\put(70,21.125){\vector(-1,0){4}}
\thicklines
\put(40,7.375){\line(1,0){26}}
\put(36,7.5){\makebox(0,0)[r]{$D(1870)^{a}$}}
\put(40,10.875){\line(1,0){26}}
\put(36,11){\makebox(0,0)[r]{$D^*(2010)^{a}$}}
\multiput(40,17.75)(3,0){9}{\line(1,0){2}}
\put(36,15){\makebox(0,0)[r]{$\Lambda_c^{}(2285)^{a}$}}
\multiput(40,22)(3,0){9}{\line(1,0){2}}
\put(36,22){\makebox(0,0)[r]{$\Sigma_c^{}(2453)^{a}$}}
\multiput(40,23.875)(3,0){9}{\line(1,0){2}}
\put(36,29){\makebox(0,0)[r]{$\Sigma^*_c(2530)^{b}$}}
\put(40,21.125){\line(1,0){26}}
\put(36,18.5){\makebox(0,0)[r]{$D_1(2420)^{a}$}}
\put(40,22.125){\line(1,0){26}}
\put(36,25.5){\makebox(0,0)[r]{$D_2(2460)^{a}$}}
\put(84,9.15){\line(1,0){26}}
\put(112,8){\makebox(0,0)[l]{$B(5279)^{a}$}}
\put(84,10.275){\line(1,0){26}}
\put(112,11.5){\makebox(0,0)[l]{$B^*(5324)^{a}$}}
\multiput(84,18.2)(3,0){9}{\line(1,0){2}}
\put(112,18.2){\makebox(0,0)[l]{$\Lambda_b(5641)^{a}$}}
\multiput(52.5,30)(0,1){3}{\makebox(0,0)[c]{$\cdot$}}
\multiput(97.5,30)(0,1){3}{\makebox(0,0)[c]{$\cdot$}}
\put(52.5,40){\makebox(0,0)[c]{$Q=c$}}
\put(97.5,40){\makebox(0,0)[c]{$Q=b$}}
\end{picture}
\end{center}
\end{figure}

Given in Fig.~1.1 are the experimentally observed mesons (solid lines)
and baryons (dashed lines) with a single charm or bottom quark. The
average masses of the ground state heavy meson doublets,
$\m^{}_D(\equiv\frac34 m^{}_{D^*} + \frac14 m^{}_D)$ and
$\m^{}_B(\equiv\frac34 m^{}_{B^*} + \frac14 m^{}_B)$, are lined up.
The hadrons can be easily lumped into approximately degenerate
doublets with $j_\pm = j_\ell \pm 1/2$. The $D^*$-$D$ splitting of
$\sim 145$ MeV is reduced to $\sim 50$ MeV in the $B^*$-$B$ multiplet,
which is consistent with the expected $1/m_Q$ behavior in the
heavy-quark limit. In Fig.~1.1, one may infer the heavy quark flavor
symmetry from the approximately same mass differences
$m_{\Lambda_c}^{}-\m_D^{}$ and $m_{\Lambda_b}^{}-\m_B^{}$.

In the Skyrme model {\it \`{a} la\/} Callan and Klebanov
(CK)\cite{CK,KK}, heavy baryons can be described by bound states of
soliton of the $SU(2)$ chiral Lagrangian and the heavy mesons
containing the corresponding heavy quark. This picture which was
originally put forward\cite{CKsm} and shown to be
successful\cite{CKsm1,CKsm2,CKsm3,SMNR} for the strange baryons was
first suggested by Rho, Riska and Scoccola\cite{RRS} to be applicable
to baryons containing one or more charm (c) and bottom (b) quarks.
The results on the mass spectra\cite{RS91} and magnetic
moments\cite{OMRS} for charmed baryons were found to be strikingly
close to the quark model description which is expected to
work better as the heavy quark involved becomes heavier.

When the bound system of the soliton and heavy mesons is quantized to
a heavy baryon of spin $j$ and isospin $i$, the mass is given by
the sum of the mass of the $SU(2)$ soliton, the meson energy and
a hyperfine splitting. For the baryons with a single heavy flavor,
the mass formula simply reads as\cite{CK,CKsm}
\beq
m_{(i,j)}^{} = M_{sol} + \omega_B
+ \frac{1}{2\I}\{cj(j+1)+(1-c)i(i+1)+c(c-1)k(k+1)\}.
\eeqlb{MF}
Here, $M_{sol}$ and $\I$ are the soliton mass and its moment of
inertia with respect to the collective isospin rotation, respectively,
and $\omega_B$ is the eigenenergy of the heavy meson bound state.
The heavy meson bound state comes as an eigenstate of the grand spin
operator $\vec{K}(\equiv \vI_h + \vJ_h$ with $\vec{J}_h$ and
$\vec{I}_h$ being the spin and isospin operator of heavy mesons,
respectively), where $k$ in Eq.~(\ref{MF}) denotes the corresponding
grand spin quantum number. The spin of the baryon $j$ can take one
of the values $|i-k|, \cdots, i+k$. The constant $c$ is defined
through
\beq
\langle k,k'_3 |\vT|k,k_3\rangle \equiv
- c \langle k,k'_3 | \vec{K} |k,k_3\rangle,
\eeq
in analogy to the Lande's g-factor in atomic physics. Here, $\vT$
is the meson field operator induced by the collective isospin rotation
and it forms the first rank tensor in the space of the grand spin
eigenstates $|k,k_3\rangle$. This ``hyperfine constant" yields the
hyperfine splittings between the heavy baryon masses and plays the
role of an {\em order parameter} for the heavy quark
symmetry.\cite{RS91,Oka} As the heavy meson masses increase, it
decreases and becomes zero in the infinite heavy meson mass limit
so that the heavy baryon masses do not depend explicitly on the
{\em spin}.

In this review, we discuss the bound state approach of the Skyrme
model to describe the heavy baryons. The plan of the review is as
follows. In the remainder of this section we briefly discuss the
successful and unsuccessful features of the straightforwardly
extended CK model which does not respect the heavy quark symmetry
explicitly. Section~2 is devoted to the construction of an effective
meson Lagrangian which satisfies both the chiral symmetry and the heavy
quark symmetry at infinite heavy quark mass limit. To construct
more realistic models for finite heavy-quark mass, one may include
$1/m_Q$ corrections to the effective theory. We refer to this approach
as ``top-down" approach. On the other hand, one may start from the CK
model of the chiral symmetry regime and go to the heavy meson mass limit.
We shall refer to this as the ``bottom-up" approach. This will indicate
whether and how the two theories can be related smoothly. Such an attempt is
discussed in Sec.~3. In Sec.~4 we review the heavy meson effective
theory in more detail. The spin and isospin operators of the heavy
meson fields are obtained and the tensor representation of the heavy
meson fields is introduced, which is very useful in calculating
physical quantities at infinite mass limit. In Sec.~5  bound states of
the soliton and heavy mesons are obtained and their quantization
procedures are presented. The heavy baryon mass spectrum so obtained
is based on the effective Lagrangian developed in the
previous sections, which contains only the interactions of pions and
heavy mesons. To include higher order interactions with pions one
may include light vector meson degrees of freedom such as $\omega$
and $\rho$ into the effective theory. Also for more realistic
descriptions, we need to include the $1/m_Q$
corrections. Recent developments in these subjects are discussed in Sec.~6
and a summary is made in Sec.~7 with some conclusions.

The mass formula (\ref{MF}) can be generalized for the
baryons with $n_i(i$=1,2) mesons of species $i$ (representing
orbital state and flavor) with energy $\omega_{B,i}$, grand spin $k_i$
and hyperfine constant $c_i$ in an approximation analogous to
the quasi-particle approximation in many-body physics with the
residual interaction ignored. The generalized one is in a slightly
more complicated form:\cite{RRS,RS91}
\beqa{1.5}{l}
\dps
m_{(i,j;j_m)} = M_{sol} + n_1 \omega_{B,1} + n_2 \omega_{B,2} \\
\dps\hskip 1.5cm
+ \left. \frac{1}{2\I} \right\{ i(i+1) + (c_1 - c_2) [c_1 j_1(j_1+1)
-c_2 j_2(j_2+1)] + c_1 c_2 j_m (j_m +1) \\
\dps\hskip 3cm
+ [(c_1+c_2)j_m(j_m+1) + (c_1-c_2)(j_1(j_1+1)-j_2(j_2+1)] \\
\dps\hskip 5cm
\left. \cdot[\frac{j(j+1) - j_m(j_m+1) - i(i+1)}{2j_m(j_m+1)}]\right\},
\eeqa
where $j_i(i$=1,2) is the total grand spin of the $n_i$ mesons
of species $i$ and $j_m$ is the total grand spin of the whole
meson system. Due to the bosonic statistics involved,
$j_i$ is restricted to its maximum value that can be obtained
by combining $n_i$ mesons; that is, $j_i = n_i k_i$.
However, since different orbital is populated and/or different
flavor is considered, this does not hold for $j_m$; here $j_m$
can take one of the values $|j_1-j_2|$, $\cdots$, $j_1+j_2$.

We show in Fig.~1.2 the resulting mass spectrum of the strange
hyperons, charmed baryons and bottom baryons quoted
from Refs.~\citenum{RRS,Oh}, which show a close resemblance to those
of experiments\cite{PDG} and quark/bag models({\sf QM}: Ref.~\citenum{QM},
{\sf BM}: Ref.~\citenum{BagM}). The parameters used in obtaining the
masses are presented in Table~1.1. Two quantities $M_{sol}$ and $1/\I$
associated with the $SU(2)$ soliton are fitted to the nucleon and $\Delta$
masses by adjusting the parameters in the pionic sector.
In {\sf SM II} and {\sf SM III}, $\omega_B^{}$'s and $c$'s are further
fitted to the experimental masses of $\Lambda$, $\Sigma$, $\Lambda_c$ and
$\Sigma_c$. In the case of {\sf SM I} and {\sf SM IV}, the experimental
meson masses $m_K^{}$ and $m_B^{}$ are used to obtain $\omega_B$
and $c$.
\setlength{\unitlength}{1mm}
\begin{figure}[p]
{\bf Figure 1.2} : The spectrum of (a) strange hyperons,
(b) charmed baryons and (c) bottom baryons
in the bound state approach.
\vskip 2mm
\begin{picture}(130,200)(-10,-100)
\ptc{30,90}{(a)}
\put(5,5){\line(0,1){80}}
\multiput(5,5)(0,8){11}{\line(1,0){2}}
\ptr{4,21}{\small\sf 1.0} \ptr{4,61}{\small\sf 1.5}
\ptl{8,85}{\small\sf GeV}
\ptc{20,10}{\small\sf SM I} \ptc{35,10}{\small\sf SM II}
\ptc{50,10}{\small\sf Exp.}
\thicklines
\pl{15,16.12} \pl{30,16.12} \pl{45,16.12} \ptl{11,16.12}{\small $N$}
\pl{15,27.88} \pl{30,30.28} \pl{45,30.28} \ptl{11,27.88}{\small $\Lambda$}
\pl{15,37.40} \pl{30,36.44} \pl{45,36.44} \ptl{11,37.00}{\small $\Sigma$}
\pl{15,39.56} \pl{30,39.56} \pl{45,39.56} \ptl{11,40.00}{\small $\Delta$}
\pl{15,45.88} \pl{30,48.12} \pl{45,46.44} \ptl{11,45}{\small $\Xi$}
\pl{15,46.60} \pl{30,50.60} \pl{45,51.80} \ptl{11,48.00}{\small $\Sigma^*$}
\pl{15,55.00} \pl{30,62.28} \pl{45,63.40} \ptl{11,55}{\small $\Xi^*$}
\pl{15,64.92} \pl{30,74.52} \pl{45,74.76} \ptl{11,64.92}{\small $\Omega$}
\thinlines
\put(25.5,16.12){\line(1,0){4}} \put(40.5,16.12){\line(1,0){4}}
\put(25.5,27.75){\line(3,2){4}} \put(40.5,30.28){\line(1,0){4}}
\put(25.5,37.40){\line(4,-1){4}} \put(40.5,36.44){\line(1,0){4}}
\put(25.5,39.56){\line(1,0){4}} \put(40.5,39.56){\line(1,0){4}}
\put(25.5,46){\line(2,1){4}}    \put(40.5,48){\line(3,-1){4}}
\put(25.5,47){\line(5,4){4}}    \put(40.5,50.7){\line(4,1){4}}
\put(25.5,55.7){\line(2,3){4}}  \put(40.5,62.3){\line(4,1){4}}
\put(25.5,65.92){\line(1,2){4}} \put(40.5,74.66){\line(1,0){4}}
\ptc{105,90}{(b)}
\put(80,5){\line(0,1){80}}
\multiput(80,5)(0,8){11}{\line(1,0){2}}
\ptl{83,85}{\small\sf GeV}
\ptc{95,10}{\small\sf SM III} \ptc{110,10}{\small\sf QM}
\ptc{125,10}{\small\sf Exp.}
\ptr{79,5}{\small\sf 2.0} \ptr{79,45}{\small\sf 2.5} \ptr{79,85}{\small\sf 3.0}
\thicklines
\pl{90,27.80} \pl{105,25.8} \pl{120,27.8}  \ptl{86,27.8}{\small$\Lambda_c$}
\pl{90,41.24} \pl{105,40.2} \pl{120,41.24} \ptl{86,41.24}{\small$\Sigma_c$}
\pl{90,44.52} \pl{105,45.8} \pl{120,47.4}  \ptl{86,44.52}{\small$\Sigma^*_c$}
\pl{90,48.20} \pl{105,43.4} \pl{120,41.8}  \ptl{86,48.20}{\small$\Xi_c$}
\pl{90,52.68} \pl{105,51.0}                \ptl{86,52.68}{\small$\Xi'_c$}
\pl{90,56.92} \pl{105,56.6}                \ptl{86,56.92}{\small$\Xi^*_c$}
\pl{90,67.88} \pl{105,63.4} \pl{120,64.2}  \ptl{86,67.38}{\small$\Omega_c$}
\pl{90,69.88} \pl{105,68.2}                \ptl{86,70.38}{\small$\Omega_c^*$}
\thinlines
\put(100.4,27.60){\line(5,-2){4}} \put(115.4,26.00){\line(5,2){4}}
\put(100.4,41.24){\line(4,-1){4}} \put(115.4,40.32){\line(4,1){4}}
\put(100.4,44.64){\line(4,1){4}} \put(115.4,45.9){\line(5,2){4}}
\put(100.4,47.8){\line(1,-1){4}} \put(115.4,43.23){\line(3,-1){4}}
\put(100.4,52.51){\line(3,-1){4}}
\put(100.4,56.76){\line(1,0){4}}
\put(100.4,67.84){\line(1,-1){4}} \put(115.4,63.45){\line(6,1){4}}
\put(100.4,69.7){\line(3,-1){4}}
\ptc{22.5,-11}{(c)}
\put(5,-80){\line(0,1){64}}
\multiput(5,-80)(0,8){9}{\line(1,0){2}}
\ptr{4,-64}{\small\sf 5.5} \ptr{4,-24}{\small\sf 6.0}
\ptl{8,-16}{\small\sf GeV}
\ptc{20,-76}{\small\sf SM IV} \ptc{35,-76}{\small\sf BM}
\thicklines
\pl{15,-66.80} \pl{30,-59.6} \ptl{11,-66.8}{\small$\Lambda_b$}
\pl{15,-51.68} \pl{30,-45.2} \ptl{11,-52.68}{\small$\Sigma_b$}
\pl{15,-50.96} \pl{30,-42.48} \ptl{11,-49.96}{\small$\Sigma^*_b$}
\pl{15,-47.84} \pl{30,-45.12} \ptl{11,-46.84}{\small$\Xi_b$}
\pl{15,-40.00} \pl{30,-33.6} \ptl{11,-42}{\small$\Xi'_b$}
\pl{15,-39.76} \pl{30,-31.04} \ptl{11,-38.26}{\small$\Xi^*_b$}
\pl{15,-28.00} \pl{30,-22.24} \ptl{11,-29}{\small$\Omega_b$}
\pl{15,-27.60} \pl{30,-19.92} \ptl{11,-26.10}{\small$\Omega^*_b$}
\thinlines
\put(25.5,-66.1){\line(2,3){4}}
\put(25.5,-50.8){\line(5,6){4}}
\put(25.5,-49.6){\line(2,3){4}}
\put(25.5,-47.5){\line(2,1){4}}
\put(25.5,-39.2){\line(5,6){4}}
\put(25.5,-38.4){\line(3,5){4}}
\put(25.5,-27.5){\line(5,6){4}}
\put(25.5,-26.7){\line(2,3){4}}
\put(95,-15){\makebox(0,0){
{\bf Table 1.1} : Parameters used in {\sf SM I, II, III} and {\sf IV}. }}
\put(95,-53){\makebox(0,0){
\renewcommand{\arraystretch}{1}
\begin{tabular}{ccccc}
\hline
  & {\sf SM I}$^{a)}$ & {\sf SM II}$^{a)}$
  & {\sf SM III}$^{a)}$ & {\sf SM IV}$^{b)}$ \\
\hline
$M_{sol}^{c)}$  & 866$^{d)}$ & 866$^{d)}$ & 866$^{d)}$ & 866$^{d)}$ \\
$1/\I^{c)}$     & 196$^{d)}$ & 196$^{d)}$ & 196$^{d)}$ & 196$^{d)}$ \\
$\omega_{B,s}^{c)}$  & 209 & 223$^{d)}$ & 223$^{d)}$ & --  \\
$\omega_{B,c}^{c)}$  & --  & --  & 1418$^{d)}$ & -- \\
$\omega_{B,b}^{c)}$  & --  & --  & --   &  4599 \\
$c_s^{}$ & 0.39 & 0.60$^{d)}$ & 0.60$^{d)}$ & -- \\
$c_c^{}$ & -- & -- & 0.14$^{d)}$ & -- \\
$c_b^{}$ & -- & -- & -- & 0.03 \\
\hline
& \multicolumn{4}{l}{ $^{a)}$ Ref.~\citenum{RRS},  $^{b)}$ Ref.~\citenum{Oh}
}\\ & \multicolumn{4}{l}{ $^{c)}$ in MeV unit, $^{d)}$ fitted. } \\
\end{tabular}
}}
\end{picture}
\end{figure}

Although qualitatively successful, there have been a few problems
in fine-tuning the Skyrme model to achieve a quantitative success.
{}From a phenomenological point of view, the genuine mesonic parameters
of the Skyrme Lagrangian leads to too deeply bound heavy baryons with
somewhat large hyperfine splittings. One may improve the situation
by modifying the symmetry breaking terms of the Lagrangian, for
example, by incorporating the different values of the decay constants
of the mesons of different flavor.\cite{RS91,OMRS,PSW} The overbinding can be
removed by increasing the ratio of the heavy meson decay constant to
the pion decay constant but this leads to a larger hyperfine
splitting,\cite{BDRS}
which is certainly at odds with the heavy quark symmetry.

One can see that treating the heavy vector mesons in the traditional
bound state approach\cite{SMNR} cannot be compatible with the
heavy quark symmetry, when straightforwardly extended.
As far as the strange flavor is concerned,
in analogy to the vector mesons $\rho$ and $\pi$, the
vector mesons $K^*$ may be integrated out via {\em an ansatz} of the form
\beq
K^*_\mu = \frac{\sqrt2}{m_K^*} A_\mu K,
\eeqlb{Kstar}
in favor of a combination of a background $A_\mu$ (see Sec.~2 for its
detailed form) and the pseudoscalar meson field $K$. This
approximation is valid, however, only when the vector meson masses
are much heavier than those of the pseudoscalar mesons.
Since the mass ratios of the pseudoscalar and vector mesons
are not small enough for c- and b-quark mesons,
as one can see in Table~1.2, the approximation of the type
Eq.~(\ref{Kstar}) can no longer be reliable.
Furthermore, Eq.~(\ref{Kstar}) suppresses the vector meson field by a factor
inversely proportional to the vector meson mass, which is 0.56, 0.25 and
0.09, respectively, for the strange, charm and bottom sector
when the explicit vector meson masses are substituted.
It goes in the opposite direction to what we have expected of the heavy quark
symmetry, according to which the role of the heavy vector mesons
should become more important as the heavy vector meson mass
becomes heavier and in the infinite mass limit,
degenerate to that of the pseudoscalar meson.

This difficulty has been neatly resolved
by Jenkins \etal\cite{JMW}, whose work has led to a burst of publications
in this field.\cite{GLM,MOPR,JM92,NRZ,GMSS,BP}
The idea is to apply
the bound-state approach to the heavy meson
effective Lagrangian\cite{BD92,Wise} where the heavy quark symmetry
{\it and} chiral symmetry are incorporated on the same footing.
As shown in detail in Ref.~\citenum{OPM1}, the heavy vector mesons play
an essential role in giving correct hyperfine splitting in heavy baryons.

A remarkable new feature that emerges from the heavy-quark symmetry is that
one can associate the structure of the hyperfine splitting with a non-abelian
Berry potential~\cite{BP}. In fact one can interpret certain baryonic
excitations as following from induced gauge fields in appropriate
flavor spaces and the vanishing of the hyperfine splitting can be identified
with the vanishing of the Berry potential, in analogy to atomic and molecular
systems. Readers interested in this matter are referred to Ref. \citenum{BP}
for details.

\begin{table} 
\renewcommand{\arraystretch}{1}
\begin{center}
{\bf Table 1.2} : Masses of the pseudoscalar and vector mesons
and their ratio. \vskip 3mm
\begin{tabular}{lcccc}
\hline
                                   & $\pi,\rho$ & $Q=s$ & $Q=c$ & $Q=b$ \\
\hline
Pseudoscalar Meson Mass ($m_P^{}$) &  138       & 498   & 1865  & 5278 \\
Vector Meson Mass ($m_{P^*}$)      &  770       & 892   & 2010  & 5325 \\
Mass Ratio ($m_P^{}/m_{P^*}$)      &  0.18      & 0.56  & 0.93  & 0.99 \\
\hline
\end{tabular} \end{center}
\end{table} 

\setcounter{equation}{0}
\section{Effective Lagrangian}
In this section, we discuss how one can construct a Lagrangian of heavy
mesons interacting with  light Goldstone bosons by incorporating
simultaneously both the chiral symmetry and the heavy quark symmetry.
We start
with a familiar meson Lagrangian which yields Klein-Gordon equations
for the spin-0 and spin-1 heavy meson fields. The elegant
4$\times$4 `$H(x)$'-matrix representation\cite{Ge92,Bjorken,Falk}
particularly designed for the heavy mesons --- which may be however
unfamiliar to some of the readers --- will be discussed in Sec.~4.2.

\subsection{Chiral Symmetry}
The part of the QCD Lagrangian density involving the
light degrees of freedom (light quarks and gluons) is
\beq
\L = - \ts\frac14 \Tr G^{\mu\nu} G_{\mu\nu}
+ \bar{q} (i {{D}\hskip-2.2mm/^c} - m_q^{}) q,
\eeqlb{LQCD}
where the quark field $q=(u,d,\cdots)^T$ is in the fundamental representation
of both the color $SU(3)_c$ and the flavor $SU(N_f)$ and
$G^a_{\mu\nu}$ is the gluon field
strength with $A_\mu^a$ in the adjoint representation
of $SU(3)_c$ ($a$=1,2,$\cdots$,8). The covariant
derivative\footnote{We will be working with various
different forms of 
covariant derivatives associated with the chiral symmetry,
the color gauge symmetry and the hidden gauge symmetry, and so on.
In order to avoid  confusion, we
distinguish them by using different notations as
$D_\mu$, $D^c_\mu$, $\D_\mu$, etc.
} 
$D^c_\mu$ is
$$
D^c_\mu = \partial_\mu + ig_s A_\mu^a\lambda^a_{},
\seqno{LQCD}{a}$$
with the strong coupling constant $g_s$
and $m_q^{}$ is the light quark mass matrix
$$
m^{}_q =\left(\begin{array}{ccc} m_u^{} & 0 & \cdots \\
0 & m_d^{} & \cdots \\ \vdots & \vdots & \ddots \end{array}\right).
\seqno{LQCD}{b}$$
For simplicity, we will work with two light flavors hereafter.
The generalization to three flavors can be done straightforwardly.

Let $q_{R,L}^{}$ be the right- and left-handed quark fields defined by
\beqa{1.5}{l}
q_L^{}=\frac12(1-\gamma_5) q, \\
q_R^{}=\frac12(1+\gamma_5) q.
\eeqa
Then the Lagrangian density can be expressed as
$$
\L=-\ts\frac14\Tr G^{\mu\nu} G_{\mu\nu}
+ \bar{q}_L^{}i{D\hskip-2.2mm/}^c q_L^{}
+ \bar{q}_R^{}i{D\hskip-2.2mm/}^c q_R^{}
- (\bar{q}_L^{} m_q^{} q_R^{} + \bar{q}_R^{} m_q^{} q_L^{}).
$$
Without the mass term, the Lagrangian is invariant under independent
left- and right-trans\-for\-ma\-tions in the flavor space:
\beq
q_R^{} \rightarrow q^\prime_R = R q^{}_R, \hskip 5mm
q_L^{} \rightarrow q^\prime_L = L q^{}_L,
\eeqlb{LR}
where $R$ and $L$ are arbitrary constant $SU(2)$ matrices.
According to Noether's theorem, such an $SU(2)_L\times SU(2)_R$
chiral symmetry leads to the (classically) conserved
left- and right-vector currents:
\beq
\ts\partial^\mu (\bar{q}_R^{} \frac12\tau^i \gamma_\mu q_R^{})=0, \hskip 5mm
\partial^\mu (\bar{q}_L^{} \frac12\tau^i \gamma_\mu q_L^{})=0, \hskip 5mm
i=1,2,3,
\eeq
where the $\tau^i$'s are the generators of the $SU(2)$ flavor in the
fundamental representation. The corresponding $SU(2)$ conserved
charges are
\beq
Q^i_R = \int\!\!d^3r q^\dagger_R \ts\frac12\tau^i q_R^{}, \hskip 5mm
\dps Q^i_L = \int\!\!d^3r q^\dagger_L \ts\frac12\tau^i q_L^{}.
\eeq

Chiral $SU(2)_L\times SU(2)_R$ symmetry is a symmetry of the
Lagrangian density {\em but not of the vacuum}. The absence of the
parity doublets in the physical spectrum suggests that the
$SU(2)_L\times SU(2)_R$ is {\em spontaneously} broken to $SU(2)_V$
leaving the Goldstone bosons, $\pi^\pm_{}$ and $\pi^0_{}$, associated
with the broken generators. In other words, the QCD vacuum can have
nonvanishing expectation value of the quark bilinear
$\bar{q}_{aR}^{} q_{bL}^{}$
\beq
\langle 0|\bar{q}_{aR}^{} q_{bL}^{} |0\rangle = \upsilon\delta_{ab},
\eeq
where $a$, $b$(=$u$, $d$) are flavor indices.
Under the $SU(2)_L\times SU(2)_R$ transformation, we see that
$$
\langle 0| \bar{q}_{aR}^{} q_{bL}^{} |0\rangle
\rightarrow \langle 0|\bar{q}_{cR}^{}R^*_{ac} L_{bd}^{} q_{dL}^{}
|0\rangle
=R^\dagger_{ca}L_{bd}^{} \langle 0|\bar{q}_{cR}^{} q_{dL}^{}
|0\rangle
=R^\dagger_{ca}L_{bd}^{} \upsilon\delta_{cd}
=\upsilon (LR^\dagger)_{ba}, $$
that is, the vacuum is invariant only when the transformation
is restricted to the vector subgroup $SU(2)_V$ where $L=R$.

The quark mass term explicitly breaks the chiral
$SU(2)_L\times SU(2)_R$ symmetry, giving the Goldstone bosons
small masses. Quark masses can be treated as perturbation since the
up and down current quark masses ($m^{}_u\approx 5$ MeV and
$m_d^{}\approx 10$ MeV)   are small compared with the
QCD scale, $\Lambda_{QCD}\sim 200$ MeV, or the chiral symmetry scale
$\Lambda_{\chi}\sim 1$ GeV.

The strong interactions at low energy can be described rigorously
in terms of effective chiral Lagrangians involving Goldstone
bosons. A powerful approach to this is chiral perturbation theory
($\chi PT$) which consists of systematic expansions in power of
derivatives and the quark mass matrix.
As is customarily done in the literature, we shall
work with the $SU(2)_L\times SU(2)_R$ chiral symmetry realized
{\em nonlinearly}. For this, we define the chiral field
\beq
U=e^{iM/f_\pi},
\eeqlb{U}
where $M$ is a $2\times 2$ matrix for the triplet of Goldstone bosons
($\pi^+$, $\pi^0$ and $\pi^-$):
$$
M = \vec\tau\!\cdot\!\vec\pi=\left(\begin{array}{cc}\!\!\pi^0\!\!&
\!\!\!\sqrt2\pi^+\!\!\\ \!\!\sqrt2\pi^-\!\!\!&\!\!-\pi^0
\!\!\end{array}\right),
\seqno{U}{a}$$
and
$f_\pi$ is the pion decay constant
$$
f_\pi = 93 \mbox{ MeV}.
\seqno{U}{b}$$
Under $SU(2)_L\times SU(2)_R$, $U$ transforms
$$U\rightarrow U^\prime=LUR^\dagger, \eqno(\mbox{\ref{U}c})$$
where $L\in SU(2)_L$ and $R\in SU(2)_R$ are global transformation.
Then the leading-order interactions of the Goldstone bosons
are described by a single parameter, the pion decay constant,
\beq
\L_M = \frac{f_\pi^2}4 \Tr(\partial_\mu U^\dagger\partial^\mu U)
+ \cdots ,
\eeqlb{LM}
where terms with higher derivatives are abbreviated by the ellipsis.
The factor $f^2_\pi/4$ gives a properly
normalized kinetic terms for the pions:
$$
\L_M = \frac12 \partial_\mu\pi^a\partial^\mu\pi^a + \cdots.
$$

Witten\cite{Wi} has observed that
the non-linear $\sigma$-model Lagrangian (\ref{LM})
exhibits two discrete symmetries
\beq
\mbox{(i)}\hskip 2mm U(\vec{r},t)\rightarrow U(-\vec{r},t), \hskip 5mm
\mbox{(ii)}\hskip 2mm U(\vec{r},t)\rightarrow U^\dagger(\vec{r},t),
\eeq
whereas QCD requires only the invariance of the
Lagrangian under $U(\vec{r},t)\rightarrow U^\dagger(-\vec{r},t)$,
that is, the combination of (i) and (ii).
The second symmetry prohibits processes that involve $n$-pseudoscalar
vertex where $n$ is an odd number,
{\eg}, $K^+ K^-\rightarrow \pi^+\pi^-\pi^0$ which is allowed
by QCD. The latter is encoded in an anomalous term, known as
 ``Wess-Zumino term"\cite{WZ},
which breaks (i) and (ii) separately but preserves their combination.
The Wess-Zumino term cannot be written as a local Lagrangian
density in $(3+1)$ dimensions but can be as a local action in
five-dimensions\cite{Wi},
\beq
\Gamma_{WZ}=-\frac{iN_c}{240\pi^2}\int_{M_5}\!\!d^5x
\ve^{\mu\nu\rho\sigma\lambda}\Tr(U^\dagger\partial_\mu U
U^\dagger\partial_\nu U U^\dagger \partial_\rho U U^\dagger
\partial_\sigma U U^\dagger\partial_\lambda U),
\eeqlb{LWZ}
where the integration is over a five-dimensional disk whose boundary is
the ordinary space-time $M_4$ and $U$ is extended so that $U(\vec{r},t,s=0)
=1$ and $U(\vec{r},t,s=1)=U(\vec{r},t)$.  This term is non-vanishing for
$N_f\geq 3$. When the soliton is built in $SU(2)$ space, this
term does not figure directly. However we shall be considering (2+1) flavors
where one flavor can be heavy in which case the dynamics
can be influenced by the Wess-Zumino term as in the Callan-Klebanov model.

Consider a heavy meson containing a heavy quark $Q$ and
a light anti-quark $\bar q$.
We will assume that the light anti-quark in the heavy meson
forms a point-like object with the heavy quark, just providing them
a color, flavor, spin and parity.
Let $P$ and $P^*_\mu$ be the operators that annihilate $J^P=0^-$ and
$1^-$ mesons, respectively. If, for example, the heavy quark is charmed,
these fields form an $SU(2)$ anti-doublets:
\beq
P=(D^0, D^+) \mbox{\hskip 5mm and \hskip 5mm} P^*=(D^{*0}, D^{*+}).
\eeq
Their conventional free field Lagrangian density is given by
\beq
{\cal L}=\partial_\mu P \partial^\mu P^\dagger-m^2_PPP^\dagger
-\frac12P^{*\mu\nu}P^{*\dagger}_{\mu\nu}+m^2_{P^*}P^{*\mu}P^{*\dagger}_\mu,
\eeq
where $P^*_{\mu\nu}=\partial_\mu P^*_\nu - \partial_\nu P^*_\mu$
is the field strength tensor of the heavy vector meson fields $P^*_\mu$,
and $m_P^{}$ and $m^{}_{P^*}$ are the masses of the heavy
pseudoscalar and vector mesons, respectively.

In order to construct a chirally invariant Lagrangian containing
$P$, $P_\mu^*$ and their coupling to the Goldstone bosons,
we need to assign to the heavy meson fields a transformation rule
with respect to the full chiral symmetry group $SU(2)_L\times SU(2)_R$.
There is a considerable freedom for doing this. The standard one
is to introduce
\beq
\xi=U^{\frac12},
\eeqlb{xi}
which transforms under $SU(2)_L\times SU(2)_R$ as
\beq
\xi \rightarrow \xi^\prime = L \xi \vartheta^\dagger
= \vartheta \xi R^\dagger,
\eeqlb{Xct}
where $\vartheta$ is a local
unitary matrix depending on $L$, $R$ and the Goldstone
fields $M(x)$. From $\xi$ we can construct an ``induced"
vector field $V_\mu$ and an ``induced" axial vector field $A_\mu$ as
\beqa{1.5}{l}
V_\mu=\frac12(\xi^\dagger\partial_\mu\xi+\xi\partial_\mu\xi^\dagger),\\
A_\mu=\frac{i}2(\xi^\dagger\partial_\mu\xi-\xi\partial_\mu\xi^\dagger),
\eeqa
which transform under chiral symmetry
\beqa{1.5}{l}
V_\mu \rightarrow V^\prime_\mu=\vartheta V_\mu\vartheta^\dagger
+\vartheta\partial_\mu\vartheta^\dagger,\\
A_\mu \rightarrow A^\prime_\mu=\vartheta A_\mu\vartheta^\dagger.
\eeqalb{AVct}
The vector field $V_\mu$ behaves as a gauge field under the local chiral
transformation, while the axial vector field transforms covariantly.
Since the light quark doublet $q$ transforms\cite{MG,georgi}
$$
\ts q = \left(\begin{array}{c}\!\!u\!\!\\\!\!d\!\!\end{array}\right)
\rightarrow q^\prime=\vartheta q,
$$
the heavy meson anti-doublets, $P$ and $P^*$ whose quark contents are
$Q\bar{q}$, transform as
\beq
P\rightarrow P^\prime=P\vartheta^\dagger, \and
P^*\rightarrow P^{*\prime}=P^*\vartheta^\dagger.
\eeqlb{Pct}

We next define a covariant derivative in terms of the vector field $V_\mu$
\beqa{1.5}{l}
D_\mu P^\dagger=(\partial_\mu+V_\mu)P^\dagger,\\
D_\mu P\equiv (D_\mu P^\dagger)^\dagger
=P(\stackrel{\leftarrow}{\partial}_\mu+V^\dagger_\mu),
\eeqa
which transforms as
$$
D_\mu P\rightarrow (D_\mu P)^\prime=(D_\mu P)\theta^\dagger.
$$
We can also construct similar equations for the vector meson fields
$P_\mu^*$.

Given the above definitions, it is an easy matter
to write down the chirally invariant Lagrangian for
$P$ and $P_\mu^*$ with couplings to the Goldstone bosons.
In terms of derivatives acting on the Goldstone boson fields, it has
the form \cite{Yan}
\beqa{1.5}{l}
{\cal L}={\cal L}_M+D_\mu PD^\mu P^\dagger-m^2_PPP^\dagger
-\frac12P^{*\mu\nu}P^{*\dagger}_{\mu\nu}+m^2_{P^*}P^{*\mu}P^{*\dagger}_\mu\\
\hskip 1cm +\fQ(PA^\mu P^{*\dagger}_\mu+P^*_\mu A^\mu P^\dagger)
+\frac12\gQ\ve^{\mu\nu\lambda\rho}(P^*_{\mu\nu}A_\lambda P^{*\dagger}_\rho
+P^*_\rho A_\lambda P^{*\dagger}_{\mu\nu}),
\eeqalb{Yan}
where $\fQ$ and $\gQ$ are the $P^*PM$ and $P^*P^*M$ coupling constants,
respectively, and the field strength tensor is
\beq
P^*_{\mu\nu}=D_\mu P^*_\nu-D_\nu P^*_\mu.
\eeq
Note that the Lagrangian contains the $P^*PM$ and $P^*P^*M$ couplings
but no $PPM$ coupling that would violate parity invariance.

One can also work with heavy-meson fields defined differently.
For instance, a suitable pair of heavy-meson fields are
$\hat{P}$ and $\hat{P}^*_\mu$ which transform as $(\bar{2}_L^{}, 1^{}_R)$
under $SU(2)_L\times SU(2)_R$; {\it viz.},
\beq
\hat{P} \rightarrow \hat{P}^\prime = \hat{P} L^\dagger, \and
\hat{P}^*_\mu \rightarrow \hat{P}^{*\prime}_\mu = \hat{P}^*_\mu L^\dagger,
\eeq
with $L\in SU(2)_L$.
These fields may appear simpler than those of Eq.~(\ref{Pct}), but their
transformation under parity is a bit more complicated.
Since parity interchanges left- and right-handed quark fields,
the parity image of the hatted heavy meson fields
must transform under $SU(2)_L\times SU(2)_R$ as $(1_L,\bar{2}_R)$;
{\it i.e.\/},
\beq
{\cal P}\hat{P}(\vec{r},t){\cal P}^{-1}
= - \hat{P}(-\vec{r},t)U(-\vec{r},t)
\stackrel{SU(2)_L\times SU(2)_R}{\longrightarrow}
- \hat{P}(-\vec{r},t)U(-\vec{r},t) R^\dagger,
\eeq
and similarly for the vector meson fields. In contrast, the unhatted fields
transform simply under parity (taking into account that the heavy meson
has negative intrinsic parity)
\beq
{\cal P} P(\vec r,t){\cal P}^{-1} = - P(-\vec{r},t).
\eeqlb{Parity}
One can easily verify that
the hatted fields are related to the unhatted ones
by
\beq
\hat{P} = P \xi^\dagger  \and
\hat{P}^*_\mu = P^*_\mu \xi^\dagger.
\eeq
The Lagrangian Eq.~(\ref{Yan}) rewritten in terms of the hatted fields
is
\beqa{1.5}{l}
\L = \L_M + \hat{D}_\mu \hat{P} \hat{D}^\mu \hat{P}^\dagger
 - m^2_P \hat{P} \hat{P}^\dagger
-\frac12 \hat{P}^{*\mu\nu} \hat{P}^{*\dagger}_{\mu\nu}
 + m^2_{P^*} \hat{P}^{*\mu} \hat{P}^{*\dagger}_\mu \\
\hskip 1cm + \frac{i}{2} \fQ
  (\hat{P} U^\dagger \partial_\mu U \hat{P}^{*\dagger}_\mu
 + \hat{P}^*_\mu U^\dagger \partial_\mu U \hat{P}^\dagger) \\
\hskip 1cm + \frac{i}{4} \gQ \ve^{\mu\nu\lambda\rho}
(\hat{P}^*_{\mu\nu} U^\dagger \partial_\lambda U \hat{P}^{*\dagger}_\rho
 + \hat{P}^*_\rho U^\dagger \partial_\lambda U \hat{P}^{*\dagger}_{\mu\nu}),
\eeqalb{Lhat}
where the ``hatted" covariant derivatives denote
$$\begin{array}{l}
\hat{D}_\mu \hat{P} = \partial_\mu \hat{P}
 + \hat{P} \frac12\partial_\mu U^\dagger U, \\
\hat{D}_\mu \hat{P}^*_\nu = \partial_\mu \hat{P}^*_\nu
+ \hat{P}^*_\nu \frac12\partial_\mu U^\dagger U.
\end{array}\seqno{Lhat}{a}$$
Note that the vector and axial vector currents
of Eq. (\ref{Yan}) do not appear in the Lagrangian (\ref{Lhat}).
There appears only the term $U^\dagger \partial_\mu U$
which transforms as an $SU(2)_L$ triplet.

\subsection{Heavy Quark Symmetry}
In the heavy quark limit ($m^{}_Q\gg \Lambda_{QCD}$), the subleading
terms in $1/m^{}_Q$ are suppressed and the heavy-quark symmetry becomes
manifest in the Lagrangian: the heavy pseudoscalar
and vector mesons become degenerate.  This can be easily understood
in quark models. In the constituent quark model,
the $P$-$P^*$ mass difference is due to the hyperfine splitting
generated by the one-gluon exchange between constituent quarks\cite{QM},
\beq
m_{P^*}^{}-m_P^{}=\frac{\kappa}{m_Qm_q},
\eeq
where $m_Q (m_q)$ is the heavy (light)-quark mass and $\kappa$ is
a constant depending on the wavefunction of the hadron.
If $m_Q$ is sufficiently large compared with $\Lambda_{QCD}$, the wavefunction
of the heavy-quark--light-anti-quark(s) bound state does not depend on the
heavy-quark mass and the constant $\kappa$ approaches an asymptotic
value $\kappa_\infty$. Although experimental information is at the
moment severely limited, the $D^*$-$D$ mass difference\cite{PDG} of $\sim 145$
MeV
and the $B^*$-$B$ splitting\cite{CUSB85} of $\sim 50$ MeV
are consistent with the expected $1/m_Q$ behavior.

As the heavy-quark mass goes to infinity,
the heavy-quark spin decouples from the rest of the strongly
interacting light-quark system, namely the ``brown muck,"
since their coupling is a relativistic effect of order $1/m_Q^{}$.
This is referred to as ``heavy-quark spin symmetry."
Furthermore, in that limit, the structure of the brown
muck in the heavy meson is independent of the heavy-quark flavor.
This is referred to  as ``heavy-quark flavor symmetry."
An analogy is found in the excitation spectrum
and transition matrix element of a hydrogen-like atom, both of
which are independent of the mass and spin of the nucleus.
The heavy-quark spin symmetry relates two coupling constants
$\fQ$ and $\gQ$ in the Lagrangian (\ref{Yan})
and the heavy-quark flavor symmetry gives their dependence
on the heavy-quark masses.
The Lagrangian (\ref{Yan}) and the axial current
\beq
A_\mu = -f_\pi \vec\tau\!\cdot\!\partial_\mu\vec\pi+\cdots,
\eeqlb{Amu}
determine, at tree order, the matrix elements for the emission of a soft pion,
\beqa{1.5}{l}
M(P^*(\ve)\rightarrow P+\pi^a(q))=(\frac12\phi^\dagger(P)\tau^a \phi(P^*))
\dps\frac{\fQ}{f_\pi}(\ve\cdot q), \\
M(P^*(\ve)\rightarrow P^*(\ve^\prime)+\pi^a(q))
=(\frac12\phi^\dagger(P^*)\tau^a\phi(P^*))\dps\frac{2\gQ}{f_\pi}(-i)
\ve^{\mu\nu\lambda\rho}p_\mu\ve_\nu^{\prime*} q_\lambda \ve_\rho,
\eeqalb{M1}
where $\phi^\dagger$ is the isospin wavefunction of the heavy meson
anti-doublets, $\ve$ the polarization vector of $P^*$,
and $p_\mu(q_\mu)$  the momentum of the heavy mesons (soft pion).
We have used that $p_\mu\approx p_\mu^\prime$ in the heavy-quark limit.
Now PCAC implies
\beqa{1.7}{l}
M(P^*\rightarrow P+\pi^a(q))_{PCAC}=\dps\frac{1}{f_\pi}
\langle P|q^\mu A_\mu^a|P^*\rangle,\\
M(P^*\rightarrow P^*+\pi^a(q))_{PCAC}=\dps\frac{1}{f_\pi}
\langle P^*|q^\mu A_\mu^a|P^*\rangle.
\eeqalb{PCAC}

Due to the heavy-quark spin decoupling, the Hilbert space of the
four states, the $0^-$-$|P\rangle$ state and the 3 spin states of
the $1^-$-$|P^*\rangle$, can be conveniently represented
in a tensor product notation:\cite{Ge92}
\beq
\ts |P, \pm\frac12, \pm\frac12 \rangle,
\eeqlb{tpn}
where the first $\pm\frac12$ is the third component of the heavy-quark
spin $s_h$, and the second $\pm\frac12$ is that of the angular
momentum of the light degrees of freedom $s_m$. In this notation,
the meson states are written as
\beqa{1.5}{c}
|P^*,+1\rangle=\sqrt{2m^{}_P}|P,+\frac12,+\frac12\rangle,\\
|P^*,0\rangle=\sqrt{2m^{}_P}\frac1{\sqrt2}(|P,+\frac12,-\frac12\rangle
+|P,-\frac12,+\frac12\rangle),\\
|P^*,-1\rangle=\sqrt{2m^{}_P}|P,-\frac12,-\frac12\rangle,\\
|P\rangle=\sqrt{2m^{}_P}\frac1{\sqrt2}(|P,+\frac12,-\frac12\rangle
-|P,-\frac12,+\frac12\rangle).
\eeqalb{tpn2}
The relative phase between the $P^*$ and the $P$ states is arbitrary,
but that between the three $P^*$ states
is fixed by the angular momentum structure. Then, the
$|P,\pm\frac12,\pm\frac12\rangle$ state can be approximated by a
product of the free heavy-quark state and a complicated brown-muck state:
\beq
|P,s_h,s_m\rangle\approx|h,s_h\rangle|\muck,i_m,s_m\rangle,
\eeqlb{Fac}
with $i_m$ the isospin of the brown muck associated with the light anti-quark.
In the heavy-quark limit, this factorization becomes exact with the
brown-muck state $|\muck$, $i_m$, $s_m\rangle$ remaining the same for all
states of Eq.~(\ref{tpn}) independently of the heavy quark spin $s_h$ and
flavor $h$.

Evaluating the matrix elements of the right-hand-side of
Eq.~(\ref{PCAC}) gives
$$\begin{array}{l}
\langle P|q^\mu A_\mu^a|P^*,0\rangle=m_P^{}(
\langle\muck,i^\prime_m,+\frac12|q^\mu A_\mu|\muck,i_m,+\frac12\rangle\\
\hskip 4cm
-\langle\muck,i^\prime_m,-\frac12|q^\mu A_\mu|\muck,i_m,-\frac12\rangle),\\
\langle P^*,+1|q^\mu A_\mu^a|P^*,+1\rangle=2m_P^{}
\langle\muck,i^\prime_m,+\frac12|q^\mu A_\mu|\muck,i_m,+\frac12\rangle.
\end{array}$$
With Wigner-Eckart theorem, the spin and isospin structure of
$A_\mu$ given by Eq.~(\ref{Amu}) leads to
\beq
\langle\muck,i_m^\prime,s_m^\prime|q^\mu A_\mu^a|\muck,i_m,s_m\rangle
=\alpha(\ts\frac12\phi^{\prime\dagger}\tau^a\phi)
\langle s_m^\prime|S^\mu|s_m\rangle q_\mu,
\eeqlb{WE}
where $S^\mu$ is the spin operator of the brown muck and
$\alpha$ is a constant independent of the heavy-quark mass.
Thus
\beqa{1.5}{l}
\dps\frac1{f_\pi}\langle P|q^\mu A_\mu^a|P^*,0\rangle
=\frac1{f_\pi}\alpha m^{}_P(\ts\frac12\phi^\dagger(P)\tau^a\phi(P^*))q_3^{},\\
\dps\frac1{f_\pi}\langle P^*\!\!,+1|q^\mu A_\mu^a|P^*\!\!,+1\rangle
=\frac1{f_\pi}\alpha m_P^{}(\ts\frac12\phi^\dagger(P)\tau^a\phi(P^*))q_3^{}.
\eeqalb{M2}
Comparing Eq.~(\ref{M2}) with Eq.~(\ref{M1}) in the rest frame of $P^*$
where the polarization vectors for the states $|P^*,+1\rangle$ and
$|P^*,0\rangle$, respectively, are $\ve(+1)=\frac{1}{\sqrt2}(0,1,+i,0)$ and
$\ve(0)=(0,0,0,1)$, we find
\beq
\ts\fQ=\alpha m^{}_P  \and \gQ=\frac12\alpha,
\eeqlb{fgQ}
that is,
$$\gQ=\frac{\fQ}{2m^{}_P}.$$
Furthermore, Eq.~(\ref{fgQ}) gives us a heavy-quark dependence of
$\fQ$ and $\gQ$:
\beq
\fQ=2m^{}_Pg  \and \gQ=g,
\eeqlb{fQgQ}
with a universal constant $g$ independent of the heavy-quark flavor.
The coupling constant $g$ can be evaluated using a nonrelativistic quark
model (NRQM). In NRQM, the factorization (\ref{Fac}) becomes
much simpler:\cite{Yan}
$$
|P,s_Q,s_\ell\rangle\!\rangle = |Q_{s_Q}\bar{q}_{s_\ell}\rangle,
$$
with $s_Q$ and $s_\ell$ the spins of the heavy quark and the light
anti-quark. (Here, $|\mbox{ }\rangle\!\rangle$ is to specify that
it is a state in the nonrelativistic quark model.)
For example, $|P\rangle$ and $|P^*,0\rangle$ appearing in
Eq.~(\ref{M2}) can be written as
\beqa{1.5}{c}
|P^*_{+\frac12},0\rangle\!\rangle=\sqrt{2m_P}\frac{1}{\sqrt2}
[|Q_{\uparrow}\bar{d}_{\downarrow}\rangle
+|Q_{\downarrow}\bar{d}_{\uparrow}\rangle ],\\
|P_{-\frac12}\rangle\!\rangle=\sqrt{2m_P}\frac{1}{\sqrt2}
[|Q_{\uparrow}\bar{u}_{\downarrow}\rangle
-|Q_{\downarrow}\bar{u}_{\uparrow}\rangle ],
\eeqa
where the arrows represent the quark spin and the subscripts
$\pm\frac12$ the third component of the isospin.
Since the axial current $A_\mu$ is defined in terms of the
light quark doublet $q$ as
$$
A_i^a\equiv\ts g_A^{ud}\frac12\bar{q}\gamma_i\gamma_5\tau^a q=\left\{
\begin{array}{ll}
g_A^{ud}u^\dagger\sigma_id &\mbox{ if $a=1+i2$}, \\
\frac1{\sqrt2}g^{ud}_A(u^\dagger\sigma_i u-d^\dagger\sigma_id)
&\mbox{ if $a=3$},\\
g_A^{ud}d^\dagger\sigma_i u &\mbox{ if $a=1-i2$},
\end{array}\right.
$$
with the axial vector coupling constant $g_A^{ud}$ of the quark\cite{MG},
the matrix element in the left-hand-side of Eq.~(\ref{M2}) can be
simply evaluated as
\beqa{1.5}{l}
\langle\!\langle P^*_{+\frac12}|A^{1-i2}_3|P_{-\frac12}\rangle\!\rangle\\
\hskip 5mm =m_P^{}g^{ud}_A\{\langle\bar{u}_{\downarrow}|d^\dagger\sigma_3 u
|\bar{d}_{\downarrow}\rangle-\langle\bar{u}_{\uparrow}|d^\dagger\sigma_3 u
|\bar{d}_{\uparrow}\rangle\}\\
\hskip 5mm=2m_P^{}g^{ud}_A,
\eeqa
which implies that $g$=$-g^{ud}_A$ [$\alpha$=$-2$ in Eq.~(\ref{WE})].
A similar calculation leads to the matrix element of the axial
vector current between the nucleon states:
\beq
\langle\!\langle N^\prime,S_3^\prime|A^a_i|N,S_3\rangle\!\rangle
=g_A^N\psi^\dagger_{N^\prime,S_3^\prime}\tau^a\sigma_i\psi_{N,S_3},
\eeq
where $\psi_{N,S_3}$ is the nucleon state ($N$=$p,n$) with  spin
$S_3$(=$\pm\frac12$). Therefore we get the familiar NRQM formula
for the nucleon axial-vector coupling constant $g_A^N$
\beq
g_A^N=\ts\frac53 g^{ud}_A.
\eeq
The experimental value $g^N_A$=1.25 determines the value for $g^{ud}_A$
\beq
g=-g_A^{ud}=-0.75.
\eeq
Higher-order corrections in $1/N_c$ to $g^{ud}_A$ are discussed in
Ref.~\citenum{Wei}.

In the case of $Q$(or $h$)=$c$, the matrix elements (\ref{M1}) determine
the decay width
\beq
\Gamma(D^{*+}_{}\rightarrow D^0_{}\pi^+_{})
 = \frac{1}{12\pi}\frac{g^2}{f^6_\pi}|\vec p_\pi|^3.
\eeq
The width for $D^{*+}_{}\rightarrow D^+_{}\pi^0_{}$
is reduced by a factor of two due to isospin symmetry.
The experimental upper limit\cite{ACCMOR} of 131 KeV on the
$D^*$ width when combined with the $D^{*+}_{}\rightarrow D^{+}_{}\pi^0_{}$
and $D^{*+}_{}\rightarrow D^0_{}\pi^+_{}$ branching ratios\cite{CLEO}
implies that
$|g|^2$\raisebox{-0.6ex}{$\stackrel{\ts <}{\sim}$} 0.5.
The nonrelativistic quark model prediction on $g$ is slightly bigger
than the experimental upper limit but is  consistent with this value.

\setcounter{equation}{0}
\section{Skyrme Model and Hidden Symmetry: ``Bottom-Up" Approach}
Thus far, we have discussed the heavy-quark chiral Lagrangian
in the heavy-quark symmetry limit with $m_Q\rightarrow\infty$.
Restricting ourselves to two massless flavors and one heavy flavor,
the leading order Lagrangian is constructed by taking into account
both $SU(2)\times SU(2)$ chiral symmetry and the heavy-quark symmetry.
For the actual hadrons with finite heavy-quark mass as relevant to
the c and b quarks, we need to include the deviation from the symmetry
limit which starts with order $1/m_Q$ corrections. In this section,
we approach the heavy-quark regime from ``below."

The Skyrme model\cite{Sk} describes a baryon as a soliton solution
of a nonlinear chiral Lagrangian of weakly interacting Goldstone
bosons. Since an exact bosonization of fermionic theories has not
yet been found in (3+1) dimensions, such a bosonic Lagrangian describing
QCD does not exist except in the limit that the number of colors is infinite.
Nonetheless the soliton approach based on approximate effective Lagrangians
has enjoyed a great success ranging from the static properties of the
baryons to the inter-nucleon interactions. (For review, see
Refs.~\citenum{CHT,ZB,Liu,Mei,SWHH,NR,elafmr}.)  We wish to show below that
this model, with a minimal complication, can provide an amazingly simple
way of constructing such a heavy meson effective Lagrangian by starting
from the $SU(3)_L\!\times\!SU(3)_R$\footnote{Since we are interested
in the hadron system which contains one heavy flavor, it is not
necessarily the conventional $SU(3)$ with $u$, $d$ and $s$ flavors.
It could be generalized to any $SU(3)$ subgroup of the full
$SU(N_f)(N_f\geq 3)$ associated with the $u$, $d$ and $h$ flavor of our
interest.} chiral limit and then climbing up, {\it \`a la\/} CK approach,
to the massive system with the symmetry breaking to $SU(2)\times U(1)$.

Suppose that we start with three massless quarks, assuming the spontaneous
breaking of chiral $SU(3)_L\times SU(3)_R$ down to the $SU(3)_V$ vector
symmetry. We write the chiral field as
\beq
U=\exp(\frac{i}{f_\pi}\sum_{a=1}^8\lambda_a\pi_a)=e^{iM/f_\pi},
\eeqlb{U3}
where $\lambda_a(a$=1,2,$\cdots$,8) is the Gell-Mann matrices for
flavor $SU(3)$ and $M$ is the $SU(3)$-valued meson field:
$$
 M = \left[\begin{array}{ccc}
\!\!\pi^0+\frac1{\sqrt3}\Phi\!\!&\!\!\sqrt2\pi^+\!\!&\!\!P^+\!\!\\
\!\!\sqrt2\pi^-\!\!&\!\!-\pi^0+\frac1{\sqrt3}\Phi\!\!&\!\!P^0\!\!\\
\!\!P^-\!\!&\!\!\bar{P}^0\!\!&\!\!-\frac{2}{\sqrt3}\Phi\!\!
\end{array}\right].
\seqno{U3}{a}$$
Here, $P^+$, $P^0$, $P^-$, $\bar{P}^0$ and $\Phi$ denote the mesons
with the quantum numbers of $\bar{h}\gamma_5 u$, $\bar{h}\gamma_5 d$,
$\bar{u}\gamma_5 h$ and $\bar{d}\gamma_5 h$ and
$\bar{u}\gamma_5u+\bar{d}\gamma_5d-2\bar{h}\gamma_5h$, respectively.
For example, if $h$=$s$, they correspond to $K^+$, $K^0$, $K^-$,
$\bar{K}^0$ and $\eta_8$. The Lagrangian for interactions among
the Goldstone bosons is given by generalizing Eq.~(\ref{LM}) to three
flavors, where the Wess-Zumino term figures crucially:
\beq
{\cal L}=\frac{f_\pi^2}{4}\Tr(\partial_\mu U^\dagger\partial^\mu U)
+\cdots+{\cal L}_{WZ}.
\eeqlb{Lsu3}

What we are interested in is the situation where the symmetry
$SU(3)_L\!\times\!SU(3)_R$ is {\it explicitly} broken to
$SU(2)_L\!\times\!SU(2)_R\!\times\!U(1)$ by an $h$-quark
mass\footnote{For simplicity, we turn off the light quark masses.},
thereby making the $P$-meson massive and its decay constant $f_P$
different from that of the pion. These two effects of symmetry
breaking can be effectively incorporated into the
Lagrangian\cite{PSW,GL} by a term of the form
\beqa{1.5}{l}
{\cal L}_{_{SB}}=\frac16 f_P^2m_P^2\Tr[(1-\sqrt3\lambda_8)(U+U^\dagger-2)]\\
\hskip 12mm+\frac1{12}(f_P^2-f_\pi^2)\Tr[(1-\sqrt3\lambda_8)
(U\partial_\mu U^\dagger\partial^\mu U
+U^\dagger\partial_\mu U\partial^\mu U^\dagger)].
\eeqalb{Lsb}
The appropriate ansatz for the chiral field is the
Callan-Klebanov (CK)-type which we shall take in the form
\beq
U=N_\pi U_P N_\pi,
\eeqlb{Uck}
where
$$
N_\pi = \exp(\frac{i}{2f_\pi}\sum_{a=1}^3\lambda_a\pi_a)
= \left[ \begin{array}{cc} \!\xi\!&\!0\! \\ \!0\!&\!1\!
\end{array}\right],
\seqno{Uck}{a}$$
$$
U_P = \exp(\frac{i}{f_\pi}\sum_{a=4}^7\lambda_a\pi_a)
=\exp(\frac{i\sqrt2}{f_\pi}\left[\begin{array}{cc}
\!\!{\bf 0}\!&\! P^\dagger\!\!\\\!\!P\!&\!0\!\!\end{array}\right]),
\seqno{Uck}{b}$$
with $SU(2)$ matrix $\xi$ defined by Eq.~(\ref{xi}), the $P$-meson
anti-doublets $P=(P^-\!\!,\bar{P}^0)$, and $P$-meson doublets
$P^\dagger=(P^+\!\!, P^0)^T$.
One can easily see that $\xi$ and $P$ transform exactly in the same way
as Eqs.~(\ref{Xct}) and (\ref{Pct}) under the embedded
$SU(2)_L\!\times\!SU(2)_R$ rigid chiral transformation
$$
 U \rightarrow U^\prime=\left[\begin{array}{cc}\!L\!&\!0\\
\!0\!&\!1\!\end{array}\right]U\left[\begin{array}{cc}\!R^\dagger\!&\!0\\
\!0\!&\!1\!\end{array}\right]
$$
with $L\in SU(2)_L$ and $R\in SU(2)_R$.

Substituting the CK ansatz (\ref{Uck}) into the Lagrangian (\ref{Lsu3})
with the symmetry breaking term (\ref{Lsb}) and expanding up to
second order in the $P$-meson field, we obtain
\beq
{\cal L}={\cal L}_M+D_\mu P D_\mu P^\dagger-M_P^2PP^\dagger
-PA_\mu^\dagger A^\mu P^\dagger
-\frac{iN_c}{4f_P^2}B_\mu(D^\mu PP^\dagger-PD^\mu P^\dagger),
\eeqlb{Lck}
where we have rescaled the $P$-meson fields as $P/\chi$ with $\chi =
f_P / f_\pi$.
The covariant derivative
$D_\mu P^\dagger$ is $(\partial_\mu+V_\mu)P^\dagger$, the
vector field $V_\mu$ and the axial-vector field $A_\mu$ are
the same as in the Lagrangian (\ref{Yan}), and $B_\mu$
is the topological current
\beq
B^\mu=\frac{1}{24\pi^2}\ve^{\mu\nu\lambda\rho}\Tr(U^\dagger\partial_\nu U
U^\dagger\partial_\lambda UU^\dagger\partial_\rho U),
\eeq
which is the baryon number current in the Skyrme model.
One can see that as far as the $P$-fields are concerned,
to the lowest order in derivative on the
Goldstone boson fields, Eq.~(\ref{Lck}) is the same as the Lagrangian
Eq.~(\ref{Yan}). Furthermore as argued by Nowak {\etal}\cite{NRZ1}, one expects
that as  the $h$ quark mass increases above the chiral scale $\Lambda_\chi$,
the Wess-Zumino term would vanish, thereby turning off the last term of
(\ref{Lck}). Thus
the two Lagrangians are indeed equivalent as far as the pseudoscalars are
concerned.

As mentioned above, going to heavy-quark systems requires the vector degrees of
freedom which become degenerate with the pseudoscalars in the infinite
quark mass limit. From a chiral Lagrangian point of view, the
vector mesons can be viewed as ``matter fields" and there are several ways
of introducing matter fields in general. If chiral symmetry is correctly
implemented, they are all equivalent in the sense that the $S$-matrix
is identical. When anomalies are involved, the situation is a bit
delicate but by now there is no conceptual
difficulty.\cite{Mei,GG,BKUYY,BKY,St,JJMPS}
Here we follow the hidden gauge symmetry (HGS) approach\cite{BKUYY,BKY}
which in our opinion offers psychologically the most powerful one.

The chiral field  $U$ in the Lagrangian (\ref{Lsu3}) transforms
$$
U \rightarrow U^\prime = L U R^\dagger \hskip 5mm
( L \in [ SU(3)_L ]_{\mbox{\scriptsize global}} \ \
R \in [ SU(3)_R ]_{\mbox{\scriptsize global}} ).
$$
The hidden gauge symmetry $SU(3)_V$ of the Lagrangian (\ref{Lsu3})
can be made apparent by rewriting $U$ in terms of two $SU(3)$ matrices
$\xi_{L,R}^{}(x)$ as
\beq
U(x)=\xi^\dagger_L(x)\cdot\xi^{}_R(x).
\eeq
The Lagrangian is invariant under
$[SU(3)_L\!\times\!SU(3)_R]_{\mbox{\scriptsize global}}\!\times
[SU(3)_V]_{\mbox{\scriptsize local}}$ transformations:
\beqa{1.5}{l}
\xi^{}_L(x) \rightarrow \xi_L^\prime(x) = h(x) \xi^{}_L(x) L^\dagger, \\
\xi^{}_R(x) \rightarrow \xi_R^\prime(x) = h(x) \xi^{}_R(x) R^\dagger,
\eeqa
with $h(x)\in[SU(3)_V]_{\mbox{\scriptsize local}}$.
The gauge connection associated with the $SU(3)_V$ local symmetry
can be written as
\beq
U_\mu=\frac12\left[\begin{array}{cc}
\!\omega_\mu+\rho_\mu\!\!&\!\!\sqrt2 P_\mu^{*\dagger}\!\\
\!\!\sqrt2 P_\mu^*\!\!&\!\!\Phi^*_\mu\!\end{array}\right].
\eeqlb{Umu}
It transforms as
$$U_\mu\rightarrow
U_\mu^\prime=h(x)U_\mu(x)h^\dagger(x)-\frac{i}{g}h(x)\partial_\mu h^\dagger(x).
\seqno{Umu}{a}$$
The covariant derivative relevant to the HGS is then
$$\D_\mu\xi^{}_{L,R}\equiv(\partial_\mu + ig_*U_\mu)\xi^{}_{L,R},
\seqno{Umu}{b}$$
with a gauge coupling constant $g_*$ to be specified later.
Up to second order in this covariant derivative, we can construct two
independent terms consistent with the
$[SU(3)_L\!\times\!SU(3)_R]_{\mbox{\scriptsize global}}\!\times
[SU(3)_V]_{\mbox{\scriptsize local}}$ symmetry and parity:
\beqa{1.5}{l}
{\cal L}_V\equiv-\frac14{f_\pi^2}\Tr[\D_\mu\xi^{}_L\xi^\dagger_L
+\D_\mu\xi^{}_R\xi^\dagger_R]^2,\\
{\cal L}_A\equiv-\frac14{f_\pi^2}\Tr[\D_\mu\xi^{}_L\xi^\dagger_L
-\D_\mu\xi^{}_R\xi^\dagger_R]^2.
\eeqa
In working at tree order, we might as well choose the unitary gauge
\beq
\xi^\dagger_L=\xi^{}_R\equiv\xi(x), \hskip 5mm \xi^2(x)=U(x).
\eeq
Then, we have
$$\begin{array}{l}
{\cal L}_A=-\frac14{f_\pi^2}\Tr[\partial_\mu\xi\xi^\dagger
-\partial_\mu\xi^\dagger\xi]^2=\frac14{f_\pi^2}\Tr(\partial_\mu
U^\dagger\partial^\mu U),\\
{\cal L}_V=f_\pi^2\Tr[g_* U_\mu-\frac12i(\partial_\mu\xi\xi^\dagger
+\partial_\mu\xi^\dagger\xi)]^2.
\end{array}$$
Thus ${\cal L}_A$ reduces to the original Lagrangian
while ${\cal L}_V$ vanishes identically with $U_\mu$ satisfying
the equation of motion
\beq
U_\mu=\frac{i}{2g_*}(\partial_\mu\xi\xi^\dagger+\partial_\mu\xi^\dagger\xi).
\eeqlb{Ueq}
Clearly nothing is gained by ``gauging" the hidden local symmetry.
The gauge field is just an auxiliary field. However the dynamics
changes dramatically if the gauge field becomes a propagating field
by acquiring a kinetic energy term. The kinetic term is higher order in
derivative and chirally invariant. Therefore a systematic chiral
expansion would naturally allow such a term. Were we to attempt to derive
a HGS Lagrangian from QCD by functional integration of the gluon and
quark fields in the presence of auxiliary vector fields, then one would
naturally encounter such a term from the fermion determinant. (This is
obvious by ``bosonizing" the Nambu--Jona-Lasinio (NJL) to an effective
bosonic Lagrangian with vector mesons implemented.)
The resulting Lagrangian is
\beq
{\cal L}_0={\cal L}_A+a{\cal L}_V-\ts\frac12\Tr(F_{\mu\nu}F^{\mu\nu}),
\eeqlb{L0}
with the field strength tensor of the vector mesons
$$F_{\mu\nu}=\partial_\mu U_\nu-\partial_\nu U_\mu+ig_*[U_\mu,U_\nu].$$
The vector meson mass $M_V$ and the
$\rho\pi\pi$ coupling constant can be read off from the Lagrangian,
\beqa{1.5}{l}
m_V^2=ag_*^2f_\pi^2,\\
g_{\rho\pi\pi}=\frac12ag_*.
\eeqa
With the KSRF relation\cite{KSRF}
\beq
m^2_\rho=2g_*^2f^2_\pi,
\eeq
and the universality of the vector-meson coupling
\beq
g_{\rho\pi\pi}=g_*,
\eeq
we can fix the arbitrary parameter $a$ to 2.

[We should parenthetically mention an interesting situation that arises
in the limit $M_V^2\rightarrow\infty$ with $g_*$ held fixed\cite{IJKOSS}.
In this limit, we have the constraint
$$
U_\mu = \frac{i}{2g_*} (\partial_\mu \xi^{}_L \xi^\dagger_L
+ \partial_\mu \xi^{}_R \xi^\dagger_R).
$$
When this constraint is substituted into the
field strength tensor $F_{\mu\nu}$, we get
$$\begin{array}{rl}
F_{\mu\nu}\!\!\!&=\dps\frac{1}{4ig_*}
[(\partial_\mu\xi^{}_L\xi_L^\dagger-\partial_\mu\xi_R^{}\xi^\dagger_R),
(\partial_\nu\xi^{}_L\xi_L^\dagger-\partial_\nu\xi_R^{}\xi^\dagger_R)]\\
&=\dps\frac{i}{4g_*}\xi^{}_L[\partial_\mu UU^\dagger,
\partial_\nu UU^\dagger]\xi^\dagger_L.
\end{array}$$
Thus, the kinetic term for the vector meson becomes the Skyrme term
in the Skyrme Lagrangian
\beq
{\cal L}^{\mbox{\small kin}}_V\stackrel{m^2_V\rightarrow\infty}
{\longrightarrow}{\cal L}_{Sk}=+\frac{1}{32g_*^2}\Tr[\partial_\mu UU^\dagger,
\partial_\nu UU^\dagger]^2,
\eeqlb{SkTerm}
with $e$=$g_*$ the Skyrme parameter\cite{Sk}.

There are two remarkable
features associated with this term. First it renders the soliton stable
against the Hobard-Derrick collapse whereas
vector mesons for {\it any} finite vector-meson mass cannot.
This means that the stabilization occurs only in the limit that the mass is
infinite. The second observation is that
when the energy density of a dense hadronic matter is computed for a
multi-skyrmion system at asymptotic
density, it is found to be identical to that of the free quark gas.
This implies
that the quartic Skyrme term possesses
short-distance physics described by asymptotically free quarks.
This may be the reason why the Skyrme quartic term can stabilize the
soliton while finite mass vectors cannot.]

The effective action should satisfy the same anomalous Ward identities
as does the underlying fundamental theory QCD.\cite{WZ} In the presence
of vector mesons, $A^\mu_{L,R}$ associated with the external gauge
transformation $U(x)\rightarrow e^{i\ve_L}U(x)e^{-i\ve_R}$ and $U_\mu$
with the hidden gauge transformation discussed so far, the Wess-Zumino
anomaly equation reads
\beq
\delta\Gamma(\xi_L,\xi_R,A_H,A_L,A_R)=-\frac{N_c}{24\pi^2}
\int_{M_4}\!\Tr[\ve_L^{}\{(dA_L^{})^2-\ts\frac12idA_L^3\}-\{L\leftrightarrow
R\}],
\eeqlb{WZeq}
where the gauge transformation $\delta$ is
$$
\delta=\delta_L(\ve_L)+\delta_R(\ve_R)+\delta_H(h).
\seqno{WZeq}{a}$$
Here the subscript $H$ stands for hidden gauge symmetry.
For convenience, we use the differential one-form notations:
$$\begin{array}{l}
A_H\equiv U_\mu dx^\mu,\\
A_{L,R}\equiv A^\mu_{L,R}dx_\mu.
\end{array}
\seqno{WZeq}{b}$$
The general solution to Eq.~(\ref{WZeq}) is given by a special solution
plus general solutions of the homogeneous equation $\delta\Gamma=0$.
The former is the Wess-Zumino action (\ref{LWZ}) (see Ref.~\citenum{Mei,BKY}
for details) and the latter, the anomaly free terms, can be made of
gauge-covariant building blocks. There are six independent forms\cite{FKTUY}
that
conserve parity but violate intrinsic parity\footnote{The intrinsic parity of
a particle is defined to be even if its parity equals
$(-1)^{\mbox{\tiny spin}}$. Considering $C$-parity, $\L_3$ and
$\L_5$ may be discarded and ${\cal L}_6$ should read $\frac{i}{2} \Tr \{
\hat F_L [ \hat L, \hat R] - \hat F_R [ \hat R, \hat L] \}$. See
Ref.~\citenum{Mei} for details.}:
\beqa{1.5}{ll}
{\cal L}_1=\Tr[\hat{L}^3\hat{R}-\hat{R}^3\hat{L}], &
{\cal L}_2=\Tr[\hat{L}\hat{R}\hat{L}\hat{R}],\\
{\cal L}_3=ig\Tr[F_H(\hat{L}^2-\hat{R}^2)],&
{\cal L}_4=ig\Tr[F_H(\hat{L}\hat{R}-\hat{R}\hat{R})],\\
{\cal L}_5=i\Tr[\hat{F}_L\hat{R}^2-\hat{F}_R\hat{L}^2], &
{\cal L}_6=i\Tr[\hat{F}_L\hat{L}\hat{R}-\hat{F}_R\hat{R}\hat{L}],
\eeqa
with the differential one-forms defined as
$$\begin{array}{l}
\hat{L},\hat{R}\equiv\D\xi^{}_{L,R}\xi^\dagger_{L,R}
=d\xi^{}_{L,R}\xi^\dagger_{L,R}-igA_H
+i\xi^{}_{L,R}A_{L,R}\xi^\dagger_{L,R},\\
F_H\equiv dA_H+igA_H^2, \\
\hat{F}_{L,R}^{}=\xi_{L,R}^{}(dA_{L,R}^{}-iA_{L,R}^2)\xi^\dagger_{L,R}.
\end{array}$$
Thus, for the intrinsic parity violation processes, we have
\beq
\Gamma=\Gamma_{WZ}[\xi^\dagger_L\xi^{}_R,A_L^{},A_R^{}]
+\sum_{i=1}^6c_i\int_{M_4}{\cal L}_i,
\eeqlb{Gwz}
with 6 arbitrary constants $c_i$, which are determined by experimental
data. Vector meson dominance (VMD) in the process like
$\pi^0\rightarrow 2\gamma$ and $\gamma\rightarrow3\pi$ is very useful
in determining the constants. In Table~3.1 a few sets of constants
used in the literature are listed.
\renewcommand\arraystretch{1.2}
\begin{table}
\begin{center}
Table 3.1 : Constants $c_i$ in Eq.~(\ref{Gwz}). ($C$=$-iN_c/240\pi^2$ and
$\alpha-\beta$=1)\\
\vskip 3mm
\begin{tabular}{c|cccccc}
\hline
 & $c_1$ & $c_2$ & $c_3$ & $c_4$ & $c_5$ & $c_6$ \\
\hline
PVMD$^*$\cite{FKTUY} & $15C\alpha$ & $15C\beta$ & 0 & $-15C$ & 0 & $-15C$ \\
CVMD$^\dagger$\cite{FKTUY} & $5C\alpha$ & $5C\beta$ & 0 & $-15C$ & 0 & $-15C$
\\
Minimal Model$^\ddagger$\cite{MKWW87} & $10C$ & $-10C$ & 0 & 0 & 0 & 0 \\
\hline
\multicolumn{7}{l}{\hskip 1cm $^*$ Partial VMD with no contact term for the
$\omega\rightarrow3\pi$ decay.}\\
\multicolumn{7}{l}{\hskip 1cm $^\dagger$ Complete VMD.}\\
\multicolumn{7}{l}{\hskip 1cm $^\ddagger$ VMD in the isoscalar channel.}
\end{tabular}
\end{center}\end{table}
\renewcommand\arraystretch{1.5}

The $[SU(3)_L\!\times\!SU(3)]_{\mbox{\scriptsize global}}$ breaking term
can be incorporated without affecting the hidden symmetry by writing
${\cal L}_{A,V}$ as \cite{BKY85}
\beqa{1.5}{l}
{\cal L}_A=-\frac14f_\pi^2\Tr\{
(\D_\mu\xi_L^{}\xi^\dagger_L+\D_\mu\xi^{}_L\ve_A^{}\xi^\dagger_R)
+(\D_\mu\xi_R^{}\xi^\dagger_R+\D_\mu\xi^{}_R\ve_A^{}\xi^\dagger_L)\}^2,\\
{\cal L}_V=-\frac14f_\pi^2\Tr\{
(\D_\mu\xi_L^{}\xi^\dagger_L+\D_\mu\xi^{}_L\ve_V^{}\xi^\dagger_R)
-(\D_\mu\xi_R^{}\xi^\dagger_R+\D_\mu\xi^{}_R\ve_V^{}\xi^\dagger_L)\}^2,
\eeqalb{Lsb1}
with the covariant derivative given by Eq.~(\ref{Umu}b). The symmetry breaking
matrices $\ve_{A,V}$ are defined by
\beq
\ve_{A,V}=c_{A,V}\ts\frac13(1-\sqrt3\lambda_8)
\eeq
with $c_{A,V}$ being constants. These additional terms lead to a
renormalization of the Goldstone boson field $M$ of Eq.~(\ref{U3}a) through
$$\sqrt{1+\ve_A^{}}M\sqrt{1+\ve_A^{}}\rightarrow M, $$
and thus modify the vector meson masses as
\beq
m_\rho^2=m_\omega^2=ag^2f^2_\pi=\frac{m_{P^*}^2}{1+c_V^{}}
=\frac{m_{\Phi^*}^2}{(1+c_V^{})^2},
\eeqlb{SBm1}
and the meson decay constants as
\beq
f_P = f_\pi\sqrt{1+c_A^{}}.
\eeq
In the case of $h$=$s$ ($P^*$=$K^*$ and $\Phi^*$=$\phi$), the relation
derived from Eq.~(\ref{SBm1})
$$
\frac{m_\phi}{m_{K^*}} = \frac{m_{K^*}}{m_\rho}
 = \sqrt{\mbox{$1+c_V^{}$}}
\sim 1.15
$$
holds within 2\% and, furthermore, $c_A^{}$=$c_V^{}$ results in
$$
f_K/f_\pi = m_{K^*}^{}/m_\rho^{}\sim 1.15,
$$
in good agreement with experimental data.
However, for the heavier flavor, neither $c_A^{}=c_V^{}$
nor $m^{}_{\Phi^*}/m_{P^*}^{}=m^{}_{P^*}/m^{}_\rho$ holds well.
(See Table~3.2.)
\renewcommand\arraystretch{1.3}
\begin{table}
\begin{center}
Table 3.2 : Symmetry breakings in the vector meson masses (in MeV)\\
and in the meson decay constants.\\
\vskip 3mm
\begin{tabular}{c|cccccccc}
\hline
$h$ & $m_{P^*}^{}$ & $\frac{m^{}_{P^*}}{m_\rho}$ & $c^{}_V$  &
$m_{\Phi^*}^{a)}$ & $m_{\Phi^*}^{b)}$ & $m_{\Phi^*}^{c)}$
& $f_P/f_\pi$ & $c_A$ \\
\hline
$s$ & \ 892 & 1.161 & 0.349 & 1020 & 1036 & 1001 & 1.22 & 0.488 \\
$c$ & 2010 & 2.617 & 5.850 & 3097 & 5260 & 2736 & 1.80 & 2.240 \\
$b$ & 5325 & 6.934 & 47.07 & 9460 & 3.7$\times10^3$ & 7491 & - & - \\
\hline
\multicolumn{9}{c}
{\hskip 1cm $^{a)}$ experimental data, $^{b)}$ Eq.~(\ref{SBm1})
and $^{c)}$ Eq.~(\ref{SBm2}).}\\
\end{tabular}
\end{center}\end{table}
\renewcommand\arraystretch{1.5}

One may introduce the symmetry breaking in a different way from
that of Eq. (\ref{Lsb1}).
Note that, while its role is the same in modifying the
meson decay constant as the second term of Eq.~(\ref{Lsb}),
the additional terms in ${\cal L}_A$ cannot be reduced to the
same form. We can rewrite the latter in terms of $\xi_L^{ }$
and $\xi_R^{ }$ by substituting $U=\xi^\dagger_L\xi^{ }_R$:
$$\begin{array}{l}
-\frac1{12}(f_K^2-f_\pi^2)\Tr\left\{(1-\sqrt3\lambda_8)
(U\partial_\mu U^\dagger\partial^\mu U+U^\dagger\partial_\mu U
\partial^\mu U^\dagger)\right\}\\
\hskip 1cm=-\frac14f_\pi^2\Tr\left\{
(\partial_\mu\xi^{}_L\xi^\dagger_L-\partial_\mu\xi^{}_R\xi^\dagger_R)
(\xi^{}_R\ve_A\xi^\dagger_L+\xi_L^{}\ve_A\xi^\dagger_R)
(\partial^\mu\xi^{}_L\xi^\dagger_L-\partial^\mu\xi^{}_R\xi^\dagger_R)
\right\}.
\end{array}$$
It suggests introducing the following symmetry breaking into the
Lagrangian\cite{Oh,BGP}
\beqa{1.5}{l}
{\cal L}_A=-\frac14f_\pi^2\Tr\left\{
(\D_\mu\xi^{}_L\xi^\dagger_L-\D_\mu\xi^{}_R\xi^\dagger_R)
(1+\xi^{}_R\ve_A^{}\xi^\dagger_L+\xi_L^{}\ve_A^{}\xi^\dagger_R)
(\D^\mu\xi^{}_L\xi^\dagger_L-\D^\mu\xi^{}_R\xi^\dagger_R)\right\},\\
{\cal L}_V=-\frac14f_\pi^2\Tr\left\{
(\D_\mu\xi^{}_L\xi^\dagger_L+\D_\mu\xi^{}_R\xi^\dagger_R)
(1+\xi^{}_R\ve_V^{}\xi^\dagger_L+\xi_L^{}\ve_V^{}\xi^\dagger_R)
(\D^\mu\xi^{}_L\xi^\dagger_L+\D^\mu\xi^{}_R\xi^\dagger_R)\right\}.
\eeqalb{Lsb2}
Compared with Eq.~(\ref{Lsb1}), Eq.~(\ref{Lsb2}) can be understood
as a linearized version of it.\footnote{It is also similar to (but not
equal to) the
$\alpha$-type symmetry breaking term discussed in Ref.~\citenum{JJPSW}.
Eq.~(\ref{Lsb2}) can be regarded as a generalization of their
$\alpha$-type term.}
Although not exact, it corresponds to eliminating
the terms with $\ve_{A,V}^2$  in the expansion of Eqs.~(\ref{Lsb1}).
Note that the disastrous $\Upsilon$ mass in Eq.~(\ref{SBm1})
is mainly due to the $c_V^2$ in the denominator.
As a consequence, ${\cal L}_V$ leads to the vector meson masses
\beq
m_{\rho,\omega}^2=ag^2_{}f^2_\pi=\frac{m_{P^*}^2}{1+c_V^{}}
=\frac{m_{\Phi^*}^2}{1+2c_V^{}},
\eeqlb{SBm2}
which satisfy Gell-Mann--Okubo mass formula
$$
m_{\Phi^*}^2 - m_{P^*}^2 = m_{P^*}^2 - m_{\rho,\omega}^2,$$
and yield a lot more reasonable masses for the vector mesons $\phi$,
$J/\psi$ and $\Upsilon$ than Eq.~(\ref{SBm1}). (See Table~3.2.)

Finally, we substitute the CK ansatz (\ref{Uck}), $U=N_\pi U_P N_\pi$
(that is, $\xi_L^\dagger=N_\pi\sqrt{U_P}$ and
$\xi^{}_R=\sqrt{U_P}N_\pi$\footnote{This
does not exactly correspond to the standard unitary gauge,
$\xi^\dagger_L$=$\xi^{}_R$, unless $N_K$=1. Different ans\"aze
correspond roughly to different gauge choices and we may think of
this as one particular gauge. However, it is plausible that
the results do not depend on these gauge choices.\cite{SMNR}}),
into the Lagrangian constructed so far in the form of
${\cal L}={\cal L}_0+{\cal L}_{an}+{\cal L}_{SB}$
with ${\cal L}_0$, ${\cal L}_{an}$ and ${\cal L}_{SB}$ given by
Eq.~(\ref{L0}), Eq.~(\ref{Gwz}) and Eq.~(\ref{Lsb2}),
respectively. The resulting Lagrangian reads
\beq
{\cal L}={\cal L}^{}_{SU(2)}+{\cal L}_{PP^*}^{n}+{\cal L}_{PP^*}^{an},
\eeqlb{Lsk}
with
\renewcommand\theequation{\arabic{section}.\arabic{equation}\alph{subeq}}
\addtocounter{equation}{-1}\setcounter{subeq}{1}
\be&&\begin{array}{l}
{\cal L}_{SU(2)}^{}=\frac14{f^2_\pi}\tr(\partial_\mu U_\pi^\dagger
\partial^\mu U_\pi^{})-\ts\frac12\tr(q_{\mu\nu}^{}q^{\mu\nu}) \\ \hskip 1.5cm
+m_\rho^2\tr(q_\mu+\dps\frac{i}{g_*}V_\mu)^2+\ts6\pi^2 ( c_1 - c_2 )
ig_* \omega_\mu B^\mu \\ \hskip 1.5cm
- \frac{1}{2} g_*^3
( c_1 + c_2) \{ -\ve^{\mu\nu\alpha\beta} \omega_\mu \tr ( A_\nu
{\overline \rho}_\alpha {\overline \rho}_\beta) + \ve^{\mu\nu\alpha\beta}
\tr ( A_\mu {\overline \rho}_\nu {\overline \rho}_\alpha {\overline
\rho}_\beta) \}
\\ \hskip 1.5cm
+ i g_* c_4 \left\{ -\ve^{\mu\nu\alpha\beta} \omega_\mu \tr
( q_{\nu\alpha} A_\beta) + \ve^{\mu\nu\alpha\beta} \tr \{ q_{\mu\nu}
( A_\alpha {\overline \rho}_\beta - {\overline \rho}_\alpha A_\beta) \}
\right\},
\end{array}\\
\addtocounter{equation}{-1}\addtocounter{subeq}{1}
&&\begin{array}{l}
{\cal L}_{PP^*}^{n}=D_\mu PD^\mu P^\dagger-m_P^2KK^\dagger
- P A_\mu A^\mu P^\dagger \\ \hskip 1.5cm
+ \frac{f_\pi^2}{f_P^2} \left[P(V_\mu-ig_* q^{}_\mu)D^\mu P^\dagger
-\D_\mu P(V_\mu-ig_* q^{}_\mu)P^\dagger\}\right] \\ \hskip1.5cm
-\frac12\left[P^{*\mu\nu}P^{*\dagger}_{\mu\nu}
-2g^2_*P^*_\mu q^{\mu\nu}_{}P^{*\dagger}_\nu\right] \\ \hskip 1.5cm
+m^2_{P^*}\dps[P^{*\mu}-\frac{\sqrt2}{m_{P^*}}PA^\mu]
[P^{*\dagger}_\mu-\frac{\sqrt2}{m_{P^*}}A_\mu P^\dagger],
\end{array}\\
\addtocounter{equation}{-1}\addtocounter{subeq}{1}
&&\begin{array}{l}
{\cal L}_{PP^*}^{an} =-\frac{iN_c}{8f_P^2}B_\mu
\{D^\mu PP^\dagger-PD_\mu P^\dagger\} \\ \hskip 1.5cm
+ ( c_1 - c_2 ) \{ - \frac{3\pi^2}{f_P^2} B_\mu ( D^\mu P P^\dagger - P
D^\mu P^\dagger ) + \dots  \} \\ \hskip 1.5cm
 - ( c_1 + c_2 ) \{ - \frac{2}{f_P^2} g_*^3 \ve^{\mu\nu\alpha\beta} \omega_\mu
P A_\nu {\overline \rho}_\alpha {\overline \rho}_\beta P^\dagger + \dots \}
\\ \hskip 1.5cm
 + i c_4 \{\frac12g^2_*\ve^{\mu\nu\lambda\rho}(P^*_{\mu\nu}A_\lambda
P^{*\dagger}_\rho - P^{*}_\mu A_\nu P^{*\dagger}_{\lambda\rho})\\ \hskip 2.5cm
-\dps{i}{ }g_*\frac{1}{f_P} \ve^{\mu\nu\lambda\rho}
(D_\mu P q^{}_{\nu\lambda} P^{*\dagger}_\rho
+P^*_\mu q^{}_{\nu\lambda} D_\rho P^\dagger)+\cdots\},
\end{array}
\ee
where \renewcommand\theequation{\arabic{section}.\arabic{equation}}
$$\begin{array}{l}
D_\mu P^\dagger = (\partial_\mu+V_\mu)P^\dagger, \\
q^{}_\mu = \frac12(\omega_\mu+\rho_\mu)=\frac12\left[\begin{array}{cc}
\!\omega_\mu+\rho^0_\mu\!\!\! & \!\! \sqrt2\rho^+_\mu\!\!  \\
\!\!\sqrt2\rho^-_\mu\!\! & \!\!\!\omega_\mu-\rho^0_\mu\!  \\
\end{array}\right], \\
q_{\mu\nu}^{} = \frac12(\omega_{\mu\nu}+\rho_{\mu\nu})
=\partial_\mu q^{}_\nu-\partial_\nu q^{}_\mu+ig_*[q^{}_\mu,q^{}_\nu], \\
P^{*\dagger}_{\mu\nu} = (\partial_\mu+ig_* q_\mu^{})P^{*\dagger}_\nu
-(\partial_\nu+ ig_* q^{}_\nu)P^{*\dagger}_\mu, \\
{\overline \rho}_\mu = \rho_\mu +i 2 V_\mu / g_*, \\
B^\mu=\frac{1}{24\pi^2}\ve^{\mu\nu\lambda\rho}\tr
(U_\pi^\dagger\partial_\nu U_\pi U_\pi^\dagger\partial_\lambda U_\pi
U_\pi^\dagger\partial_\rho U_\pi).
\end{array}\seqno{Lsk}{d}$$

One may  check that Eq.~(\ref{Lsk}) contains all the terms of
Eq.~(\ref{Yan}). Explicitly, we have
\beqa{1.5}{l}
{\cal L} =\dps\frac{f_\pi^2}{4}\tr(\partial_\mu U_\pi^\dagger
\partial^\mu U_\pi)+\frac{1}{32g_*^2}\tr[U^\dagger_\pi\partial_\mu U_\pi,
U^\dagger_\pi\partial_\nu U_\pi]^2 \\
\hskip 1cm + D_\mu P D_\mu P^\dagger - m_P^2 P P^\dagger
-\frac12 P^{*\mu\nu} P^{*\dagger}_{\mu\nu} + m^2_{P^*} P^{*\mu}
P^{*\dagger}_\mu \\ \hskip 1cm
-\sqrt2 m_{P^*}^{} (PA^\mu P^{*\dagger}_\mu + P^*_\mu A^\mu P^\dagger)
+\frac{i}{2} c_4 g_*^2 \ve^{\mu\nu\lambda\rho} (P^*_{\mu\nu}A_\lambda
P^{*\dagger}_\rho
+P^*_\lambda A_\rho P^{*\dagger}_{\mu\nu}) + \dots,
\eeqalb{Lhml}
where we have replaced the light vector meson fields $\rho_\mu$ and
$\omega_\mu$ by $i2V_\mu/g_*$ and $(c_1-c_2)i6 \pi^2 B_\mu/g_*f_\pi^2$
respectively
and kept the leading order terms in $m^{}_P$ and $m_{P^*}^{}$ with
a single  derivative on the pion fields.
Comparing Eq.~(\ref{Lhml}) with Eq.~(\ref{Yan}), we get
two relations:
$$\fQ=-\sqrt2m_{P^*}, \and \gQ= ic_4g_*^2. $$
The first relation implies (see Eq.~(\ref{fQgQ})) that the $\gQ$ value is
\beq
\gQ=-\frac{1}{\sqrt2}\simeq -0.71,
\eeq
which is quite close to $\gQ=-0.75$ evaluated with the NRQM in Sec.~2.
If one assumes that the VMD works in the heavy meson sector
for which $c_4$ is fixed to $iN_c/16\pi^2$, we could obtain $g_*$
in the heavy quark limit: {\it viz.\/}
\beq
g_*=\sqrt{\frac{16\pi^2}{\sqrt2 N_c}}\simeq 6 \hskip 1.5cm
(\mbox{with $N_c$=3}).
\eeq
It is intriguing that this is so close to $g^{}_*$=$g^{}_{\rho\pi\pi}$(=6.11)
found in the light meson sector.

\setcounter{equation}{0}
\section{Effective Field Theory for Heavy Mesons}
\subsection{Heavy-Quark Effective Theory}
At this point, we digress a bit to discuss in a simple picture\cite{Ge92,FI}
how the heavy quark symmetries arise in the hadronic processes involving
a heavy quark. Consider two hadrons $A$ and
$B$, each of which is made of a single heavy quark of
mass $m_Q^A$ and $m_Q^B$, respectively, and the light degrees of freedom. If
the
heavy quark masses are much larger than the scales of
QCD interactions, $m_Q^A, m_Q^B\gg \Lambda_{QCD}$, then in the rest
frame of the heavy quark, how QCD distributes the light degrees of
freedom around the static heavy quark is independent of the heavy
flavor; {\it i.e.,\/} as far as the color charge is concerned, the
light degrees of freedom do not know the difference. This feature induces a
new $SU(N_h)$ flavor symmetry for $N_h$ flavors of heavy quark. By
boosting one can extend the heavy flavor symmetry to any heavy quarks
of the same {\em velocity}. This is the reason
why the velocity (not the momentum)
has a physical significance in the heavy quark theory. Furthermore,
the heavy quark spin decouples as $1/m_Q$ when the heavy quark mass
becomes infinitely large.

The heavy system that we are considering here is a QCD bound state
of a heavy quark and light quarks (and/or light anti-quarks), where
the heavy quark carries most, but not all, of the momentum. Consider,
for example, a heavy meson moving with a 4-velocity $v^\mu$
($v_\mu v^\mu$=1, $v^0$$>$0). Due to its huge momentum, its evolution
is classical and the 4-momentum of the bound state has the form of
\beq
P^\mu_{bs}=m_P^{}v^\mu,
\eeq
where the heavy meson mass $m_P^{}$ is essentially the same as the
heavy quark mass $m_Q$
$$
m_P^{}\approx m_Q^{} ,
$$
and the difference is expected to be independent of $m_Q$. Let the
small momentum of the light degrees of freedom and the residual
momentum of the heavy quark be $q^\mu$ and $k^\mu$, respectively.
Then, we can write the momentum of the heavy quark as
\beq
p_{hq}^\mu=P_{bs}^\mu-q^\mu=m_Q^{}v^\mu+k^\mu.
\eeq
The 4-velocity of the heavy quark can be defined as
$$v^\mu_{hq}\equiv\frac{p^\mu_{hq}}{m_Q}=v^\mu+\frac{k^\mu}{m_Q}, $$
which is the same as that of the bound state in the heavy quark limit:
$$v^\mu_{hq}\rightarrow v^\mu \hskip 5mm\mbox{as}\hskip 5mm
m_Q\rightarrow\infty.$$
When the heavy system is scattered by low-energy QCD interactions
into a new state of the same heavy quark, the momentum conservation
requires that
$$P^\mu=m_P^{}v^{\prime\mu}+q^{\prime\mu}.$$
As $m_P^{}\rightarrow\infty$ for a fixed momentum transfer $q^{\prime\mu}$,
we must have $v^\mu_{}=v^{\prime\mu}_{}$. That is, in the limit of
infinite mass, the velocity of the heavy quark is unchanged by the
QCD interactions, independently of
what the light degrees of freedom do within its typical
energy scale $\sim\Lambda_{QCD}$. Furthermore, two heavy quarks with
different velocities do not communicate with each other, because their
momenta are infinitely different and we stay at the energy scale of the
light degrees of freedom. This
``{\em velocity superselection rule}''\cite{Ge90}
makes it much easier to do physics with a heavy quark.

We first consider the dynamics of a free heavy quark in the effective theory.
In the full theory, the heavy quark field $h$ has the Lagrangian density
\beq
\L_{hq}=\bar{h}(x)(i\partial\hskip-1.85mm/-m_Q)h(x),
\label{Lhq}\eeq
in position space.
For the heavy quark of velocity $v$, in evaluating the action written
as an integral in momentum space as
\beq
\int\!\!\frac{d^4p}{(2\pi)^4}\bar{h}(-p)(\sp-m_Q)h(p),
\label{Ahq}\eeq
with $a\hskip-1.8mm/$ denoting $a_\mu\gamma^\mu$,
the relevant region of the integral is a cell of
$p^\mu=m_Qv^\mu+k^\mu$ with
$k^\mu$\raisebox{-0.6ex}{$\stackrel{\ts <}{\sim}$}$\Lambda_{QCD} \ll m_Q$
(see Fig.~4.1)
\begin{figure}[p]
\beginpicture
\setcoordinatesystem units <1cm,1cm> point at -5 2
\setplotarea x from -7.5 to 7.5, y from 0 to 3.4
\put{Figure 4.1: \hskip 0.2cm Momentum space for the functional integral
of eq.(\ref{Ahq}). The $+$ light cone}[c] at 0 0.7
\put{is divided up into cells of width $\sqrt{m_Q\Lambda}$, {\it i.e.\/}
$\Delta v\approx\sqrt{\Lambda/m_Q}$.}[c] at 0 0.2
\setcoordinatesystem units <1cm,0.66cm> point at -5 -5
\setplotarea x from -5 to 5, y from -5 to 5
\setquadratic
\plot
-4.8  4.903   -4.2  4.317   -3.6  3.736   -3.0  3.162   -2.4  2.600
-1.8  2.059   -1.2  1.562   -0.6  1.166    0.0  1.000    0.6  1.166
 1.2  1.562    1.8  2.059    2.4  2.600    3.0  3.162    3.6  3.736
 4.2  4.317    4.8  4.903 /
\setdashes \plot
-4.8 -4.903   -4.2 -4.317   -3.6 -3.736   -3.0 -3.162   -2.4 -2.600
-1.8 -2.059   -1.2 -1.562   -0.6 -1.166    0.0 -1.000    0.6 -1.166
 1.2 -1.562    1.8 -2.059    2.4 -2.600    3.0 -3.162    3.6 -3.736
 4.2 -4.317    4.8 -4.903 / \setsolid\setlinear
\unitlength=1mm
\def\garo#1{\put {\line (1,0){6}} [Bl] at #1}
\garo{4.32 5.503} \garo{4.32 4.303} \garo{3.72 4.917} \garo{3.72 3.717}
\garo{3.12 4.336} \garo{3.12 3.136} \garo{2.52 3.762} \garo{2.52 2.562}
\garo{1.92 3.200} \garo{1.92 2.000} \garo{1.32 2.659} \garo{1.32 1.459}
\garo{0.72 2.162} \garo{0.72 0.962} \garo{0.12 1.766} \garo{0.12 0.566}
\garo{-0.48 1.600} \garo{-0.48 0.400} \garo{-1.08 1.766} \garo{-1.08 0.566}
\garo{-1.68 2.162} \garo{-1.68 0.962} \garo{-2.28 2.659} \garo{-2.28 1.459}
\garo{-2.88 3.200} \garo{-2.88 2.000} \garo{-3.48 3.762} \garo{-3.48 2.562}
\garo{-4.08 4.336} \garo{-4.08 3.136} \garo{-4.68 4.917} \garo{-4.68 3.717}
\garo{-5.28 5.503} \garo{-5.28 4.303}
\unitlength=6.6mm
\def\sero#1#2{\put {\line (0,1){#1}} [Bl] at #2}
\sero{1.200}{4.8 4.303} \sero{1.786}{4.2 3.717}
\sero{1.781}{3.6 3.136} \sero{1.774}{3.0 2.562}
\sero{1.762}{2.4 2.000} \sero{1.741}{1.8 1.459}
\sero{1.697}{1.2 0.962} \sero{1.596}{0.6 0.566}
\sero{1.366}{0.0 0.400} \sero{1.366}{-0.6 0.400}
\sero{1.200}{-5.4 4.303} \sero{1.786}{-4.8 3.717}
\sero{1.781}{-4.2 3.136} \sero{1.774}{-3.6 2.562}
\sero{1.762}{-3.0 2.000} \sero{1.741}{-2.4 1.459}
\sero{1.697}{-1.8 0.962} \sero{1.596}{-1.2 0.566}
\put{describes heavy quarks}[c] at 0 3.7
\put{describes heavy anti-quarks}[c] at 0 -3.7
\put{$+$ light cone}[c] at 0 2.7
\put{$-$ light cone}[c] at 0 -2.7
\endpicture
\end{figure}
and thus the relevant $h(p)$ is nearly  on mass shell;
$(\sp-m_Q)h(p)=O(1/m_Q)\approx 0$, which implies that
$$(\sv-1)h(p)\approx-\frac{k\hskip-1.75mm/}{m_Q}h(p)\approx 0.$$
All other quantities of order $m_Q$ give a large contribution to the action
and so make a small contribution to the integral.
In a cell around $p^\mu=m_Q^{}v^\mu$, define the heavy quark field
$h_v(k)=h(p)-O(1/m_Q)$ satisfying exactly
\beq
\sv h_v=h_v.
\label{hv}\eeq
Then, in terms of the residual momentum $k=p-m_Qv$, the Lagrangian
in the cell looks like
$$\bar{h}_v(\sp-m_Q)h_v=\bar{h}_v{k\hskip-1.75mm/}h_v
=\ts\frac12\bar{h}_v\{\sv,{k\hskip-1.75mm/}\}h_v
=\bar{h}_v v^\mu k_\mu h_v.$$
Now, as $m_Q\rightarrow\infty$, the cell gets closer together in velocity
space, but the size of each cell in momentum space grows. Each cell
becomes a mini-Lagrangian relevant only for the heavy quark field with the
corresponding velocity:
\beq
\int_{cell}\!\frac{d^4k}{(2\pi)^4}\bar{c}_v(-k)v^\mu k_\mu c_v(k).
\label{Lhq2}\eeq
The velocity superselection rule is equivalent to the statement
that $h_v$ and $h_{v'}$ are independent fields for $v^\mu\neq v^{\prime\mu}$:
they correspond to different cells on the mass shell hyperboloid.
The Lagrangian is a sum over all $v$:
\beq
\L_{hq}=\sum_{\vec v}\L_{hq}[h_v].
\eeq

The corresponding Lagrangian for the heavy quark field in position
space can be easily read off from Eq.~(\ref{Lhq2})
\beq
\L_{hq}[h_v]=i\bar{h}_v(x)v\cdot\partial h_v(x).
\label{Lhq3}\eeq
The same Lagrangian could also be obtained by substituting\cite{Ge90}
\beq
h(x)=\frac{1+\sv}2 e^{-im^{}_Q v\cdot x}h_v(x)+O(1/m_Q^{})
\label{hv2}\eeq
directly into Eq.~(\ref{Lhq}). It projects away the negative
light cone which describes anti-quarks and eliminates the trivial
dependence on the heavy quark mass.
With this effective Lagrangian, the Feynman propagator
for the heavy quark field simplifies to
\beq
\frac{1+\sv}{2}\frac{1}{v\!\cdot\!k + i\epsilon},
\eeq
which can be consistently derived by inserting $p^\mu=m_Q^{}v^\mu+k^\mu$
into the Feynman propagator of the full theory and by keeping the
leading order terms in $m_Q$ in both the numerator and the denominator:
$$\frac{\sp+m_Q}{p^2_{}-m^2_Q+i\epsilon}
\approx\frac{m_Q^{}\sv+m_Q^{}}{2m_Q^{}(v\!\cdot\!k)+i\epsilon}
=\frac{1+\sv}{2}\frac{1}{(v\!\cdot\!k)+i\epsilon}.$$
In the rest frame, we have
$$\frac{1+\gamma^0}{2} \frac{1}{k^0+i\epsilon},$$
whose Fourier transform is
$$\propto\delta^3(\vec{r}-\vec{r}')\Theta(t-t').$$
This just describes the particle sitting still, propagating in time
along its classical trajectory.

QCD interactions can be easily incorporated by imposing the
color gauge symmetry (we recall that the covariant derivative with
the color gauge field was denoted $D^c_\mu$):
\beq
\L_{hq}[h_v]=i\bar{h}_v v^\mu D^c_\mu h_v.
\eeq
Here, the heavy-quark symmetry is manifest: the Lagrangian does
not depend on the heavy quark mass (heavy-quark flavor symmetry)
and does not contain any $\gamma$-matrices (heavy-quark spin symmetry);
that is, by Eqs.~(\ref{hv}) and (\ref{hv2}) we have ignored the variation
of the spinor within the cell and eliminated the trivial heavy quark
mass dependence from the theory.  Let us identify explicitly
the symmetries of the heavy quark Lagrangian with two heavy flavors, say
$c$ and $b$:
\beq
\L_c[c_v]+\L_b[b_v]
=i\bar{c}_v v^\mu {D^c}_\mu c_v
+i\bar{b}_v v^\mu {D^c}_\mu b_v
=i\bar{h}_v v^\mu {D^c}_\mu h_v,
\eeq
where we have put the two fields together into an 8-component field:
\beq
h_v\equiv\left(
\begin{array}{c}\!\!c_v\!\!\\\!\!b_v\!\!
\end{array}\right).
\label{hcb}\eeq
In the rest frame, $v^0$=1, $\vec v$=0, the Lagrangian is
$$\L_c[c_0]+\L_b[b_0]=i\bar{h}_0 {D^c}^0 h_0.$$
The Isgur-Wise heavy-quark symmetry is the $SU(4)$ spin-flavor symmetry
characterized by the transformations whose generators are
\beq
P_0\sigma_j, \hskip 3mm P_0\eta_j, \hskip 3mm P_0\sigma_j\eta_k,
\hskip 5mm\mbox{ for $j,k$=1,2,3},
\label{Iwsu4}\eeq
where $P_0$ and $\sigma_j(j$=1,2,3), respectively, are combinations of the
$4\times 4$ Dirac-matrices
$$P_0\equiv\ts\frac12(1+\gamma^0)  \and \dps
\vec\sigma\equiv\left(\begin{array}{cc}\!\!\vec\sigma\!&\!0\!\!\\
\!\!0\!&\!\vec\sigma\!\!\\\end{array}\right), $$
and $\eta_j(j$=1,2,3) is an additional set of Pauli matrices
that implements the $SU(2)$ rotation between $c$ and $b$ subspaces.

\subsection{Heavy Meson Theory}
Returning to the Lagrangian (\ref{Yan}) for the heavy mesons,
we proceed in a way analogous to the heavy-quark Lagrangian.
First, in analogy to Eq.~(\ref{hv2}),
we introduce new heavy meson fields for each velocity:
\beqa{1.4}{l}
\dps P=e^{-iv\cdot x \m^{}_P}\frac{1}{\sqrt{2\m_P^{}}}P_v,\\
\dps P^*_\mu=e^{-iv\cdot x\m^{}_P}\frac{1}{\sqrt{2\m_P^{}}}P^*_{v\mu},
\eeqalb{HQF}
where $\m \equiv \frac14(m^{}_P + 3m^{}_{P^*})$ is the averaged
mass of the heavy mesons. Note that in the infinite $m_Q$ limit we
have $m_P = m_{P^*}$.
The factor $1/\sqrt{2\m_P}$ is introduced for later convenience.
Substituting Eq.~(\ref{HQF}) into the Lagrangian (\ref{Yan}) and keeping
the leading order terms in $\m_P^{}$, we obtain a heavy-flavor independent
Lagrangian
\beqa{1.2}{l}
\L_v = \L_M  + iv^\mu(D_\mu P_v^{} P_v^\dagger
 - D_\mu P_{v\nu}^* P_v^{*\nu\dagger})    \\ \hskip 1cm
 +\frac12 g (P_v^{} A^\mu P^{*\dagger}_{v\mu} + P^*_{v\mu} A^\mu P^\dagger_v)
-i\frac12 g\ve^{\mu\nu\lambda\rho}v_\mu P^*_{v\nu}
        A_\lambda P^{*\dagger}_{v\rho}.
\eeqalb{Lv}
Here, we have used Eq.~(\ref{fQgQ}) for the coupling constants
in the heavy quark limit and the constraint that
$v\cdot P^*_v=0$, and replaced $P_v D_\mu P^\dagger$
by $-D_\mu P_v P^\dagger_v$, dropping the trivial total derivative
term $\partial_\mu(P_v P^\dagger_v)$.

A special care should be taken in this procedure.
The Lagrangian (\ref{Yan}) defines the canonical momenta conjugate
to the fields $P$, $P^\dagger$, $P^{*i}$ and
$P^{*i\dagger}(i$=1,2,3)\footnote{$P^{*0}$ and $P^{*0\dagger}$ are not
dynamical variables and do not have
conjugate momenta.}:
\beqa{1.2}{ll}
\dps \Pi^\dagger\equiv\frac{\delta\L}{\delta\dot{P}}=D_0 P^\dagger, &
\dps \Pi\equiv\frac{\delta\L}{\delta\dot{P}^\dagger}=D_0 P, \\
\dps \Pi^{*i\dagger}\equiv\frac{\delta\L}{\delta\dot{P}^{*i}}
=P^{*0i\dagger}, &
\dps \Pi^{*i}\equiv\frac{\delta\L}{\delta\dot{P}^{*i\dagger}}
=P^{*0i}
\eeqalb{pi}
where $\dot{a}$ denotes $\partial_0 a$.
Their equal-time commutation relations are
\beqa{1.3}{c}
[P_{i^{}_3}(t,\vec{r}),
\Pi^\dagger_{i^\prime_3}(t,\vec{r}\,^\prime)]
=[P^\dagger_{i^{}_3}(t,\vec{r}),
\Pi_{i^\prime_3}(t,\vec{r}\,^\prime)]
=i \delta_{i^{}_3i^\prime_3} \delta^3(\vec{r}-\vec{r}\:^\prime),\\ {}
[P^{*j}_{i^{}_3}(t,\vec{r}),
\Pi^{*k\dagger}_{i^\prime_3}(t,\vec{r}\,^\prime)]
=[P^{*j\dagger}_{i^{}_3}(t,\vec{r}),
\Pi^{*k}_{i^\prime_3}(t,\vec{r}\,^\prime)]
=i \delta^{jk} \delta_{i^{}_3i^\prime_3} \delta^3(\vec{r}-\vec{r}\,^\prime),
\eeqalb{CR}
where the subscript $i^{}_3(i^\prime_3)$ is introduced
to denote explicitly the isospin of the fields.
On the other hand, the Lagrangian (\ref{Lv}) defines the canonical momenta
conjugate to the fields $P_v$ and $P^{*i}_v$
\beqa{1.3}{l}
\dps \Pi^\dagger_v\equiv\frac{\delta\L_v}{\delta\dot{P}_v}
=iv^0 P^\dagger_v, \\
\dps \Pi^{*i\dagger}_v\equiv\frac{\delta\L_v}{\delta\dot{P}^{*i}_v}
=iv^0 P^{*i\dagger}_v.
\eeqalb{piv}
Their commutation relations
\beqa{1.2}{c}
[P_{vi_3}(t,\vec{r}),
\Pi^\dagger_{vi^\prime_3}(t,\vec{r}\,^\prime)]
= [P_{vi^{}_3}(t,\vec{r}),
iv^0 P^\dagger_{vi^{}_3}(t,\vec{r}\,^\prime)]
=i \delta_{i_3i^\prime_3} \delta^3(\vec{r}-\vec{r}^\prime),\\ {}
[P^{*j}_{vi^{}_3}(t,\vec{r}),
   \Pi^{*k\dagger}_{vi^\prime_3}(t,\vec{r}\:^\prime)]
=[P^{*j}_{vi^{}_3}(t,\vec{r}),
iv^0 P^{*k\dagger}_{vi^\prime_3}(t,\vec{r}\:^\prime)]
=i \delta^{jk} \delta_{i_3i^\prime_3} \delta^3(\vec{r}-\vec{r}^\prime),
\eeqalb{CRv}
appear to be inconsistent with what we would have obtained by inserting
Eq.~(\ref{HQF}) naively into Eq.~(\ref{CR}) with (\ref{pi}), which
results in commutation relations that differ by a factor of 2.

To understand what is going on, consider the free pseudoscalar
meson field. (The same is true for vector mesons as well as for
interacting fields.) The meson field operator $P$ can be expanded
in terms of the classical eigenmodes (plane wave solutions) as
\beq
P(x)=\int\frac{d^3p}{\sqrt{(2\pi)^3(2\omega_p)}}
(e^{-ip\cdot x} a(p) + e^{+ip\cdot x} b^\dagger(p)),
\eeqlb{P}
where $\omega_p=\sqrt{m^2_P+\vec{p}^2}$ and
$a(p)$ $[b^\dagger(p)]$ is the annihilation (creation) operator of
heavy meson (its anti-particle) which satisfies the commutation relations
\beqa{1.2}{c}
[a(p),\; a^\dagger(p')]
=[b(p),\; b^\dagger(p')]
=\delta^3(\vec{p}-\vec{p}\:^\prime),\\
\mbox{all the other commutators vanish.}
\eeqa
For simplicity, we are ignoring here the isospin
structure of the meson field.
In the cell around $p^\mu=m_P^{}v^\mu$ of our interest,
the anti-particles are projected out by Eq.~(\ref{HQF}) and hence
the field operator $P_v$ is expanded only in terms of the meson particle
states,
\beq
P_v(x)=\int\frac{d^3k}{\sqrt{v^0(2\pi)^3}}e^{-ik\cdot x} a(k).
\eeqlb{Pv}
The integral is over the cell  $k^\mu$
\raisebox{-0.6ex}{$\stackrel{\ts <}{\sim}$}
$\Lambda_{QCD} \ll m_P\approx m_Q$
and we have used that $\omega_p\approx m_P^{} v^0$ in that region.
This simple exercise explains the factor-2 problem in the commutation
relations. In using Eq.~(\ref{HQF}), one has to keep in mind that it
eliminates the anti-particles from the theory.
Note that the plane wave solutions in Eq.~(\ref{Pv}),
$\varphi_{v\vec k}(\vec r,t) \equiv [v^0(2\pi)^3]^{-1/2}e^{-ik\cdot x},$
are normalized as
$$v_0\!\! \int \!\!d^3r
 ( \varphi^*_{v\vec{k}}\; \varphi_{v\vec{k}\,^\prime})
=\delta^3(\vec{k}-\vec{k}\,^\prime), $$
while those in Eq.~(\ref{P}),
$\varphi_{\vec p}(\vec r,t) \equiv [(2\pi)^32\omega_p]^{-1/2}e^{-ip\cdot x},$
are normalized as
$$i\!\!\int\!\!d^3r(\varphi^*_{\vec{p}}\; \dot{\varphi}_{\vec{p}\,^\prime}
 - \dot{\varphi}^*_{\vec{p}}\; \varphi_{\vec{p}\,^\prime})
=\delta^3(\vec{p}-\vec{p}\,^\prime). $$

In the rest frame, $v$=(1,0,0,0), the Lagrangian (\ref{Lv})
is further simplified to
\beq
\L_0=\L_M + i(D_0 P_v^{} P^\dagger_v
 + D_0\vec{P} ^*_v\!\cdot\!\vec{P} ^{*\dagger}_v)
-g(P_v^{}\vec{A}\!\cdot\!\vec{P} ^{*\dagger}_v
+\vec{P} ^*_v\!\cdot\!\vec{A}\:P^\dagger_v)
-ig\vec{P} ^*_v\cdot\vec{A}\times\vec{P} ^{*\dagger}_v.
\eeqlb{Lv0}
One can easily identify the heavy-quark spin symmetry with this Lagrangian.
In analogy with Eq.~(\ref{tpn2}),
we rewrite $P_v$ and $P^{*i}_v(i$=1,2,3)\footnote{In the rest frame
$P^{*0}_v$ is identically zero due to the condition that $v\!\cdot\!P^*_v=0$.}
as combinations of the four fields
$P_{s_Q s_\ell}(s_Q, s_\ell = \uparrow, \downarrow)$ as
\beqa{1.2}{l}
\hskip 1mm P_v^{} = \frac{1}{\sqrt2} (\Pud - \Pdu),\\
P_v^{*1} = \frac{1}{\sqrt2} (\Puu - \Pdd),\\
P_v^{*2} = \frac{i}{\sqrt2} (\Puu + \Pdd),\\
P_v^{*3} = \frac{1}{\sqrt2} (\Pud + \Pdu),
\eeqalb{Pud}
where $s_Q$ and $s_\ell$ denote the heavy quark spin and the total
angular momentum of the light degrees of freedom, respectively.
Then, Lagrangian (\ref{Lv0}) can be written in terms of these
new fields,
\beq
\L_0 = \L_M + \L_{0\uparrow}+\L_{0\downarrow},
\eeqlb{Lud}
where
$$\begin{array}{l}
\L_{0\uparrow}=
 i(D_0\Puu\Puu^\dagger + D_0\Pud\Pud^\dagger) \\ \hskip 1cm
+g( \Puu A^0\Puu^\dagger + \Puu A^+\Pud^\dagger
    +\Pud A^-\Puu^\dagger - \Pud A^0\Pud^\dagger),
\end{array} \seqno{Lud}{a}$$
and
$$\begin{array}{l}
\L_{0\downarrow}=
 i(D_0\Pdu\Pdu^\dagger + D_0\Pdd\Pdd^\dagger) \\ \hskip 1cm
+g( \Pdu A^0\Pdu^\dagger + \Pdu A^+\Pdd^\dagger
     +\Pdd A^-\Pdu^\dagger - \Pdd A^0\Pdd^\dagger),
\end{array} \seqno{Lud}{b}$$
with $A^\pm \equiv (A^1 \pm i A^2)$ and $A^0\equiv A^3$.
Here $\L_{0\uparrow}$ contains only the fields with heavy quark spin
up and $\L_{0\downarrow}$ only those with spin down.
Then the Lagrangian is completely separated into two identical copies for each
heavy quark spin $s_h$, so it is symmetric under the transformations
\beq
\left(\begin{array}{c} \!\! P_{\uparrow s_\ell} \!\! \\
\!\! P_{\downarrow s_\ell} \!\! \end{array}\right)
\rightarrow
\left(\begin{array}{c} \!\! P_{\uparrow s_\ell}^\prime \!\! \\
\!\! P_{\downarrow s_\ell}^\prime \!\! \end{array}\right)
=i\vec\epsilon\cdot\vec\sigma
\left(\begin{array}{c} \!\! P_{\uparrow s_\ell} \!\! \\
\!\! P_{\downarrow s_\ell} \!\! \end{array}\right),
\hskip 5mm \mbox{($s_\ell$=$\uparrow$ and $\downarrow$)}
\eeq
with a set of Pauli matrices $\vec\sigma$ acting on the spinor fields
$\left(\begin{array}{c} \!\! P_{\uparrow s_\ell} \!\! \\
\!\! P_{\downarrow s_\ell} \!\! \end{array}\right)$
$(s_\ell$=$\uparrow$ and $\downarrow$).

In terms of the field operators $P_{s_hs_\ell}$,
the corresponding heavy-quark spin operators $\vec S_Q$ are defined by
\beqa{1.2}{l}
S_Q^3 = \frac12\int d^3r ( \Puu\Puu^\dagger + \Pud\Pud^\dagger
      - \Pdu\Pdu^\dagger - \Pdd\Pdd^\dagger ), \\
S_Q^+ \equiv S^1_h + i S^2_h = -\int d^3r
      ( \Pdu\Puu^\dagger + \Pdd\Pud^\dagger ), \\
S_Q^- \equiv S^1_h - i S^2_h = -\int d^3r
      ( \Puu\Pdu^\dagger + \Pud\Pdd^\dagger ).
\eeqalb{Sh}
Using the equal time commutation relations
for the field operators $P_{s_Q s_\ell}$\footnote{They are
$$[P_{s_Q^{} s_\ell^{}, i_3^{}}(t,\vec{r}),\;
 P^\dagger_{s'_Q s'_\ell, i'_3}(t,\vec{r}\,')]
=\delta_{s_Q^{} s'_Q} \delta_{s_\ell^{} s'_\ell}
 \delta_{i_3^{} i'_3} \delta^3(\vec{r}-\vec{r}\,'),$$
in the rest frame ($v_0$=1).},
one can easily show that they satisfy the correct $SU(2)$ spin algebra
$$ [S_Q^i,\; S_Q^j]=i\ve^{ijk} S_Q^k, \hskip 1cm
   [S_Q^3,\; S_Q^\pm]=\pm S_Q^\pm,
\seqno{Sh}{a}$$
and the correct commutation relations with the fields; {\it e.g.\/},
$$\ts [S_Q^3,\; \Puu]=-\frac12\Puu, \hskip 4mm
[S_Q^+,\; \Puu]=\Pdu, \hskip4mm \cdots.
\seqno{Sh}{b}$$
These relations might appear strange. However they represent
correct spin operators; one should keep in mind that
 $\Puu$ is the {\em annihilation operator} of the
$s_Q$=$+\frac12$ and $s_\ell$=$+\frac12$ particle.
Rewriting the heavy quark spin operators
in terms of $P_v^{}$ and $P^{*i}_v$, we have
\beq
\vec{S}_Q = \ts \frac12 \!\!\dps \int \!\! d^3r
\:[\: i( \vec{P}_v^{*} \times \vec{P}_v^{*\dagger} )
+( P^{}_v \vec{P}^{*\dagger}_v - \vec{P}^{*}_v P^\dagger_v)\:].
\eeq
Note that they are not simply one-half of the spin operators
of the heavy meson field\footnote{The 
Lagrangian (\ref{Yan})
is invariant under the infinitesimal Lorentz transformation
$$\begin{array}{l}
x^\mu\rightarrow {x^\mu}^\prime=x^\mu+{\epsilon^\mu}\!_\nu x^\nu,
\mbox{ with $\epsilon^{\mu\nu}=-\epsilon^{\nu\mu}$}\\
P(x)\rightarrow P^\prime(x^\prime)=P(x),\\
P^*_\alpha(x)\rightarrow P^{*\prime}_\alpha(x^\prime)=
P^*_\alpha(x)+{\epsilon_\alpha}\!^\beta P^*_\beta(x)=P^*_\alpha(x)
+\frac12\epsilon^{\mu\nu}{(S_{\mu\nu})_\alpha}\!^\beta P^*_\beta(x),
\end{array}$$
with
$$
(S^{\mu\nu})_{\alpha\beta}={g^\mu}\!_\alpha{g^\nu}\!_\beta
-{g^\mu}\!_\beta{g^\nu}\!_\alpha
$$
and gives a conserved angular momentum
$$ M_{\rho\sigma}=\int\!d^3r\M_{0\rho\sigma}, $$
where
$$\begin{array}{l}
\M^{\mu\rho\sigma}=(x^\rho\T^{\mu\sigma}-x^\sigma\T^{\mu\rho})\\
\hskip 1.5cm{\dps+P^{*\beta\dagger}(S^{\sigma\rho})_{\alpha\beta}
\frac{\partial\L}{\partial(\partial_\mu P^{*\dagger}_\alpha)}
+\frac{\partial\L}{\partial(\partial_\mu P^{*\dagger}_\alpha)}
(S^{\sigma\rho})_{\alpha\beta}P^{*\beta\dagger}}
\end{array}$$
with the canonical energy-momentum tensor $\T^{\mu\nu}$.
The indices $\rho$ and $\sigma$ run from 1 to 3.
[The operators with $\rho$(or $\sigma$)=0 are for the Lorentz boost,
in which case the second term vanishes due to
$\partial\L/\partial\dot{P}^*_0=0$.]
The first part is the orbital angular momentum carried by
the $P$ and $P^*$ mesons and the second part the spin
angular momentum carried by $P^*$ mesons. Explicitly, the spin angular
momentum can be written as
$$\begin{array}{l}
S^i \equiv \frac12\ve^{ijk}S^{jk} = -\frac12\ve^{ijk}\int\!d^3r
 (\Pi^{*\alpha} (S^{jk})_{\alpha\beta} P^{*\beta\dagger}
 + P^{*\beta} (S^{jk})_{\alpha\beta} \Pi^{*\alpha\dagger} \\
\hskip 1.6cm = - \ve^{ijk} \int\!d^3r
 ( \Pi^{*j} P^{*k\dagger} + P^{k*} \Pi^{*j\dagger} ).
\end{array}$$
In the limit  $m_{P^*} \rightarrow \infty$,
substitution of Eq.~(\ref{HQF}) results in Eq.~(\ref{Svm}) as
the leading order term in meson mass. } 
\beq
\vec{S} = i \!\! \int \!\! d^3r
\vec{P}_v^{*} \times \vec{P}_v^{*\dagger}.
\eeqlb{Svm}
The spin of the light degrees of freedom, $\vec S_\ell$, can be
defined by $\vec S_\ell=\vec S-\vec S_Q$: {\it viz.\/}
\beq
\vec{S}_\ell = \ts \frac12 \!\!\dps \int \!\! d^3r
\:[\: i( \vec{P}_v^{*} \times \vec{P}_v^{*\dagger} )
-( P^{}_v \vec{P}^{*\dagger}_v - \vec{P}^{*}_v P^\dagger_v)\:].
\eeq
To be consistent, they can be rewritten in terms of the fields operator
$P_{s_hs_\ell}$ as
$$\begin{array}{l}
S_\ell^3 = \frac12\int d^3r ( \Puu\Puu^\dagger - \Pud\Pud^\dagger
         + \Pdu\Pdu^\dagger - \Pdd\Pdd^\dagger ), \\
S_\ell^+ \equiv S^1_\ell + i S^2_\ell = -\int d^3r
      ( \Pud\Puu^\dagger + \Pdd\Pdu^\dagger ), \\
S_\ell^- \equiv S^1_\ell - i S^2_\ell = -\int d^3r
      ( \Puu\Pud^\dagger + \Pdu\Pdd^\dagger ).
\end{array}$$

Such expressions can be written in a compact form by taking a
4$\times$4 matrix representation.\cite{Ge92,Bjorken,Falk}
Let $H(x)$ be a 4$\times$4 matrix field defined by
\beq
H(x)=\frac{1+\sv}{2}(\gamma_5 P_v^{}-\gamma_\mu P^{*\mu}_v),
\label{Hx}\eeq
with the conventional Dirac $\gamma$-matrices.
For example, the Lagrangian density (\ref{Lv}) can be rewritten simply
as\cite{Wise}
\beq
\L_v = \L_M - iv_\mu\Tr(D^\mu H\bar{H})
 - g\Tr(H\gamma_5 A_\mu\gamma^\mu\bar{H}),
\eeqlb{LvHx}
where
$$
\bar{H}=\gamma_0 H^\dagger\gamma_0.
\seqno{LvHx}{a}$$
The parity transformations (\ref{Parity}) are
$$
{\cal P} H(\vec{r},t) {\cal P}^{-1} = \gamma^0 H(-\vec{r},t)\gamma^0.
\seqno{LvHx}{b}$$
Explicitly, in the rest frame, $H(x)$ has the structure
$$
H(x)=\left[\begin{array}{cccc}
\!0\! & \!0\! & \!\!+\Pud\!\! & \!\!+\Puu\!\! \\
\!0\! & \!0\! & \!\!-\Pdd\!\! & \!\!-\Pdu\!\! \\
\!0\! & \!0\! & \!0\! & \!0\! \\
\!0\! & \!0\! & \!0\! & \!0\!
\end{array}\right],
\seqno{LvHx}{c}$$
that is,
$$
H_{13}=+\Pud, \hskip 3mm H_{14}=+\Puu, \hskip 3mm
H_{23}=-\Pdd, \hskip 3mm H_{24}=-\Pdu,
\seqno{LvHx}{d}$$
and all the other 12-components vanish. The spin operators
can be summarized as
\beqa{1.2}{l}
\vec S = - \int\!d^3r \Tr([ \frac12 \vec\sigma,\, H]\bar{H}), \\
\vec S_Q = - \int\!d^3r \Tr( \frac12 \vec\sigma H\bar{H}), \\
\vec S_\ell = + \int\!d^3r \Tr( H \frac12 \vec\sigma \bar{H}),
\eeqalb{Sops}
with the 4$\times$4 spin matrices defined by
$$
\sigma^i \equiv \ts\frac{1}{4} i\ve^{ijk} [\gamma^j,\; \gamma^k].
$$

The Hamiltonian of the system can be obtained by taking the
Legendre transform of Lagrangian (\ref{LvHx}):
\begin{eqnarray}
H &=& \int\!\!d^3r\; \left\{
\left( \frac{\delta\L_v}{\delta\dot{U}_{\iota\kappa}}
\dot{U}_{\iota\kappa} +\mbox{h.c.} \right)
+\frac{\delta\L_v}{\delta\dot{H}_{\rho\sigma}}
\dot{H}_{\rho\sigma} - \L_v \right\} \nonumber \\
&=& \int\!\! d^3r\; \left\{ \left(\ts\frac14 f_\pi^2
\Tr(\dot{U}^\dagger\dot{U}+\vec\nabla U^\dagger\!\cdot\!\vec\nabla U)
+\cdots\right) \right. \nonumber \\
&&\hskip 2cm\left. - i\vec{v}\!\cdot\!\Tr(\vec{D} H\bar{H})
+g \Tr(H\gamma_5 A_\mu\gamma^\mu \bar{H}) \right\},
\label{Hamiltonian}\end{eqnarray}
where $\iota$ and $\kappa$ run from 1 to 2 for the 2$\times$2
matrix $U$, and $\rho$ and $\sigma$ run from 1 to 4 for the $4\times$4
matrix $H$.

We derived the above Lagrangian (\ref{LvHx}) starting
from a traditional meson Lagrangian by using the heavy quark
symmetry. Originally, it was constructed by
Wise\cite{Wise} directly from the heavy quark symmetry.

\setcounter{equation}{0}
\section{Heavy Baryons as Skyrmions}
\subsection{Heavy-Meson-Soliton Bound State}
The non-linear Lagrangian ${\cal L}_M$ supports
a classical soliton solution
\beq
U_0(\vec r) = \exp[i\vec\tau\cdot\hat{r}F(r)],
\eeqlb{CSC}
with the boundary conditions
$$F(0)=\pi \and F(r)\stackrel{r\rightarrow\infty}{\longrightarrow}0,
\seqno{CSC}{a}$$
which, due to its nontrivial topological structure,
carries a winding number identified as the baryon number
$$\begin{array}{rl}
B\!\!\! &= - \dps \frac{1}{24\pi^2}\int d^3r\; \ve^{ijk}
 \Tr(U_0^\dagger\partial_i U_0^{} U_0^\dagger\partial_j U_0^{}
     U_0^\dagger\partial_k U_0^{}) \\  \hskip 1cm
&=\dps -\frac{2}{\pi} \int^\infty_0\!\!\!r^2dr
\frac{\sin^2\!F}{r^2}F^\prime =1,
\end{array}\seqno{CSC}{b}$$
and a finite mass
$$ M_{sol} = 4\pi \int^\infty_0\!\!\!r^2dr
\frac{f_\pi^2}{2} \left( F^{\prime 2}+2\frac{\sin^2\!F}{r^2}\right),
\seqno{CSC}{c}$$
with $F^\prime=dF/dr$.
With $\L_M$ alone, however,
the solution is unstable against scale transformations;
under the transformation $U(\vec {r})\rightarrow U(\lambda\vec r)$,
the mass scales as $M_{sol}\rightarrow \lambda^{-1}M_{sol}$.
That is, it collapses into a pointlike and zero-energy configuration.
The stabilization can be established simply by adding a quartic term
(\ref{SkTerm}) to the Lagrangian, which scales like $r^{-4}$
or by incorporating the repulsion mediated by vector mesons at short distance.
As mentioned above, the Skyrme quartic term can be considered as the
infinite mass limit of the $\rho$ exchange.
(See Sec.~3)
In Fig.~5.1 is presented a typical wave function $F(r)$ stabilized
by the Skyrme term.\cite{ANW,JR}
\begin{figure}
\beginpicture
\setcoordinatesystem units <2cm,1.6cm> point at 1.75 1
\setplotarea x from 0 to 7.5, y from 0 to 3.5
\put {Figure 5.1: Typical hedgehog profile function $F(r)$ (stabilized}
 at 3.75 0.3
\put { by the Skyrme term, $r$ in unit of $(ef_\pi)^{-1}$).} [l] at 2 0.0
\setcoordinatesystem units <2cm,1.6cm> point at 0 0
\setplotarea x from 0 to 4, y from 0 to 3.5
\axis bottom ticks in numbered from 1 to 4 by 1 unlabeled short quantity 21 /
\axis left ticks in numbered from 1 to 3 by 1
  unlabeled short from 0.2 to 3.4 by 0.2 /
\axis top ticks in unlabeled long from 1 to 3 by 1 short quantity 21 /
\axis right ticks in unlabeled long from 1 to 3 by 1
  short from 0.2 to 3.4 by 0.2 /
\put {$F(r)$} at 0.1 3.7
\put {$r$} at 3.5 -0.2
\setquadratic
\plot  
 0.0 3.142  0.2 2.745  0.4 2.381  0.6 2.024  0.8 1.701  1.0 1.414  1.2 1.164
 1.4  .951  1.6  .777  1.8  .637  2.0  .527  2.2  .440  2.4  .372  2.6  .318
 2.8  .275  3.0  .240  3.2  .211  3.4  .187  3.6  .167  3.8  .150  4.0  .135 /
\setdashes
\plot  
  .0 1.996  0.2 1.931  0.4 1.767  0.6 1.523  0.8 1.255  1.0  .998  1.2  .772
 1.4  .586  1.6  .441  1.8  .332  2.0  .253  2.2  .195  2.4  .152  2.6  .121
 2.8  .097  3.0  .080  3.2  .066  3.4  .055  3.6  .046  3.8  .039  4.0  .034 /
\plot  
  .0 1.996  0.2 1.955  0.4 1.859  0.6 1.703  0.8 1.527  1.0 1.344  1.2 1.156
 1.4  .966  1.6  .783  1.8  .620  2.0  .486  2.2  .381  2.4  .301  2.6  .240
 2.8  .194  3.0  .159  3.2  .131  3.4  .110  3.6  .093  3.8  .079  4.0  .068 /
\setsolid
\put {$\dps\sim\frac{\beta}{r^2}$} [bl] at 3 0.4
\put {$\sim \pi-\alpha r$} [bl] at 0.25 2.8
\arrow <7pt> [.2,.67] from 1.1 1.9 to 0.8 1.55
\put {$-F'(r)$} [l] at 1.15 1.9
\arrow <7pt> [.2,.67] from 0.8 0.8 to 1.0 0.95
\put {$\dps\frac{\sin\!F}{r}$} [r] at 0.75 0.8
\endpicture
\end{figure}

The next step is to find the eigenstates of the heavy meson fields
interacting with the static potentials
\beqa{1.5}{l}
V^\mu=(V^0,\vec V)=(0,-i\dps(\vec\tau\times\hat{r})
\frac{\sin^2(F/2)}{r}), \\
A^\mu=(A^0,\vec A)=(0,\frac12\dps[ \frac{\sin\!F}{r}\vec{\tau}
 + (F^\prime - \frac{\sin\!F}{r}) \hat{r}\vec\tau\!\cdot\!\hat{r}]),
\eeqalb{AV}
given by the $B$=1 soliton configuration (\ref{CSC}) sitting at the origin,
with a focus on meson-soliton bound state(s).
In the rest frame, the Lagrangian takes the form
\beq
L_0 = -M_{sol} + \dps\int\!\!d^3r\left\{
-i\Tr(\partial_0 H\bar{H})
+ g\Tr(H\vec A\cdot\vec\sigma\bar{H})\right\},
\eeqlb{L0msb}
from which follows the equation of motion
\beq
i\partial_0 h(\vec{r},t) = \ve h(\vec{r},t)
 = h(\vec{r},t) [g \vec A\cdot\vec\sigma],
\eeqlb{EoM}
for the wavefunction $h(\vec{r},t)$ of the classical eigenmode
with an eigenenergy $\ve$.

In the ``hedgehog" configuration (\ref{CSC}) -- and consequently in the
static potentials (\ref{AV}), the isospin and
the angular momentum are correlated in such a way that neither of them
is separately a good quantum number, but their sum (the grand spin) is;
\beq
\vec K=\vec J+\vec I\equiv (\vec L+\vec S)+\vec I.
\eeq
Thus, the equation of motion (\ref{EoM}) is invariant under rotations
in $K$-space and the eigenstates are classified by the quantum numbers
($k;k_3$).\footnote{More specifically, 
the grand spin operator can be written as
$$ \vec K=\vec L+\vec S_\ell+\vec S_h+\vec I \equiv \vec K_\ell+\vec S_h, $$
and the heavy quark spin symmetry implies that the heavy quark spin decouples.
That is, the equation of motion is invariant separately under the heavy
quark spin rotations and under rotations by $\vec K_\ell$, the grand spin
of the light degrees of freedom. For comparison with the
conventional bound-state
approach\cite{CK} in the Skyrme model, we will work with the grand
spin $\vec{K}$
(not the grand spin of the light degrees of freedom, $\vec{K}_\ell$).
}
The wavefunctions of the heavy meson eigenmodes can be written
as the product of a radial function $h(r)$ and the eigenfunction of the
grand spin $\K_{kk_3}(\hat r)$:
\beq
h(\vec r,t)=\sum_{i} \alpha_i h^i_{k}(r)\K^i_{kk_3}(\hat{r})e^{-i\ve t},
\eeq
where the sum is over the possible ways of constructing the
eigenstates of the same grand spin and parity by combining the eigenstates
of the spin, isospin and orbital angular momentum,
and the expansion coefficients
$\alpha_i$ are normalized by $\sum_i|\alpha_i|^2=1$.
We assume that both the soliton and the heavy mesons are
infinitely heavy so that in the lowest energy state
they are on top of each other
at the same spatial point,\footnote{Note that 
the potentials take their minimum values
at the origin. (See Fig.~5.1.)} 
just propagating in time. That is, the radial functions $h_k^i(r)$ of the
lowest energy eigenstate can be approximated by a delta-function-like one,
say $h(r)$, which is strongly peaked at the origin and normalized as
$\int r^2 dr |h(r)|^2 = 1$.
Thus, the wavefunction is normalized as
\beq
-\int\!\!d^3r \Tr(h\bar{h})=1,
\eeq
with the orthonormalized
eigenfunction $\K^i_{kk_3}(\hat r)$ of the grand spin
\beq
\int\!\!d\Omega \Tr\left( \K^i_{kk_3}(\hat{r})
 \bar{\K}^{i'}_{k'k'_3}(\hat{r}) \right)
=-\delta_{ii'}\delta_{kk'}\delta_{k_3k'_3}.
\label{Knorm}\eeq

By integrating out the radial part of the equation of motion (\ref{EoM}),
we obtain
\beq
\ve\K_{kk_3}(\hat{r}) = \K_{kk_3}(\hat{r})
[\ts \frac12 gF^\prime(0) (2\sdr\tdr-\sdt)],
\eeqlb{EoM1}
with $\K_{kk_3} \equiv \sum_i c_i\K^i_{kk_3}$.
Here, we have used that $F(r)\sim\pi+F^\prime(0)r$ and consequently
$\vec{A}\cdot\vs \sim \frac12 F^\prime(0) (2\sdr\tdr-\sdt)$
near the origin. Now, our problem is reduced to finding $\K_{kk_3}$.
First, we construct the grand spin eigenstates $\K^i_{kk_3}(\hat{r})$
by combining the eigenstates of the spin, isospin and orbital
angular momentum.

Equation~(\ref{Sops}) implies that the quantum mechanical spin operators
acting on the wavefunctions are of the form
\beqa{1.5}{ll}
\vec{S}\{h\} =[\frac12\vec\sigma, h] & \mbox{spin of the heavy-meson},
\\
\vec{S}_h\{h\}=\frac12\vec\sigma h & \mbox{heavy-quark spin}, \\
\vec{S}_\ell\{h\}=-h\frac12\vec\sigma & \mbox{spin of the brown muck}.
\eeqa
The minus signs in Eq.~(\ref{Sops}) originate from the
normalization of the $H$-field. Note that the heavy {\em flavor} number
operator
$$ N_f = 2 f_h \int\!\!d^3r (P_v^{} P^\dagger_v
+ \vec{P}^{*}_v\cdot\vec{P}^{*\dagger}_v ), $$
with the heavy quark flavor $f_h(f_c \equiv +1,\; f_b \equiv -1)$
is expressed in terms of $H$ field as
$$ N_f =  - f_h \int\!\!d^3r\; \Tr(H\bar{H}). $$
Since we are working with the heavy mesons of spin 0 and 1,
the corresponding eigenstates $(s;s_3)$ of the spin operator $\vec S$
can be readily written down,
\beqa{1.5}{l}
(0;0) = [\frac{1}{2\sqrt2}(1+\gamma^0)] \gamma_5, \\
(1,+1) = [\frac{1}{2\sqrt2}(1+\gamma^0)]
    [\frac{-1}{\sqrt2}(\gamma^1+i\gamma^2)], \\
(1;0) = [\frac{1}{2\sqrt2}(1+\gamma^0)] \gamma^3, \\
(1;-1) = [\frac{1}{2\sqrt2}(1+\gamma^0)]
    [\frac{1}{\sqrt2}(\gamma^1-i\gamma^2)],
\eeqalb{Sbasis}
with the states normalized as
$$-\Tr\{(s;s_3)\overline{(s';s'_3)}\}=\delta_{ss'}\delta_{s_3s'_3}.
\eqno(\mbox{\ref{Sbasis}a})$$
The overall factor $[\frac12(1+\gamma^0)]$ is to incorporate
the fact that these states are for the heavy mesons (not for their
anti-particles).
The explicit eigenstates of the spin operators $\vec{S}_h$
and $\vec{S}_\ell$ are listed in Table~5.1.
\begin{table}
\begin{center}
{\bf Table 5.1} : Eigenstates of heavy-quark spin operators $\vec{S}_h$ and \\
the spin operator of the light degrees of freedom $\vec{S}_\ell$. \vskip 2mm
\begin{tabular}{l}
\hline
$(s_h=\frac12;+\frac12|s_\ell=\frac12;+\frac12) =
  [\frac{1}{2\sqrt2}(1+\gamma^0)]
    [\frac{-1}{\sqrt2}(\gamma^1+i\gamma^2)]$ \\

$(s_h=\frac12;+\frac12|s_\ell=\frac12;-\frac12) =
  [\frac{1}{2\sqrt2}(1+\gamma^0)]
    [\frac{1}{\sqrt2}(\gamma^3+\gamma^5)]$ \\

$(s_h=\frac12;-\frac12|s_\ell=\frac12;+\frac12) =
  [\frac{1}{2\sqrt2}(1+\gamma^0)]
    [\frac{1}{\sqrt2}(\gamma^3-\gamma^5)]$ \\

$(s_h=\frac12;-\frac12|s_\ell=\frac12;-\frac12) =
  [\frac{1}{2\sqrt2}(1+\gamma^0)]
    [\frac{1}{\sqrt2}(\gamma^1-i\gamma^2)]$ \\
\hline
\end{tabular}\end{center}
\end{table}
The relative phase between the three states $(s=1;s_3)$ and
the one between the four states
$(s_h=\frac12;\pm\frac12|s_\ell=\frac12;\pm\frac12)$
are fixed by the angular momentum structure; {\it i.e.}, they satisfy
$$\begin{array}{l}
[\frac12\sigma_{\pm},\: (s=1;\mp1)]=\sqrt2(s=1;0), \\
\frac12\sigma_{\pm}(s_h=\frac12;\mp\frac12|s_\ell;s_{\ell 3})
=(s_h=\frac12;\pm\frac12|s_\ell;s_{\ell 3}), \\
(s_h=\frac12;s_{h3}|s_\ell;\mp\frac12)(-\frac12\sigma_{\pm})
=(s_h=\frac12;s_{h3}|s_\ell;\pm\frac12).
\end{array}\eqno(\mbox{\ref{Sbasis}c})$$
However, one may still choose an arbitrary phase for the state (0;0)
without altering the final results.\footnote{The 
eigenstate of the $(0;0)$ state, $\frac1{2\sqrt2}(1+\gamma^0)\gamma_5$,
corresponds to the following convention for addition of two spins
in non-relativistic quantum mechanics:
$$\begin{array}{l}
(s=0;0) = \frac1{\sqrt2}
[(s_h=\frac12;+\frac12|s_\ell=\frac12;-\frac12)
 -(s_h=\frac12;-\frac12|s_\ell=\frac12;+\frac12)], \\
(s=1;+1) = (s_h=\frac12;+\frac12|s_\ell=\frac12;+\frac12), \\
(s=1,0) = \frac1{\sqrt2}
[(s_h=\frac12;+\frac12|s_\ell=\frac12;-\frac12)
 +(s_h=\frac12;-\frac12|s_\ell=\frac12;+\frac12), \\
(s=1;-1) = (s_h=\frac12;-\frac12|s_\ell=\frac12;-\frac12).
\end{array}$$
If one had introduced a phase $\eta$ to the state $(0;0)$ as
$(0;0)=\frac1{2\sqrt2}(1+\gamma^0)\eta\gamma_5$, then he/she would
have obtained the $(s_h=\frac12;+\frac12|s_\ell=\frac12;-\frac12)$
state to be $\frac1{\sqrt2}[(s=1;0)+\eta^*(s=0;0)]$,
while the fact that it is $[\frac{1}{2\sqrt2}(1+\gamma_0)]
    [\frac{1}{\sqrt2}(\gamma^3+\gamma^5)]$ remains unchanged.} 

The invariance of the Lagrangian (\ref{Yan}) under the isovector
transformations (\ref{Xct}) and (\ref{Pct}) with $L=R=\vartheta=V$
gives  us the isospin operator
\beq
\vec I=\vec I_M+\vec I_h,
\eeqlb{Iop}
with $\vec I_M$ the isospin operator
acting on the Goldstone boson fields
$$ \vec I_M = i\!\! \int\!\!d^3r \frac{f_\pi^2}{2}
 \Tr\{\ts\frac12\vec\tau(U^\dagger\partial_0 U + U\partial_0 U^\dagger)\},
\eqno(\mbox{\ref{Iop}a})$$
and $\vec I_h$ on the heavy meson fields
$$ \vec I_h = i\!\! \int\!\!d^3r \left\{\ts
  \Pi[-\frac14(\xi^\dagger\vec\tau\xi
 + \xi\vec{\tau}\xi^\dagger)] P^\dagger
 + \Pi^{*i}[-\frac14(\xi^\dagger\vec\tau\xi
 + \xi\vec{\tau}\xi^\dagger)] P^{*i\dagger}
     + \mbox{(h.c)} \right\}.
\seqno{Iop}{b} $$
Note that the covariant couplings to the Goldstone bosons
contribute to the isospin operator. The isospin operator of
the free heavy mesons can be obtained by turning off the couplings;
{\it i.e.}, $\xi=1$.

In the heavy mass limit, $\vec{I}_h$ can be rewritten
in terms of $P_v$ and $P_v^*$ as
$$ \vec I_h = v_0\int\!\!d^3r \ts
   \left\{ P_v [-\frac14(\xi^\dagger\vec\tau\xi
 + \xi\vec{\tau}\xi^\dagger)] P^\dagger_v
 + P^{*i}_v [-\frac14(\xi^\dagger\vec\tau\xi
 + \xi\vec{\tau}\xi^\dagger)] P^{*i\dagger}_v \right\}, $$
and in terms of the $H$-field as
\beq
\vec I_h = - v_0 \int\!\!d^3r \ts \Tr \left\{ H
[-\frac14(\xi^\dagger\vec\tau\xi
 + \xi\vec{\tau}\xi^\dagger)] \bar{H} \right\}.
\eeq
This implies that, in the rest frame and for the free heavy mesons,
the quantum mechanical isospin operators acting on the wavefunction
are
\beq
\vec I_h\{h\}=h(\vec{r},t) (-\ts\frac12 \vec\tau).
\eeqlb{Iop1}
The strange-looking minus sign comes from the fact that the heavy meson
fields form {\em isospin-anti-doublets}.  The minus sign is also essential
to make the isospin operator defined by Eq.~(\ref{Iop1}) obey the
correct commutation relations of $SU(2)$-algebra:
$$\ts [I^i_h,\: I^j_h] \{h\} =
h [-\frac12\tau^j,-\frac12\tau^i] =
i\ve^{ijk} h(-\frac12\tau^k) =i\ve^{ijk}I^k_h\{h\}. $$
Let the two eigenstates of the isospin operators be $\tp_\pm$ which satisfy
\beqa{1.5}{l}
I_h^3 \{\tp_\pm\}=\tp_\pm(-\frac12\tau^3)=\pm\frac12\tp_\pm, \\
I_h^\mp \{\tp_\pm\}=\tp_\pm(-\tau^\mp)=\tp_\mp.
\eeqa
Explicitly, we have
\beqa{1.5}{l}
\tp_{+}=(0,-1)\sim\mbox{isospin state of $\bar{d}$},\\
\tp_{-}=(+1,0)\sim\mbox{isospin state of $\bar{u}$}.
\eeqa

First, we combine the orbital angular momentum and the isospin.
Let the resulting spherical spinor harmonics be
$\Y_{\lambda,\ell,\lambda_3}(\hat{r})$ which are the eigenfunctions of
$\vec\Lambda (\equiv \vec L + \vec I_h)$:
\beq
\Y_{\lambda,\ell,\lambda_3} = \sum_{i_3,m}
\ts (\ell,m,\frac12,i_3|\lambda,\lambda_3) Y_{\ell m}(\hat{r})\tp_{i_3},
\eeq
where $(\ell_1,m_1,\ell_2,m_2|j,m)$ is the Clebsch-Gordan coefficient.
Since we are interested in the lowest energy eigenmode of positive parity,
we can restrict the angular momentum
$\ell$ to be 1.\footnote{The heavy 
mesons have negative intrinsic parity. In general, the differential
equations for the radial functions $h^i_k(r)$ have the centrifugal
term with a singularity $\ell_{\eff}(\ell_{\eff}+1)/r^2$ near the origin.
It requires for the radial functions to behave as
$h^i_k(r) \sim r^{\ell_{\eff}}$ near the origin. Here, $\ell_{\eff}$ is the
``effective" angular momentum\cite{CK}, which is related to the usual
angular momentum $\ell$ as
$$
\ell_{\eff} = \left\{ \begin{array}{ll}
  \ell+1, & \mbox{if $\lambda=\ell+1/2$}, \\
  \ell-1, & \mbox{if $\lambda=\ell-1/2$}.
\end{array} \right.
$$
Due to the vector potential from the soliton configuration
$\vec{V}(\sim i(\hat{r}\times\vec{\tau})/r$, near the origin),
the singular structure of $\vec{D}^2=(\vec{\nabla}-\vec{V})^2$
is altered from $\ell(\ell+1)/r^2$ of the usual $\vec\nabla^2$ to
$\ell_{\eff}(\ell_{\eff}+1)/r^2$. Thus, the states with $\ell_{\eff}=0$
can have most strongly peaked radial functions and become the lowest
eigenstate. Note that $\ell_{\eff}=0$ can be achieved only when $\ell=1$.
}
 With $\ell$=1, two $\lambda$ values, $\frac12$ and $\frac32$, are possible,
so we have $\Y_{\frac12,1,\lambda_3}(\hr)$ and
$\Y_{\frac32,1,\lambda_3}(\hr)$. Explicitly, they are
\beqa{1.5}{l}
\Y_{\frac12,1,\pm\frac12}(\hr)
 = +\sqrt{\frac{1}{4\pi}}\tp_{\pm}(\tdr),\\
\Y_{\frac32,1,\pm\frac32}(\hr)
 = \mp\sqrt{\frac{1}{24\pi}} \tp_{\pm} \tilde{O}_\pm (\tdr), \\
\Y_{\frac32,1,\pm\frac12}(\hr)
 = -\sqrt{\frac{1}{8\pi}} \tp_{\pm} \tilde{O}_0 (\tdr)
 = \pm\sqrt{\frac{1}{8\pi}} \tp_{\mp} \tilde{O}_\pm (\tdr),
\eeqa
where $\tilde{O}_i \equiv (\tau^i - 3\hr^i\tdr)$ with
$\tilde{O}_\pm = \tilde{O}_1 \pm i\tilde{O}_2$ and $\tilde{O}_0=\tilde{O}_3$.
The explicit factorization of $(\tdr)$ is for later convenience.
Note that the term $[2\sdr\tdr-\sdt]$ in the equation of motion (\ref{EoM1})
can be simply expressed as $(\tdr)(\sigma\!\cdot\!\tau)(\tdr)$.

Next, we combine the spin $\vec S$ and $\vec\Lambda$. We will restrict our
consideration to the $k=\frac12$ state, which is expected to be the lowest
energy state from our experience of the bound-state approach in
the Skyrme model.\cite{CK} Since we have $s$=0, 1 and
$\lambda$=$\frac12$, $\frac32$, we can construct three different
grand spin states of $k=\frac12$: $\K^{(i)}_{k,k_3}(\hr)(i$=1,2,3).
Explicitly,
\beqa{1.5}{l}
\K^{(1)}_{\frac12,\pm\frac12}(\hr)=[\frac1{2\sqrt2}(1+\gamma^0)]
 \gamma_5 \tp_\pm(\tdr)\sqrt{\frac{1}{4\pi}}, \\
\K^{(2)}_{\frac12,\pm\frac12}(\hr)=[\frac1{2\sqrt2}(1+\gamma^0)]
 \sqrt{\frac13} \tp_\pm[\vec\gamma\!\cdot\!\vec\tau](\tdr)
 \sqrt{\frac{1}{4\pi}}, \\
\K^{(3)}_{\frac12,\pm\frac12}(\hr)=[\frac1{2\sqrt2}(1+\gamma^0)]
 \sqrt{\frac16}\tp_\pm [\vec\gamma\!\cdot\!\vec\tau
 - 3\vec\gamma\!\cdot\!\hr \tdr](\tdr) \sqrt{\frac{1}{4\pi}}.
\eeqalb{Ki}

The eigenstates $\K_{\frac12,\pm\frac12}(\hr)$ of the equation of motion
(\ref{EoM1}) can be expanded in terms of these bases:
\beq
\K_{\frac12,\pm\frac12}(\hr)
 = \sum_{i=1}^3 c_i \K^{(i)}_{\frac12,\pm\frac12}(\hr),
\eeqlb{ES}
with the expansion coefficients given by the solution of the
secular equation
$$ \sum_{j=1}^3\M_{ij}c_j=-\ve c_i,
\seqno{ES}{a}$$
where the matrix elements $\M_{ij}$ are defined by
$$\begin{array}{l}
\M_{ij} = \dps \int\!d\Omega \Tr\{\K^{(i)}(\hr)
[\ts\frac12gF^\prime(0)(2\sdr\tdr-\sdt)] \bar{\K}^{(j)}(\hr)\} \\
\hskip 1cm =\dps \int\!d\Omega \Tr\{\K^{(i)}(\hr)(\tdr)
[\ts\frac12gF^\prime(0)(\sdt)](\tdr)\bar{\K}^{(j)}(\hr)\}.
\end{array}\eqno(\mbox{\ref{ES}b})$$
Here again, the minus sign in Eq.~(\ref{ES}a) is due to the fact that the
basis states $\K^{(i)}_{\frac12,\pm\frac12}(\hr)$ are
normalized as (\ref{Knorm}). With the explicit form of
$\K^i_{\frac12,\pm\frac12}(\hr)$ given by Eq.~(\ref{Ki}),
the matrix elements come out to be
\beq
\M=-gF^\prime(0)\left(\!\!\begin{array}{ccc}
 0 \!&\!\! \frac{\sqrt3}{2} \!&\!\! 0  \\
 \frac{\sqrt3}{2} \!&\!\! -1 \!&\!\! 0  \\
 0 \!&\!\! 0 \!&\!\! +\frac12
\end{array}\!\!\right),
\eeqlb{M}
which is independent of $k_3$(=$\frac12$ or $-\frac12$).
All the matrix elements between $\K^{(3)}_{\frac12,\pm\frac12}(\hr)$ and
$\K^{(1,2)}_{\frac12,\pm\frac12}(\hr)$ vanish. This vanishing can be easily
seen from Eq.~(\ref{ES}b) and the fact that $(\K^{(3)}\tdr)$ and
$(\K^{(1,2)}\tdr)$
are $\ell$=2 and 0 states, respectively, while $\sdt$ cannot
have matrix elements between states of different $\ell$.
The secular equation (\ref{ES}a) with these matrix elements
yields three eigenstates as listed in Table~5.2. Since $g<0$ and
$F^\prime(0)<0$ (in the case of the
baryon-number-1 soliton solution), we have a
heavy-meson-soliton bound state of binding energy $\frac32\gF$.
The two {\em unbound} eigenstates with positive eigenenergy
$+\frac12\gF$ are not consistent with the strongly peaked radial
functions. They are improper solutions of Eq.~(\ref{EoM1}).
\begin{table}
\begin{center}
{\bf Table 5.2} : Eigenvalues and eigenfunctions of $k^\pi=\frac12^+$ states.
\vskip 2mm
\begin{tabular}{ccccc}
\hline
$\ve$ & $c_1$ & $c_2$ & $c_3$ & $\K_{\frac12,\pm\frac12}(\hr)$ \\
\hline
$-\frac32gF^\prime(0)$ & $\frac12$ & $-\frac{\sqrt3}{2}$ & 0 &
$[\frac{1}{2\sqrt2}(1+\gamma^0)] \frac12\tp_\pm
[\gamma_5-\vec\gamma\!\cdot\!\vec{\tau}](\tdr)\sqrt{\frac{1}{4\pi}}$ \\
$+\frac12gF^\prime(0)$ & $\frac{\sqrt3}{2}$ & $\frac12$ & 0 &
$[\frac{1}{2\sqrt2}(1+\gamma^0)] \frac{1}{2\sqrt3}\tp_\pm
[3\gamma_5+\vec\gamma\!\cdot\!\vec\tau](\tdr)\sqrt{\frac{1}{4\pi}}$ \\
$+\frac12gF^\prime(0)$ & 0 & 0 & 1 &
$[\frac{1}{2\sqrt2}(1+\gamma^0)] \frac{1}{\sqrt6}\tp_\pm
[\vec\gamma\!\cdot\!\vec\tau-3\vec\gamma\!\cdot\!\hr\tdr](\tdr)
\sqrt{\frac{1}{4\pi}}$ \\
\hline
\end{tabular}
\end{center}
\end{table} 

The result is independent of the phase choice for the spin basis $(s=0;0)$.
If we had given a phase $\eta$ to that state, we would have obtained
$\M_{12}=-\frac{\sqrt3}{2}\eta gF^\prime(0)$
and $\M_{21}=-\frac{\sqrt3}{2}\eta^* gF^\prime(0)$, while the others remain
unchanged.  But the secular equation must yield the same eigenenergies
and the same eigenstates as in Table~5.2 independently
of the phase $\eta$.
In Refs.~\citenum{MOPR,NRZ}, an ansatz $\K^{NRZ}_{\frac12,\pm\frac12}(\hr)=
[\frac{1}{2\sqrt2}(1+\gamma^0)] \frac12\tp_\pm
[i\eta\gamma_5-\vec\gamma\!\cdot\!\hr](\tdr)\sqrt{\frac{1}{4\pi}}$
was taken as the wavefunction of the lowest energy state. However,
it is not the eigenstate of the equation of motion (\ref{EoM}). It is
just one orthonormal basis of the three possible states of $k=\frac12$ and
$\ell=1$. The ground state energy of Refs.~\citenum{MOPR,NRZ},
$\frac34g_H F^\prime(0)(1+\frac12i(\eta-\eta^*))$\footnote{The error
of factor 3 committed in Ref.~\citenum{NRZ} is corrected. Note that in
Ref.~\citenum{NRZ} a different sign convention is adopted for the
coupling constant; {\it i.e.\/}, $g_H=-g$.},
should be understood as one of the matrix elements similar to
Eq.~(\ref{M}) and its $\eta$-dependence has no physical significance.

Let $h_\n$ be the wavefunctions for the eigenmodes classified
by the set of quantum number $\n=\{k,k_3,\cdots\}$.\footnote{The ellipsis
denotes other quantum numbers, such as parity, radial quantum number etc.}
Then, the heavy-meson field $H(x)$ can be expanded in terms of these eigenmodes
as
\beq
H(x)=\sum_\n h_\n^{}(\vec{r})e^{-i\ve_\n^{} t} a_\n^{},
\eeq
with the meson annihilation (creation) operator
$a_\n^{}$ $(a_\n^\dagger)$ corresponding to the
eigenstate satisfying the commutation relations
$$ [a_\n^{}, a^\dagger_{\{m\}}]=\delta_{\n\{m\}},
\and [a_\n^{}, a_{\{m\}}^{}] = [a_\n^\dagger, a_{\{m\}}^\dagger]=0. $$
We note that anti-particle creation operators do not figure in this expansion.
Focusing on the lowest energy eigenstate (heavy-meson--soliton bound state)
for simplicity, we write $H(x)$ explicitly as
\beq
H(x) = e^{-i\ve_{bs}t} (h_{+} a_{+} + h_{-} a_{-}) + \cdots
\eeqlb{Hxpnd}
where the subscript $\pm$ denotes that it is associated with the
bound state of $k_3=\pm\frac12$ and $\ve_{bs}=-\frac32 gF^\prime(0)$.
The heavy-meson--soliton bound state of the grand spin $k_3=\pm\frac12$
can be described by the Fock state
$a_{\pm}^\dagger|0\rangle$ with the vacuum state $|0\rangle$
defined by $a_\n|0\rangle=0$ for any $\n$.

Substituting Eq.~(\ref{Hxpnd}) into the heavy-quark spin operator
of (\ref{Sops}) and after normal-ordering, we obtain an interesting result:
\beqa{1.5}{l}
S_h^3 = \frac12 a_{+}^\dagger a_{+}^{}
      - \frac12 a_{-}^\dagger a_{-}^{} + \cdots, \\
S_h^+ = a_{+}^\dagger a_{-}^{} + \cdots, \\
S_h^- = a_{-}^\dagger a_{+}^{} + \cdots.
\eeqa
This shows that
the grand spin of the state can be identified as the
heavy-quark spin. This is the spin-grand spin
transmutation.\cite{NRZ}
Finally the Hamiltonian (\ref{Hamiltonian}) has a diagonal form of
\beqa{1.5}{rcl}
H_{\vec{v}=0} &=& \dps M_{sol}
-g \int\!\!d^3r \Tr(H\vec{A}\!\cdot\!\vec\sigma\bar{H}) \\
&=& M_{sol} +
\ve_{bs}(a_{+}^\dagger a_{+}^{} + a_{-}^\dagger a_{-}^{}) +\cdots.
\eeqa

\subsection{Collective Coordinate Quantization}
What we have obtained so far is the heavy-meson--soliton bound state which
carries a baryon number and a heavy flavor.
Therefore up to this order, baryons containing a heavy quark such as
$\Lambda_Q$, $\Sigma_Q$ and $\Sigma_Q^*$ are degenerate in mass.
However, to extract {\em physical\/} heavy baryons of correct spin and isospin,
we have to go to the next order in $1/N_c$,
while remaining in the same order in $m_Q$; namely, $O(m_Q^0 N_c^{-1})$.
This can be done by quantizing the zero modes associated with the
degeneracy under simultaneous $SU(2)$ rotation of the soliton
configuration together with the heavy meson fields:
$$
\xi^{}_0 \rightarrow C \xi^{}_0 C^\dagger \and
H \rightarrow H C^\dagger, $$
with an arbitrary constant $SU(2)$ matrix $C$ and $\xi^2_0 \equiv U_0$.
The rotation matrix becomes a dynamical variable
 when the $SU(2)$ collective variables are endowed time dependence as
\beq
\xi(\vec{r},t) = C(t) \xi^{}_0(\vec{r}) C^\dagger(t) \and
H(\vec{r},t) = H_{\bff}(\vec{r},t) C^\dagger(t),
\eeqlb{CV}
and then the quantization is done by elevating the collective variables
to the corresponding quantum mechanical operators.
In Eq.~(\ref{CV}), $H_{\bff}$ refers to the heavy meson field in the (isospin)
co-moving frame, while $H(\vec{r},t)$ refers to that in the
laboratory frame, {\it i.e.}, the heavy-quark rest frame.
Substituting Eq.~(\ref{CV}) and keeping terms up to $O(m_Q^0 N_c^{-1})$,
we obtain the Lagrangian as
\beq
L = L^1 + L^0 + L^{-1},
\eeqlb{Lcol}
where $L^{q}$ denotes the Lagrangian of order $m_Q^0 N_c^{q}$: \viz,
\addtocounter{equation}{-1}
$$ L^1 = - M_{sol},
\seqno{Lcol}{a}$$
$$ L^0 = \int\!d^3r \left\{ -i\Tr(\partial_0 H_{\bff}\bar{H}_{\bff})
  + g\Tr(H_{\bff}\,\vec{A}\!\cdot\!\vec\sigma\,\bar{H}_{\bff}) \right\},
\seqno{Lcol}{b}$$
$$ L^{-1} = \ts \frac12{\cal I}\omega^2 + \int\!d^3r \left\{
- \ts\frac14\Tr\{ H_{\bff} \,
[-\frac14(\xi_0^\dagger\tdo \xi + \xi_0\tdo \xi_0^\dagger)]
\,\bar{H}_{\bff}\}
   \right\}.
\seqno{Lcol}{c}$$
The ``angular velocity" $\vec\omega$  of the collective
rotation is defined by
$$ C^\dagger\partial_0 C \equiv \ts \frac12 i\vec\tau\!\cdot\!\vec\omega,
\seqno{Lcol}{d}$$
and ${\cal I}$ is the moment of inertia of the rotating soliton
$$ {\cal I}=\ts\frac{8\pi}{3}f^2_\pi\!\dps\int^\infty_0\!\!\!r^2dr
   (\sin^2 F+\cdots),
\seqno{Lcol}{e}$$
where the ellipsis stands for contributions from higher-derivative terms
(such as the quartic term etc.).
Note that $V_0$ does not vanish for the ``rotating" soliton, so once
again the covariant derivative $D_0$ plays a non-trivial role
as in Eq.~(\ref{Iop}b).

Given the Lagrangian (\ref{Lcol}) that describes dynamics up to order
$O(m_Q^0N_c^{-1})$, one has the equation of motion
consistent to that order:
\beq
\ts i\partial_0 H_{\bff}
= H_{\bff}\,\left\{ g\vec{A}\!\cdot\!\vec{\sigma}
  -\frac14(\xi_0^\dagger\tdo \xi + \xi_0\tdo \xi_0^\dagger)] \right\}.
\eeq
Note that the last ``Coriolis" term in the equation of motion
couples the fast and slow degrees of freedom.\cite{Rho91}
Although the heavy mesons are infinitely heavy, their angular
momentum and the isospin are associated with the light constituents.
Thus, we may take those light degrees of freedom of the heavy meson
fields as  ``fast" variables
and the collective rotation as ``slow" variables.
Note further that the scale of the eigenenergies $|\ve_n|$
of the heavy mesons is much greater than that of the
rotational velocity; $|\ve_n|\gg |\omega|$.

A generally accepted procedure of handling these different scales is as
follows. We first solve the equation of motion for fast degrees of freedom with
slow degrees of freedom ``frozen." In this way, we get ``snap-shot" pictures
of the fast motion. Next we solve the equation of motion for slow degrees
of freedom taking into account the ``relic" of the fast motion that
has been ``integrated out," in a manner completely analogous
to the incorporation of Berry phases\cite{SW}.
In Ref.~\citenum{NRZ}, it is shown that the ``Coriolis effect" on the
heavy-meson--soliton system does not induce\cite{BP} any non-trivial Berry
phase
as far as the $k^P=\frac12^+$ multiplets are concerned.
It is also analogous to the ``strong-coupling limit" of the particle-rotor
model\cite{BM} in nuclear physics, where the coupling between the
rotating ``core" and the particle is much stronger than the perturbation
of the single-particle motion by a Coriolis interaction. Here, the
roles of the particle and the rotor are played by the bound heavy-mesons
and the rotating soliton configuration.
Thus, we may make the assumption that the bound heavy mesons rotate
together with the soliton core in the {\em unchanged eigenmodes}.
It enables us to expand the $H_{\bff}(x)$ in terms of the classical
eigenmodes obtained in Sec.~5.1 as
\beq
H_{\bff}(x)=\sum_\n h_\n^{}(\vec{r})e^{-i\ve_\n^{}t} a_\n^{}
=e^{-i\ve_{bs}^{}t} ( h_{+}^{}a_{+}^{} +  h_{-}^{}a_{-}^{}) + \cdots,
\eeq
and to describe the heavy-meson soliton bound state by the Fock state
$|\pm\frac12\rangle_{bs} \equiv a^\dagger_{\pm}|0\rangle$.

Taking the Legendre transform of the Lagrangian (\ref{Lcol})\footnote{One
can get the same Hamiltonian by substituting Eq.~(\ref{CV})
into the Hamiltonian (\ref{Hamiltonian}) and keeping terms up
$O(m_Q^0N_C^{-1})$.}
we obtain the Hamiltonian as
\beqa{1.5}{rl}
H_{\vec{v}=0}\!\!\! &= \dps \int\!d^3r \left\{
\frac{\delta\L}{\delta(\dot{H}_{\bff,ab})} \dot{H}_{\bff,ab}\right\}
+\frac{\delta L}{\delta\omega_i}\omega_i - L \\
 & \dps = M_{sol} - g\!\int\!d^3r\,
\Tr(H_{\bff}\,\vec{A}\!\cdot\!\vec{\sigma}\,\bar{H}_{\bff})
+ \frac{1}{2{\cal I}}[\vec J_R-\vec\Phi(\infty)]^2 ,
\eeqalb{Hcol}
where $\vec J_R$ is the canonical momenta conjugate to the collective
variables $C(t)$:
$$ J^i_R \equiv \frac{\delta L^{\rot}_0}{\delta\omega^i}
= {\cal I}\omega^i + \Phi^i(\infty),
\seqno{Hcol}{a} $$
with $\vec\Phi(\infty)$ defined by
$$ \vec\Phi(\infty) \equiv
- \dps \int\! d^3r\Tr \left\{ H_{\bff}
\ts\frac14(\xi^\dagger_0\vt \xi + \xi_0 \vt \xi_0^\dagger)]
\bar{H}_{\bff} \right\},
\seqno{Hcol}{b} $$
whose expectation value with respect to the state $|\pm\frac12\rangle_{bs}$
is the Berry phase {\it in the heavy-quark symmetry limit}
associated with the collective rotation.
It can be easily shown that this Berry phase vanishes
identically; {\it viz.\/},
$$\begin{array}{rl} \langle \vec\Phi(\infty) \rangle \!\!\! &\equiv \ts
\,^{}_{bs}\langle\pm\frac12| - \dps \int\! d^3r
\Tr(H_{\bff}
\ts\frac14(\xi^\dagger_0\vt \xi + \xi_0 \vt \xi_0^\dagger)]
\bar{H}_{\bff}) |\ts\pm\frac12\rangle^{}_{bs} \\
& =-\frac12\dps\int\!\! d \Omega \;\Tr(h_{\pm}^{}
(\vt\cdot\hr)\vt(\vt\cdot\hr) \bar{h}_{\pm}) = 0.\end{array}$$
Note that $\vec\Phi(\infty)$ is {\em the isospin operator of heavy mesons
(in the body fixed frame) modulo the sign}.

With the collective variable introduced as in Eq.~(\ref{CV}), the
isospin of the fields $U(x)$ and $H(x)$ is entirely shifted to $C(t)$.
To see this, consider the isospin rotation
$$ U \rightarrow {\cal A} U {\cal A}^\dagger, \hskip 8mm
   H \rightarrow H {\cal A}^\dagger, $$
with ${\cal A}\in SU(2)_V$, under which the collective variables and fields
in body-fixed frame transform as
\beq
C(t) \rightarrow {\cal A} C(t), \hskip 8mm
H_{\bff}(x) \rightarrow H_{\bff}(x).
\eeq
The $H$-field is isospin-blind in the (isospin) co-moving frame.
The conventional Noether construction gives the isospin of the system,
\beq
I^a = \ts \frac12 \Tr ( \tau^a C \tau^b C^\dagger) \,
[ {\cal I} \omega^b + \Phi^b(\infty) ] = D^{ab}(C) J_R^b,
\eeq
where $D^{ab}(C)$ is the adjoint representation of the $SU(2)$ associated
with the collective variables $C(t)$. One may also obtain the same
isospin by substituting Eq.~(\ref{CV}) into Eq.~(\ref{Iop}) and keeping
terms up to $O(m_Q^0N_c^{-1})$:
$$\begin{array}{rl}
I^i \!\!\! &= \int\!\!d^3r\; \left\{ \left ( i \frac12 {f_\pi^2}
 \Tr [ \ts\frac12 \tau^i ( U^\dagger \partial_0 U
+ U \partial_0 U^\dagger)] + \cdots\right)
+ \Tr [ H (\frac12\tau^i) \bar{H}] \right\} \\
& = \ts\frac12 \Tr(\tau^i C \tau^j C^\dagger)
\, [{\cal I} \omega^j + \Phi^j(\infty)].
\end{array}$$

Under the spatial rotation, with the help of the $K$-symmetry,
the fields transform as
$$\begin{array}{l}
U(\vec{r},t) \rightarrow
e^{i\vec\alpha\cdot\vec{L}} U(\vec{r}^\prime,t)
= C(t){\cal B}^\dagger U_0(\vec{r}) {\cal B} C^\dagger(t), \\
H(\vec{r},t) \rightarrow
e^{i\vec\alpha\cdot\vec\sigma/2} H(\vec{r}^\prime,t)
                 e^{-i\vec\alpha\cdot\vec\sigma/2}
= [e^{i\vec\alpha\cdot\vec\sigma/2} H_{\bff}(\vec{r}^\prime,t)
     e^{-i\vec\alpha\cdot(\vec\sigma+\vec\tau)/2}] {\cal B}^\dagger C(t),
\end{array}$$
with $\vec{r}^\prime = \exp(i\vec\alpha\cdot\vec{L}) \vec{r}$
and ${\cal B}=\exp(i\vec\alpha\cdot\vec\tau/2)\in SU(2)$.
This means that the spatial rotation acts on the collective
variables and $H$-fields in the body fixed frame as
\beq \begin{array}{l}
C(t) \rightarrow C(t) {\cal B}^\dagger,  \\
H_{\bff}(x) \rightarrow
e^{i\vec\alpha\cdot\vec{L}} e^{i\vec\alpha\cdot\vec\sigma/2}
  H_{\bff}(x) e^{-i\vec\alpha\cdot(\vec\sigma+\vec\tau)/2}.
\end{array} \eeq
Therefore, we see that the spin of the $H_{\bff}(x)$ is the grand spin,
\ie, the spin--grand-spin transmutation.
The $H_{\bff}(x)$ becomes isospin-blind; that is, the isospin
of the $H$-field is transmuted into the part of the spin in the isospin
co-moving frame. Applying the Noether theorem to the Lagrangian (\ref{Lcol}),
we obtain the spin of the system explicitly as
\beq
\vec{J} = \vec{J}_R + \vec{K}_{\bff},
\eeq
with the grand spin of the heavy meson fields
(in the body-fixed coordinate system)
$$ \vec{K}_{\bff} = \int\!\!d^3r\;
\ts \Tr \left\{ \{\vec{L} H_{\bff} + [\frac12\vs,H_{\bff}]
    + H_{\bff}(-\frac12\vt)\}\bar{H}_{\bff} \right\}. $$

We should point out that
the heavy-quark spin symmetry of the Lagrangian under the transformation
\beq
H(x) \rightarrow e^{i\vec\alpha\cdot\vec\sigma/2} H(x)
=[e^{i\vec\alpha\cdot\vec\sigma/2} H_{\bff}(x)] C(t),
\eeq
has nothing to do with the collective rotation. The heavy-quark spin
operator remains unchanged in the isospin co-moving frame:
\beq
\vec S_Q^{} =-\int\!\!d^3r\;\Tr(\ts\frac12\vec\sigma H\bar{H})
 = \dps -\int\!\!d^3r\;\Tr(\ts\frac12\vec\sigma H_{\bff}\bar{H}_{\bff}).
\eeq

Upon canonical quantization, the collective variables become the quantum
mechanical operators; the isospin ($I$), the spin ($J$) and the
spin of the rotor ($J_R$) discussed so far become the corresponding
operators $\tilde{I}^i$, $\tilde{J}^i$ and $\tilde{J}_R$, respectively.
We distinguish those operators associated with the collective coordinate
quantization by a tilde.
The eigenfunctions of the rotor-spin operator $\tilde{J}^i_R$ are
the Wigner $D$-functions
\beq
\sqrt{2I+1} D^{(I)}_{MK}(C),
\eeqlb{WigD}
with $M,K=-I,-I+1,\cdots,I$ and
$$ \tilde{J}_R^i\tilde{J}^i_R D^{(I)}_{MK}(C) = I(I+1) D^{(I)}_{MK}(C),
\seqno{WigD}{a}$$
$$ \tilde{J}_R^3 D^{(I)}_{MK}(C) = -K D^{(I)}_{MK}(C),
\seqno{WigD}{b}$$
$$ D^{3i}\tilde{J}_R^i D^{(I)}_{MK}(C) = M D^{(I)}_{MK}(C).
\seqno{WigD}{c}$$

Now, in terms of these eigenfunctions and the bound-heavy-meson state
$|\pm\frac12\rangle_{bs}$, the heavy baryon state of isospin $i_3$ and spin
$s_3$ containing a heavy quark can be constructed as
\beqa{1.5}{l}
\dps |\Sigma_Q^{}; i_3,s_3\rangle = \sqrt{3} \sum_{m}
\ts (1,s_3-m,\frac12,m|\frac12,s_3) D^{(1)}_{1,-s_3+m}(C)|m\rangle_{bs},\\
\dps |\Sigma^*_Q; i_3,s_3\rangle = \sqrt{3} \sum_{m}
\ts (1,s_3-m,\frac12,m|\frac32,s_3) D^{(1)}_{1,-s_3+m}(C)|m\rangle_{bs},\\
\dps |\Lambda_Q^{}; 0,s_3\rangle = D^{(0)}_{0,0}(C) |s_3\rangle_{bs},
\eeqalb{HBS}
where $(j_1,m_1,j_2,m_2|j,m)$ is the usual Clebsch-Gordan coefficient.
To give correct quantum numbers to the baryons, we have quantized the
rotor as a boson.\cite{SW10}
They are the eigenstates of the collective Hamiltonian
(\ref{Hcol}) with the eigenenergies
\beqa{1.5}{l}
\dps E_{\Sigma_Q^{}} = E_{\Sigma^*_Q}
 = M_{sol} + \ve_{bs}^{} + \frac{11}{8\I}, \\
\dps E_{\Lambda_Q^{}} = M_{sol} + \ve_{bs}^{} + \frac{3}{8\I}.
\eeqa
Here, we have {\em exactly} evaluated the expectation value
of the operator $\vec\Phi^2(\infty)$ with respect to the Fock state
$|m\rangle_{bs}(m=\pm\frac12)$ and obtained\cite{OPM3}
\beq
{^{}_{bs}\langle m|\vec\Phi^2(\infty)|m\rangle_{bs}} = \frac34,
\eeqlb{Exact}
instead of taking into account only the bound state
contribution
\beq
{^{}_{bs}\langle m|\vec\Phi^2(\infty)|m\rangle_{bs}}
= \sum {^{}_{bs}\langle m | \vec{\Phi} | n \rangle \cdot
 \langle n | \vec{\Phi} m \rangle_{bs} }
\approx |^{}_{bs}\langle m|\vec\Phi(\infty)|m\rangle_{bs}|^2 = 0.
\eeqlb{Apprx}
One can show Eq.~(\ref{Exact})
by carrying out the summation in Eq.~(\ref{Apprx}) over the
complete set of energy eigenstates.

With vanishing Berry phase, the Hamiltonian
depends only on the rotor-spin so that $\Sigma_Q$ and $\Sigma^*_Q$
become degenerate as expected from the heavy-quark symmetry.
In order to compare the results with experimental
heavy baryon masses, we have to add the heavy meson masses subtracted so far
from the  eigenenergies. The mass formulas to be compared with data are
\beqa{1.3}{l}
m^{}_{\Sigma_Q^{}} = m^{}_{\Sigma^*_Q}
 = M_{sol} + \m_P^{} - \frac32 \gF + \dps\frac{11}{8\I}, \\
m^{}_{\Lambda_Q^{}}
 = M_{sol} + \m_P^{} - \frac32 \gF + \dps\frac{3}{8\I},
\eeqalb{HBmass1}
where $\m_P^{}$ is the weighted average of the heavy meson multiplets;
$\m_P^{}=\frac14(3m^{}_{P^*}+m^{}_P)$.
In the case of $Q=c$, we have $\m_P^{}=1975$~MeV.
The $SU(2)$ quantities $M_{sol}$ and
${\cal I}$ are obtained from the nucleon and $\Delta$
masses\cite{ANW}\footnote{ As discussed elsewhere\cite{elafmr}, we could
do better by calculating the $O(N_c^0)$ Casimir energy which is of the order
of $-1/2$ GeV and fitting the $N$ and $\Delta$ spectrum to obtain the
parameters of the $SU(2)$ sector. Unfortunately the Casimir energy calculation
is not yet sufficiently accurate enough to be quantitatively useful at
present. We believe that the
procedure used here is not really satisfactory and could certainly be improved
upon when the Casimir calculation is put under control.}:
\beq
M_{sol}=866\mbox{~MeV}, \and 1/{\cal I}=195\mbox{~MeV}.
\eeq
Finally, the unknown value of $\gF$ is adjusted to fit the
observed value of the $\Lambda_c$ mass,
$$m^{}_{\Lambda^{}_c}=2285\mbox{ MeV}=M_{sol}+\m_P^{}-\frac32 \gF , $$
which implies that
\beq
\gF = 419 \mbox{ MeV}.
\eeq
This set of parameters leads to a prediction on the
$\Lambda_b$ mass and the average mass of the
$\Sigma^{}_Q$-$\Sigma^*_Q$ multiplets,
$\overline{m}^{}_{\Sigma^{}_Q} [\equiv
\frac13(2m^{}_{\Sigma^*_Q}+m^{}_{\Sigma^{}_Q})]$,
\beqa{1.5}{l}
m_{\Lambda^{}_b} = M_{sol} + \overline{m}_B - \frac32\gF
+3/8\I = 5623 \mbox{ MeV} , \\
\overline{m}^{}_{\Sigma^{}_c} = M_{sol} + \overline{m}_D
- \frac32\gF + 11/8\I = 2480 \mbox{ MeV} .
\eeqa
These are comparable with the experimental masses of $\Lambda_b$
(5641 MeV) and $\Sigma_c$ (2453 MeV)\cite{PDG}
 and $\Sigma^*_c$ (2530 MeV)\cite{SKAT}.
Furthermore, with the Skyrme Lagrangian (with the quartic term for
stabilization), the wavefunction has a slope
$F^\prime(0)\sim -2ef_\pi\approx -690$ MeV\footnote{In order 
to yield $M_{sol}=866$~MeV and $1/{\cal I}=195$~MeV, the Skyrme
parameter $e$ and the meson $f_\pi^{}$ are adjusted to be 5.45 and 63MeV,
respectively. See Ref.~\citenum{ANW}. }
near the origin, which implies $g\sim -0.61$.
This is also consistent with the experimental limit $g^2<0.5$
and with the nonrelativistic quark model prediction $g=-\frac34$.

\subsection{Alternative Approach}
Up to now, we have discussed how one can obtain the heavy baryon states
containing a heavy quark, $\Sigma_Q^{}$, $\Sigma^*_Q$ and $\Lambda_Q^{}$,
as heavy-meson--soliton bound states treated in the standard way:
a heavy-meson-soliton bound state is
first found and then quantized by rotating the {\em whole} system
in the collective coordinate quantization scheme. This amounts to
proceeding systematically in a decreasing order in $N_c$; that is,
in the first step, only terms up to $N_c^0$ order are considered
and in the next step, terms of order $1/N_c$ order and so on.
In this way of proceeding, the heavy mesons first lose their quantum numbers
(such as the spin and isospin), with only the grand spin preserved.
The good quantum numbers are recovered when the
whole system is quantized properly.

An alternative approach adopted in Ref.~\citenum{JMW} is more
natural in the ``top-down" approach. In this approach, the  soliton
is first quantized to produce the light baryon states such as nucleons and
$\Delta$'s with correct quantum numbers. Then,
the heavy mesons with explicit spin and isospin are
coupled to the light baryons to form heavy baryons as a bound state.
Compared with the traditional one which is a ``body-fixed" approach,
this approach may be interpreted as a  ``laboratory-frame" approach.

To start with, we redefine the meson fields so that the
Lagrangian density is expressed in terms of $U$ instead of $\xi$ which
has a coordinate singularity.
The new heavy meson fields $P^\prime_v$ and $P^{*\prime}_{v\mu}$ defined by
\beq
H^\prime =\frac{1+\sv}{2}
(\gamma_5 P^\prime_v-\gamma^\mu P^{*\prime}_{v\mu})  = H \xi,
\eeq
transform under chiral $SU(2)_L\times SU(2)_R$ (\ref{Pct}) as
\beq
H^\prime \rightarrow H^\prime R^\dagger.
\eeq
The Lagrangian density (\ref{LvHx}) now reads
\beq
\L = \L_M -iv^\mu \Tr(\partial_\mu H^\prime\bar{H}^\prime)
+{\textstyle\frac{i}{2}}v^\mu \Tr(H^\prime U^\dagger \partial_\mu U
   \bar{H}^\prime)
-{\textstyle\frac{i}{2}}g\Tr(H^\prime\gamma_5 U^\dagger
   \partial_\mu U \gamma^\mu\bar{H}^\prime).
\eeq
As was discussed in Sec. 2, the parity transformation for the primed heavy
meson fields is a little more complicated:
\beq
H^\prime(t,\vec{r}) \rightarrow \gamma^0 H^\prime(t,-\vec{r})\gamma^0
U^\dagger(t,-\vec{r}),
\label{parity}\eeq
compared with that for the unprimed fields
$$ H(t,\vec{r})\rightarrow \gamma^0 H(t,-\vec{r})\gamma^0. $$
Note that in the background Goldstone boson field configuration of
a soliton located at the same spatial point as the heavy meson,
the factor of $U^\dagger$ becomes $-1$, whereas $U^\dagger=1$
for a meson infinitely far from the soliton. This relative minus
sign is the source of the parity flip that gives positive
parity heavy-meson-soliton bound states.

The (approximate) soliton solution of the $SU(2)_L\times SU(2)_R$ chiral
Lagrangian $\L_M$
$$ U(\vec{r},t)=C(t) U_0(\vec{r}) C^\dagger(t), $$
is substituted into the Lagrangian (\ref{LvHx}). In the rest frame,
up to order $1/N_c$, the Lagrangian reads
\beq\begin{array}{l}
L_0 = - M_{sol} + \frac12{\cal I}\omega^2 \\
\hskip 1cm - \dps\int\!\!d^3r\;
     \left\{i\Tr(\partial_0 H^\prime\bar{H}^\prime)
     +\frac{i}{2}\Tr(H^\prime U^\dagger\partial_0 U \bar{H}^\prime)
     +\frac{ig}{2}\Tr(H^\prime\gamma_5
\vec\gamma\cdot[U^\dagger\vec{\nabla}U]\bar{H}^\prime)\right\}.
\end{array}\label{L0prime}\eeq
The $1/N_c$ terms with time derivative on $U$ are neglected as we
are assuming the large-$N_c$ limit at which the soliton is very heavy.

First collective quantization of the soliton leads to the light baryon
states $|s,s_3;i,i_3\}_{lb}$
with the wavefunctions given by the Wigner $D$-functions $D^{(I)}_{MK}(C)$:
\beq \begin{array}{ll}
\phi^N_{s=\frac12,s_3;i=\frac12,i_3}(C)
=\sqrt{2} D^{(\frac12)}_{i_3,-s_3}(C) & \mbox{for nucleons}, \\
\phi^\Delta_{s=\frac32,s_3;i=\frac32,i_3}(C)
=\sqrt{4} D^{(\frac32)}_{i_3;-s_3}(C) & \mbox{for deltas}.
\end{array} \eeq
In the large-$N_c$ limit, these states are degenerate.
Next, the interaction Hamiltonian
\beq
H_I=-\frac{ig}{2}\int\!\!d^3r\;\Tr(H^\prime\gamma_5 \vec\gamma\cdot
[U^\dagger\vec\nabla U]\bar{H}^\prime),
\eeq
determines the potential energy of a configuration with a baryon (soliton)
sitting at the origin and the heavy mesons at position $\vec{r}$.
Assuming that in attractive channels the potential energy is
minimized at $\vec{r}=\vec{0}$ where the heavy meson and the
baryon soliton coincide, we can reduce the problem of determining the
bound-state spectrum to finding the eigenvalues of the potential operator
at the origin.  Near the origin, we have
$$\begin{array}{ll}
U^\dagger\vec\nabla U \!\!
&= \dps i C(t)\{\tdr\,\hat r (F^\prime-\frac{\sin2F}{2r})
+\vec\tau\,\frac{\sin2F}{2r}
+\vec\tau\!\times\!\hat{r}\,\frac{\sin^2F}{r}\} C^\dagger(t) \\
&= i F^\prime(0) C(t)\vec\tau C^\dagger(t).
\end{array}$$
The interaction energy is therefore
\beq\begin{array}{l}
V_I(0)=\frac12 gF^\prime(0)\dps\int\!\!d^3r\;\Tr(H^\prime\gamma_5
\vec\gamma\cdot [C(t)\vec\tau C^\dagger(t)]\bar{H}^\prime)\\
\hskip 1cm = 2g F^\prime(0) S_\ell^a I_h^b D^{ba}(C),
\end{array}\label{VIorg}\eeq
where $S_\ell$ and $I_h$ are the spin operator of the light degrees of
freedom and the isospin operator acting on the heavy meson states,
respectively, and we have used that
$$ C(t)\tau^a C^\dagger(t)=\tau^b D^{ba}(C),
\eqno(\mbox{\ref{VIorg}a})$$
and that
$$ S^a_\ell I^b_h = - \int\!\!d^3r \Tr \ts (H^\prime \{-\frac12\sigma^a\}
\{-\frac12\tau^b\} \bar{H}^\prime).
\eqno(\mbox{\ref{VIorg}b})$$
(See the discussions in Sec.~4.2 and 5.1 on the spin and
isospin operators.)

Note that the potential operator is invariant under rotation
by the total angular momentum operator
of the light degrees of freedom (both soliton and light anti-quarks in
heavy meson combined)
$J_\ell^a(\equiv S^a_\ell + \tilde{J}^a_R)$
and under rotation by the total isospin operator
$I^a\equiv I^a_h + \tilde{I}^a_R$. Furthermore, it is completely
independent of the heavy quark spin as required by the heavy-quark spin
symmetry. Thus, the eigenstates of the potential operator can be
classified by the corresponding quantum numbers $j_\ell$, $j_{\ell 3}$,
$i$, $i_3$ and $s_{h3}$; {\em viz.},
$|j_\ell,j_{\ell 3};i,i_3\rangle\!\rangle|
\mbox{$s_h$=$\frac12$},s_{h3}\rangle$.
Let the eigenstates of $\vec{J}_\ell$ and $\vec{I}$
be denoted by $|j_\ell,j_{\ell 3};i,i_3\rangle$,
constructed by combining the solitonic light baryon states
$|s_R^{},s_{R3}^{};i_R^{},i_{R3}^{}\}$ and the light degrees of freedom
in the heavy meson states $|s^{}_\ell,s^{}_{\ell 3}; i_h^{}, i^{}_{h3})$:
\beq
|j_\ell,j_{\ell 3};i,i_3\rangle =
\sum_{s^{}_{R3},i^{}_{R3}} \begin{array}{l}
\renewcommand{\arraystretch}{0.8}
(s^{}_R, s_{R3}^{}, s_\ell, s_{\ell3} | j^{}_\ell, j^{}_{\ell 3})
(i^{}_R, i_{R3}^{}, i_h^{}, i^{}_{h3} | i, i_3) \\
\hskip 1cm \cdot
|s^{}_R, s^{}_{R3}; i^{}_R, i^{}_{R3}\}
|s^{}_\ell, s^{}_{\ell3}; i^{}_h, i^{}_{h3}),
\renewcommand{\arraystretch}{1.5}\end{array}\eeq
with the appropriate Clebsch-Gordan coefficients.
Since we have the light baryon states of $i=s=1/2.$ $3/2$, $\cdots$,
and $i_h=s_\ell^{}=1/2$ for the light degrees
of freedom of the heavy mesons, we can construct the states
with $j_\ell, i=0,1,\cdots$. Hereafter, unless necessary, we will not specify
the third components explicitly  and simply denote
the basis states as $|j_\ell, i\rangle$, as the eigenstates
are degenerate in $j_{\ell 3}$ and $i_3$.
Finally the eigenstates of the potential operator are found by
diagonalizing the potential matrix calculated with the
states $|j_\ell, i\rangle$ as basis states.

When the basis is truncated with only nucleon-heavy meson products,
it is straightforward to show that
\beq
V_I(0)=-\ts\frac23 gF^\prime(0) S_\ell^a I_h^b \tilde{J}_R^a \tilde{I}^b_R
=-\frac23 gF^\prime(0) (J_\ell^2-3/2)(I^2-3/2).
\label{VI0}\eeq
This implies that the basis states $|j_\ell, i\rangle$ themselves are
the approximate eigenstates with eigenenergies
$-\frac23 gF^\prime(0)
[j_\ell(j_\ell+1)-3/2][i(i+1)-3/2]$.\footnote{More
precisely, this should be interpreted as matrix
elements of the potential operator.}
In getting this result, we have used the fact
that $D^{ab}(C)$ can be written in terms of
the $D$-functions, namely,
\beq
D^{3,3}(C)=D^{(1)}_{0,0}(C), \hskip 5mm
\mp\frac1{\sqrt2}[D^{1,3}(C) \pm i D^{2,3}(C)]=D^{(1)}_{\pm1,0}(C),
\eeq
and that
\beq
\int\!\!dC D^{(j_3)*}_{m_1^\prime m_1}(C)
D^{(j_2)}_{m_2^\prime m_2}(C) D^{(j_1)}_{m_1^\prime m_1}(C)
= \frac{1}{2j_3+1}
(j_1 m_1^\prime j_2 m_2^\prime | j_3 m_3^\prime)
(j_1 m_1 j_2 m_2 | j_3 m_3),
\eeq
with the Clebsch-Gordan coefficients
$(\cdot\cdot\cdot\cdot|\cdot\cdot)$.
Specifically, the expectation values with respect to the nucleon states
are obtained as
\beq
\ts \{ \frac12,s^\prime_3;\frac12,i^\prime_3| D^{ab}(C)
|\frac12,s^{}_3;\frac12,i^{}_3\}
=-\frac43 \{ \frac12,s^\prime_3;\frac12,i^\prime_3|
\tilde{J}_R^a \tilde{I}_R^b |\frac12,s^{}_3;\frac12,i^{}_3\}.
\eeq

\begin{table}
\begin{center}
{\bf Table 5.3:} Eigenstates and Eigenenergies of $V_I(0)$.\vskip 2mm
\begin{tabular}{ccc}
\hline
States & Energies neglecting the $\Delta$ & Energies including the $\Delta$ \\
\hline
$|0,\frac12,0\rangle$ & $-\frac32 gF^\prime(0)$ & $-\frac32 gF^\prime(0)$ \\
$|1,\frac12,1\rangle, |1,\frac32,1\rangle$ &
  $-\frac16 gF^\prime(0)$ & $-\frac32 gF^\prime(0)$ \\
$|1,\frac12,0\rangle$ & $+\frac12 gF^\prime(0)$ & $+\frac12 gF^\prime(0)$ \\
$|0,\frac12,1\rangle, |0,\frac32,1\rangle$ &
  $+\frac12 gF^\prime(0)$ & $+\frac12 gF^\prime(0)$ \\
\hline
\end{tabular}
\end{center}
\end{table} 

The eigenstate of $V_I(0)$ with definite isospin and spin can be
obtained by combining the heavy quark spin to the total angular
momentum for the light degrees of freedom. Let these eigenstates
be denoted by $|i,j,j_\ell\rangle$, where $i$ is the total isospin,
$j$ the total spin ($\vec{J}=\vec{J}_\ell+\vec{S}_h$) and $j_\ell$
the angular momentum of the light degrees of freedom.
Given in Table~5.3 are the eigenstates and
eigenvalues of $V_I(0)$ in the truncated basis.
Only the $|0,\frac12,0\rangle$, $|1,\frac12,1\rangle$ and
$|1,\frac32,1\rangle$ states are bound. For the case of $h=c$,
these states have the right quantum numbers $i$, $j$, $j_\ell$ and
the parity\footnote{The spatial wave functions 
of the eigenstates of Eq.~(\ref{VI0}) carry zero orbital angular momentum.
The parity of the meson-soliton bound state is thus even, because
the primed heavy-meson fields are odd under parity at infinity
where they are free, and are even under parity at the origin,
as noted below Eq.~(\ref{parity}). The unprimed heavy-meson fields
have a simple transformation law under parity, and do not have a
relative minus sign between the parity at infinity and parity
at the origin. However, in the $\xi$ basis, the wavefunction
of the bound state contains a factor of the form $\tdr$ near
the origin and it is the state of $\ell=1$.
(See Sec.~5.1)} 
to be $\Lambda_c$, $\Sigma^{}_c$ and $\Sigma^*_c$, respectively.

In the large-$N_c$ limit, the $N$ and $\Delta$ are degenerate
and the space of the basis states should be enlarged to
include products of $\Delta$-baryons with the heavy-mesons.
There appear two states in the $j_\ell=i=1$  channel,
obtained by combining the light degrees of
freedom of the heavy mesons with either the $N$ ($s_R^{}=i_R^{}=1/2$)
or the $\Delta$ ($s_R^{}=i_R^{}=3/2$) states. These states will be
distinguished by explicitly specifying the light baryon states as
$|1,1;N\rangle$ and $|1,1;\Delta\rangle$, where the first label refers
to the isospin and the second to the spin of the light degrees of freedom.
One can evaluate the interaction Hamiltonian Eq.~(\ref{VI0})
in the $|1,1;N\rangle$ and $|1,1;\Delta\rangle$ basis as
\beq
V_I(0)= -\frac{gF^\prime(0)}{6}\left( \begin{array}{cc} 1 & 4\sqrt2 \\
4\sqrt2 & 5 \end{array}\right).
\eeq
Diagonalizing it, we can obtain two $|1,1\rangle\!\rangle$ eigenstates
of $V_I(0)$ as
\beq\begin{array}{l}
|1,1\rangle\!\rangle_{-} =\sqrt{\frac13}\:|1,1;N\rangle
+ \sqrt{\frac23}\:|1,1;\Delta\rangle, \\
|1,1\rangle\!\rangle_{+} =\sqrt{\frac23}\:|1,1;N\rangle
- \sqrt{\frac13}\:|1,1;\Delta\rangle,
\end{array}\eeq
with eigenenergies $-3gF^\prime(0)/2$ and $gF^\prime(0)/2$, respectively.
In the large $N_c$ limit, we see that the states $|0,0\rangle\!\rangle$
and $|1,1\rangle\!\rangle_0$ are degenerate. Again, the former,
when combined with the heavy quark spin, is the spin-1/2 $\Lambda_Q$
baryon, and the state $|1,1\rangle\!\rangle_0$ when combined with the
heavy quark spin is the degenerate multiplet of $\Sigma_Q$
and the spin 3/2 $\Sigma^*_Q$: \viz,
\beqa{1.5}{l}
|\Sigma_Q^*,\frac32\rangle = |1\ 1\ 1\rr_{-}|\uparrow\rangle_Q, \\
|\Sigma_Q, \frac12\rangle = \sqrt{\frac23} |1\ 1\ 1\rr_{-}|\downarrow\rangle
    - \sqrt{\frac13}|1\ 1\ 0\rr_{0} |\uparrow\rangle, \\
|\Lambda_Q, \frac12\rangle = |0\ 0\ 0\rr |\uparrow\rangle.
\eeqa
Here, the state $|\ \rangle_Q$ denotes the heavy quark spin state
and $|i\ j_\ell\ m\rr$ represents the state of the light degrees of
freedom $|i\ j_\ell\rr$ with $j_{\ell 3}$=$m$.

In the large $N_c$-limit, $\Lambda_Q$ and $\Sigma_Q$ are also degenerate.
The degeneracy is lifted when the terms of
order $1/N_c$ so far neglected are taken into account.
There are two $1/N_c$-order terms in Eq.~(\ref{L0prime}). The first
is the rotational kinetic energy of the soliton that lifts the
$N$-$\Delta$ degeneracy. This term has a coefficient of order $N_c$
(the moment of inertia) in the Lagrangian, and has two time
derivatives each of which brings a factor of $1/N_c$ suppression
so that it produces an energy splitting of order $1/N_c$.
It contributes additional masses, $3/8{\cal I}$ and $15/8{\cal I}$,
to the nucleon and delta masses, respectively, and thus it yields
the $N$-$\Delta$ mass splitting of $3/2{\cal I}$.
The second is the term with $V_0=\frac12 U^\dagger\partial_0 U$,
having a coefficient of order $N_c^0$ and one time derivative.
For the $SU(2)$ soliton, however, $U^\dagger\partial_0 U$ vanishes
at the origin where the heavy meson is bound.
When the additional masses of the light baryon states
are included, the interaction energy of the $\Lambda_h$ state,
$|0,\frac12,0\rangle$, is modified to $-3gF^\prime(0)/2 + 3/8{\cal I}$.
Thus the potential matrix in the $|1,1;N\rangle$ and $|1,1;\Delta\rangle$
channel is of the form\cite{JMW,GLM}
\beq
V_I(0)= -\frac{gF^\prime(0)}{6}\left( \begin{array}{cc} 1 & 4\sqrt2 \\
4\sqrt2 & 5 \end{array}\right) + \frac{3}{8{\cal I}}
\left(\begin{array}{cc} 1 & 0 \\ 0 & 5 \end{array}\right).
\eeq
To first order in $1/{\cal I}$, the eigenvalues come out as
\beq E_{-}=-\frac{3gF^\prime(0)}{2} + \frac{11}{{8\cal I}}, \hskip 5mm
E_{+}=+\frac{gF^\prime(0)}{2} + \frac{7}{8{\cal I}}.
\eeq
The eigenenergy $E_{-}$ leads to the same heavy baryon
masses as Eq.~(\ref{HBmass1}).
The corresponding eigenstate is
\beq
|1,1\rr_{\epsilon} = a |1,1; N\rangle + b |1,1;\Delta\rangle
= |1,1\rr_{-} + \epsilon |1,1\rr_{+},
\eeqlb{ES11}
where
$$ \ts a = \sqrt{\frac13} + \sqrt{\frac23}\epsilon, \hskip 5mm
b = \sqrt{\frac23} - \sqrt{\frac13}\epsilon,
\seqno{ES11}{a}$$
with
$$ \epsilon = \frac{1}{2\sqrt2 \I \gF}.
\seqno{ES11}{b}$$

\setcounter{equation}{0}
\section{Further Developments}
In this section, we  discuss briefly some of more recent developments in the
heavy-meson-soliton bound state approach to heavy baryons, in particular,
the role of light vector mesons and finite mass corrections.
Because of space limitation, we shall have to leave out
various applications of this model, \eg,
the heavy-baryon Isgur--Wise function\cite{JMW2}, the SU(3) extension
of the background soliton\cite{MSS}, exotic\cite{OPM2} and
excited\cite{OPM3} states of heavy baryons, etc.

\subsection{Light Vector Mesons}
So far we have considered interactions among the heavy mesons
and the light pseudoscalar mesons to first order in derivative
expansion involving the latter.
It is known that in low-energy hadron physics, the vector mesons saturate
the next-to-leading-order counter terms (dimension-four operators)
in chiral Lagrangians and improve
substantially the description of light meson and baryon dynamics.
It is therefore natural to expect that introducing light vector mesons
would have non-trivial effects on the interaction of heavy particles
with light ones\cite{SS,UGVA}. For example, the semileptonic
$D\rightarrow K^*$ transition appears to dominate over
$D\rightarrow K\pi$.\cite{PDG} Indeed it has been recently shown that
light vector mesons play an important role in heavy-meson--soliton
bound states\cite{GMSS}: the $\rho$-meson contribution to the binding
energy is found to be 60\% as large as that of the pion while
the $\omega$-meson contribution is 40\% as large. The sign of the coupling
constants involved is not yet determined and hence one cannot say whether
their contributions are attractive or repulsive.\cite{MSS,SS,UGVA,UGVA2}
In this section, we choose one possible case as an illustration.
We consider the possibility that the $\omega$ contributes repulsively and
$\rho$ attractively\cite{GMSS} with the Lagrangian developed in Sec.~3.
For the case that both $\omega$ and $\rho$ mesons contribute
attractively, see Ref.~\citenum{MSS}.

To proceed, we first modify Lagrangian (\ref{LvHx}) so as to incorporate
the light vector mesons $\rho$ and $\omega$. One may do this [in the $SU(2)$
sector where there is no anomaly] either by the external-gauging of
the flavor (the
massive Yang-Mills method)\cite{SS} or by the hidden local symmetry
approach.\cite{UGVA} Here we follow Ref.~\citenum{SS} which uses the
former approach.
Let the vector and axial vector mesons be linear combinations
of the fields $A_\mu^L$ and $A_\mu^R$ which transform under the
chiral transformation Eqs.~(\ref{Xct}) and (\ref{Pct}) as
\beq
A_\mu^L \rightarrow A^{\prime L}_\mu=L A_\mu^L L^\dagger, \hskip 5mm
A_\mu^R \rightarrow A^{\prime R}_\mu=R A_\mu^R R^\dagger.
\eeq
We integrate out the axial vector mesons\cite{KS} by writing
$A_\mu^L$ and $A_\mu^R$ in terms of the physical vector field
$\rho_\mu[= \frac12 (\omega_\mu \mbox{\bf 1}
+\vec\tau\cdot\vec\rho_\mu)]$\footnote{Here, 
we include the isoscalar $\omega$ meson in the flavor symmetry group
$U(2)$.} 
as
\beq
\begin{array}{l}
\dps A_\mu^L = \xi \rho_\mu \xi^\dagger
    + \frac{i}{g_*} \xi \partial_\mu \xi^\dagger, \\
\dps A_\mu^R = \xi^\dagger \rho_\mu \xi
    + \frac{i}{g_*} \xi^\dagger \partial_\mu \xi,
\end{array}
\eeq
with the vector meson coupling constant $g_*$.
One can see that $\rho_\mu$ transforms as
\beq
\rho_\mu \rightarrow \rho^\prime_\mu
 = \vartheta \rho_\mu \vartheta^\dagger
+\frac{i}{g_*} \vartheta\partial_\mu \vartheta^\dagger,
\eeqlb{Rct}
and its field strength tensor
$F_{\mu\nu}(\rho)\equiv \partial_\mu \rho_\nu - \partial_\nu \rho_\mu
-i g_* [\rho_\mu, \rho_\nu]$ as
\beq
F_{\mu\nu}(\rho)\rightarrow F^\prime_{\mu\nu}(\rho)
=\vartheta F_{\mu\nu}(\rho) \vartheta^\dagger.
\eeq
Now, the ``minimal" chiral Lagrangian of light pseudoscalars
and vectors\cite{KS} corresponding to the normal sector of
$\L_M$ in the Lagrangian (\ref{LvHx}) is
\beq\begin{array}{l}
\L^\prime_M = -\frac12\Tr[F_{\mu\nu}(\rho) F^{\mu\nu}(\rho)]\dps
+ \frac{m_\rho^2}{4a}(1+a)\Tr(A_\mu^L A^{\mu L} + A_\mu^R A^{\mu R})\\
\hskip 2cm
\dps - \frac{m^2_\rho}{2a}(1-a)\Tr(A^L_\mu U A^{\mu R} U^\dagger),
\end{array}\eeq
where $m_\rho$ is the light vector meson mass and $a$ is a dimensionless
constant defined by $a=(m_\rho/f_\pi g_*)^2$.
The ``magic" value $a$=2 again gives the KSRF relation.
Note that this Lagrangian contains the kinetic and interaction terms of the
pseudoscalar mesons; {\it i.e.},
$\frac14 f^2_\pi\Tr(\partial_\mu U \partial^\mu U^\dagger),$
and that it is identical to the upper $SU(2)$ part of
Eq.~(\ref{L0}) in unitary gauge $\xi^\dagger_L=\xi^{}_R=\xi$
in the hidden gauge symmetry approach; {\it viz.\/},
$$\begin{array}{l}
\dps \frac{m_\rho^2}{4a}\Tr(A_\mu^L A^{\mu L} + A_\mu^R A^{\mu R}
\mp 2 A_\mu^L U A^{\mu R} U^\dagger) \\
\hskip 1cm = \frac14 f_\pi^2 g^2_*
  \Tr[\xi^\dagger A_\mu^L \xi \mp \xi A_\mu^R \xi^\dagger]^2 \\
\hskip 1cm = \frac14 f_\pi^2 g^2_* \dps
  \Tr[ (\rho_\mu + \frac{i}{g_*}\partial_\mu \xi^\dagger \xi)
   \mp (\rho_\mu + \frac{i}{g_*}\partial_\mu \xi \xi^\dagger)]^2 \\
\hskip 1cm = -\frac14 f_\pi^2 \Tr[\D_\mu\xi^\dagger \xi
\mp \D_\mu\xi \xi^\dagger]^2,
\mbox{ with $\D_\mu\xi = (\partial_\mu - ig_*\rho_\mu)\xi$.}
\end{array}$$

The anomalous-parity sector of $\L_M$ for the light mesons
is given by\cite{MKWW87,MKW87}
\beq\begin{array}{l}
\dps \L^\prime_{WZ} = \frac{N_c g_*}{2} \omega_\mu \ve^{\mu\nu\lambda\kappa}
\frac{1}{24\pi^2} \Tr(U^\dagger\partial_\nu U U^\dagger\partial_\lambda U
 U^\dagger\partial_\kappa U) \\
\hskip 1cm \dps
 + \dps\frac{N_c g^2_*}{64\pi^2}\ve^{\mu\nu\lambda\kappa} \omega_{\mu\nu}
\Tr \left\{ i\vec{\tau}\cdot\vec{\rho}_\lambda
(U^\dagger\partial_\kappa U + \partial_\kappa UU^\dagger)
  +\frac{g_*}2 \vec{\tau}\cdot\vec{\rho}_\lambda U^\dagger
  \vec{\tau}\cdot\vec{\rho}_\kappa U\right\},
\end{array}\label{LpWZ}\eeq
with the number of colors $N_c$(=3) and
$$\omega_{\mu\nu}=\partial_\mu\omega_\nu - \partial_\nu\omega_\mu.
\eqno(\mbox{\ref{LpWZ}a})$$
The first term of $\L^\prime_{WZ}$ gives the $\omega$-coupling
to the topologically conserved baryon number current $B^\mu$
with the $\omega NN$ coupling\cite{AN} given by universality,
$$g_{\omega\pi\pi\pi} = g_{\omega NN} = g_{\omega}
= \ts\frac12 N_c g_* \cong 9
\eqno(\mbox{\ref{LpWZ}b})$$
and the second term describes $\omega\rho\pi$ interactions.

The rest of the Lagrangian (\ref{LvHx}) can be generalized by
replacing the covariant derivative $D_\mu$ appearing there by
a suitably extended form. At this point, one has a choice.
As can be seen from (\ref{AVct}) and (\ref{Rct}), both the ``vector
field" $V_\mu$ built of the pseudoscalar fields, and the vector-meson
field $\rho_\mu$ transform in the same way. Therefore, a generalized
covariant derivative can be defined as
\beq\begin{array}{l}
D^\prime_\mu H = H[\stackrel{\leftarrow}{\partial}_\mu
 + i\alpha g_* \rho_\mu^\dagger + (1-\alpha) V^\dagger_\mu], \\
D^\prime_\mu \bar{H} = [\partial_\mu - i \alpha g_* \rho_\mu
 + (1-\alpha) V_\mu ] \bar{H},
\end{array}\eeq
with a dimensionless parameter $\alpha$ which reflects the extent
to which the two pseudoscalars emitted in relative $P$-wave come from an
intermediate vector state, with $\alpha=1$ representing ``vector-meson
dominance." Chiral symmetry alone cannot
fix the value of  $\alpha$ and one has to fix it from experiments.

Another chiral-invariant interaction that may be important is
\beq
\L_{PP^*\rho}^{} =
i\beta \Tr [H\gamma^\mu \gamma^\nu F_{\mu\nu}(\rho) \bar{H}],
\label{PP*v}\eeq
where $\beta$ is an unknown coupling constant. This Lagrangian can be
gotten from the heavy-mass limit of the ordinary Lagrangian given
in terms of $P$ and $P^*_\mu$:
$$\L_{PP^*\rho}^{} = -2\beta_1 P^*_\mu F^{\mu\nu}(\rho) P^{*\dagger}_\nu
 - \beta_2 \ve^{\mu\nu\lambda\kappa}
(P F_{\mu\nu} P^{*\dagger}_{\lambda\kappa}
+P^*_{\lambda\kappa} F_{\mu\nu} P^\dagger). $$
The heavy quark symmetry implies $\beta_1/M_{P^*}^{}=\beta_2=2\beta$.
This higher-dimension term is needed to provide the trilinear coupling
among $D$, $D^*_s$ and $K^*$ (for three light flavors)\cite{SS,UGVA},
which gives the $D^*_s$ pole contribution to $D\rightarrow
K^*$ decay. In Ref.~\citenum{UGVA}, the trilinear coupling constant $\beta$
is determined by comparing the calculated $D\rightarrow K^*$ semileptonic
weak decay with the data from the E653 Collaboration\cite{E653}
\beq
\beta/g_* = -(0.63 \pm 0.17) \mbox{GeV}^{-1}.
\eeq
One may add a term similar to (\ref{PP*v}) in which $F_{\mu\nu}(\rho)$
is replaced by
$F_{\mu\nu}(V)=\partial_\mu V_\nu - \partial_\nu V_\mu +[V_\mu, V_\nu]$:
$$\L_{PP^*\rho}^\prime =
i\beta^\prime \Tr [H\gamma^\mu \gamma^\nu F_{\mu\nu}(V) \bar{H}]. $$
In the spirit of (light) vector-meson dominance, we would expect this
term to be less important.

To sum up, the leading terms of the chiral invariant heavy-meson
Lagrangian written in terms of the doublet $H$ field are
\beq
\L = \L^\prime_M + \L^\prime_{WZ} - iv^\mu \Tr(D^\prime_\mu H\bar{H})
     - g \Tr(H\gamma_5 A_\mu \gamma^\mu\bar{H})
     + i \beta \Tr (H\gamma^\mu \gamma^\nu F_{\mu\nu}(\rho) \bar{H}).
\eeq
This Lagrangian contains two parameters, $\alpha$ and
$\beta$, in addition to the coupling constant $g$. For simplicity, we have
not incorporated the chiral symmetry breaking pion mass term.

The Lagrangian $\L^\prime_M+\L^\prime_{WZ}$ for the light mesons
supports a classical soliton solution with the hedgehog ansatz
\beq\begin{array}{c}
U^{}_c(\vec{r})=\exp[i\tdr F(r)], \\
\omega^{\mu=0}_c(\vec{r}) = \omega(r), \hskip 5mm
\omega^{\mu=i}_c(\vec{r}) = 0 , \\
\dps \rho^{\mu=i,a}_c(\vec{r})
 = \frac{1}{g_* r}\varepsilon^{ika} \hat{r}^k G(r), \hskip 5mm
\rho^{\mu=0,a}_c(\vec{r})=0,
\end{array}\label{HA}\eeq
with three radial functions: $F(r)$, $G(r)$ and $\omega(r)$.
Here, $i$=1,2,3
denote the three space components of the vector fields and $a$=1,2,3
are the isospin indices.
Substituting the hedgehog ansatz (\ref{HA}) into the Lagrangian
$\L_M^\prime+\L^\prime_{WZ}$, one gets the static energy functional
for the soliton mass
\beq\begin{array}{l}
\dps E[F,G,\omega] = 4\pi\int^\infty_0\!\! r^2 dr \left\{
\frac{f_\pi^2}{2} \left[ F^{\prime 2} + \frac{2\sin^2 F}{r^2}\right]
 + \frac{4f^2_\pi}{r^2}(G+1-\cos F)^2 \right.\\
\dps\hskip 4cm
 - (\ts f^2_\pi g^2_* \omega^2 + \frac12\omega^{\prime 2})
\dps + \frac{G^{\prime 2}}{g_*^2 r^2} + \frac{G^2(G+2)^2}{2g^2_* r^4} \\
\dps\hskip 4cm \left.
+ \frac{3g_*}{4\pi^2} \omega F^\prime \frac{\sin^2F}{r^2}
+ \frac{3g}{16\pi^2}  \omega^\prime \frac{G(G+2)}{r^2}\sin 2F\right\}.
\end{array}\eeq
Coupled equations of motion for $F(r)$, $G(r)$ and $\omega(r)$
can be derived by functionally minimizing the static energy:
\beq\begin{array}{l}
\dps F^{\prime\prime} = -\frac{2}{r} F^\prime + \frac{1}{r^2}
 [4(G+1)\sin\!F - \sin\!2F] \\
\dps\hskip 2cm
+ \frac{3g_*}{8\pi^2 f^2_\pi}
 \frac{\omega^\prime}{r^2}[-2\sin^2\!F + G(G+2)\cos\!2F], \\
\dps G^{\prime\prime} = 2g_*^2 f_\pi^2 (G + 1 - \cos\!F)
 + \frac{1}{r^2}G(G+1)(G+2) \\
\dps\hskip 2cm
+  \frac{3g^3_*}{16\pi^2} \omega^\prime(G+1) \sin\!2F, \\
\dps \omega^{\prime\prime} = -\frac{2}{r}\omega^\prime
 + 2f_\pi^2 g_*^2 \omega - \frac{3g_*}{4\pi^2 r^2} F^\prime \sin^2\!F \\
\dps \hskip 15mm
 + \frac{3g_*}{8\pi^2 r^2}[G(G+2) F^\prime \cos\!2F + G^\prime(G+1)\sin\!2F].
\end{array}\label{FGweq}\eeq
To ensure a singularity-free baryon number $B=n$ solution and finiteness of the
energy, we impose the boundary conditions on
$F(r)$, $G(r)$ and $\omega(r)$:
\beq\begin{array}{lll}
F(0)=n\pi, & G(0)=-[1-(-1)^n], & \omega^\prime(0)=0, \\
F(\infty)=0, & G(\infty)=0, & \omega(\infty)=0.
\end{array}\label{FGwbc}\eeq
The stability of the soliton solution is assured
by the repulsion generated by the vector mesons at short distance
(without any additional term like the Skyrme term).

Classical eigenmodes of the heavy mesons
moving under the static potentials provided by the classical soliton
configuration sitting at the origin are given by the equation of motion
\beq
i\partial_0 h(\vec{r},t) = h(\vec{r},t) \left\{
g\vec{\sigma}\cdot\vec{A} + \ts \frac12 \alpha g_* \omega(r)
 + i \beta \gamma^i\gamma^j F^{ij}(\rho) \right\},
\label{heq}\eeq
where
$$ \vec{A} = \ts\frac12  \dps\left[\frac{\sin\!F}{r} \vec{\tau}
+ (F^\prime - \frac{\sin\!F}{r})\hat{r} \tdr \right],
\eqno(\mbox{\ref{heq}a})$$
$$ \gamma^i\gamma^j F^{ij}(\rho)
\dps = \frac{i}{g_*r} \left[ -G^\prime \sdt
+ \frac{1}{r}(rG^\prime - G ( G + 2 ) ) \sdr \tdr \right].
\eqno(\mbox{\ref{heq}b})$$
Near the origin where the heavy mesons are expected to be strongly peaked,
the profile functions have the asymptotic structures
\beq\begin{array}{c}
F(r) = \pi + r F^\prime(0)
   + \frac16 r^3 F^{\prime\prime\prime}(0) + \cdots, \\
\omega(r) = \omega(0) + \frac12 r^2\omega^{\prime\prime}(0)+\cdots, \\
G(r) = - 2 + \frac12 r^2 G^{\prime\prime}(0)
  + \frac1{24} r^4 G^{\prime\prime\prime\prime}(0) + \cdots,
\end{array}\eeq
so that the potentials can be expanded as
\beq \begin{array}{l}
\omega = \omega(0) + O(r^2), \\
\vec{\sigma}\cdot\vec{A}
 = \frac12 F^\prime(0) (2\sdr\tdr - \sdt) + O(r^2), \\
\gamma^i\gamma^j F^{ij}(\rho)
\dps = \frac{i}{g_*} G^{\prime\prime}(0)(2\sdr\tdr - \sdt) + O(r^2).
\end{array}\eeq
Note that $\vec\sigma\cdot\vec{A}$ and $\gamma^i\gamma^j F^{ij}(\rho)$
have the same structure at the origin. The problem becomes now exactly
the same as that in Sec.~5.1 except that we have an overall energy
shift by an amount of $\frac12\alpha\omega(0)$ and that
$\frac12g F^\prime(0)$ appearing there is now replaced by
$\frac12gF^\prime - \beta G^{\prime\prime}(0)/g_*$:
the eigenenergy of the bound state is
\beq
\ve_{bs} = - \ts \frac32 gF^\prime(0)
  + \dps 3\frac{\beta}{g_*} G^{\prime\prime}(0)
  - \ts\frac12\alpha g_*\omega(0),
\eeqlb{Ebs}
while the corresponding eigenfunction remains the same as given in
Table~5.2. As for the other two ``unbound" states, the $g$ and $\beta$
terms are both multiplied by $-\frac13$ while the $\alpha$ term remains
unchanged.

Now, our problem is to calculate the derivative of the radial functions
$F(r)$, $G(r)$ and
$\omega(r)$  that figure in the bound state energy (\ref{Ebs}).
These are found by solving the coupled differential equations
(\ref{FGweq}) subject to the boundary conditions (\ref{FGwbc}).
The solution depends on the structure and parameters of the light meson
Lagrangian.

Instead of going into details,  we shall be content with
a rough estimate of
the contribution of the vector mesons to the heavy meson binding
energy to see their nontrivial roles.
In the limit $m_\rho \rightarrow \infty$, $\rho_\mu$ is
constrained to
\beq
\rho_\mu = \frac{i}{g_*} V_\mu,
\eeqlb{Rho}
which gives
$$ G(r) = -1 + \cos F(r), \seqno{Rho}{a}$$
and consequently
$$ G^{\prime\prime}(0) = [F^\prime(0)]^2.
\seqno{Rho}{b}$$
As for the $\omega$-meson radial function $\omega(r)$, it can be
integrated to give
\beq
\omega(r) =  \int^\infty_0 \!\! r^{\prime 2}dr' G(r,r')
\left[ \frac{3 g_*}{4\pi^2}
\frac{\sin^2 F(r')}{r^{\prime 2}} F'(r') \right],
\eeqlb{Omega}
with the help of the radial Green's function
$$ G(r,r') =
\frac{\ts e^{-m_\omega |r-r'|} - e^{-m_\omega(r+r')}}{ 2 m_\omega r r'}.
\seqno{Omega}{a} $$
Here, we have neglected the $\omega$-$\rho$ coupling term.
To leading order in $m_\omega$, Eq.~(\ref{Omega}) becomes
$$ \begin{array}{rcl}
\omega(0) &=& \dps \int^\infty_0 dr'
\left[\frac{e^{-m_\omega r'}}{r'} \right]
\left[ \frac{3 g_*}{4\pi^2}
\frac{\sin^2 F(r')}{r^{\prime 2}} F'(r') \right]  \\ & \sim &
\dps \frac{3g_*}{m_\omega^2 4\pi^2}[F'(0)]^3.
\end{array} \seqno{Omega}{b}$$
Substituting Eqs.~(\ref{Rho}b) and (\ref{Omega}b) into Eq.~(\ref{Ebs}) with
$F'(0) \sim -0.7$ GeV yields
\beq
\ve_{bs} = 1.05 g + 1.47 \frac{\beta}{g_*} + 0.75 \alpha .
\eeq
Examining the individual terms with $g=-\frac34$, $\beta/g_* = -0.67$
and $\alpha = 1$ (vector meson dominance), we see that the
``attractive"\footnotemark \addtocounter{footnote}{-1}
$\rho$-meson and ``repulsive"\footnote{One should  
perhaps not conclude that
the effect of the $\rho$-meson is truly repulsive  and that of $\omega$-meson
truly attractive.
First of all, their effects are strongly dependent on the coupling
constant $\beta/g_*$
and the parameter $\alpha$. In Ref.~\citenum{UGVA2}, $\beta/g_*$ is
found to have a positive sign and in Ref.~\citenum{GMSS} $\alpha=-2$
is taken for a better fit. Furthermore, the vector mesons contribute
to the energy of the heavy meson bound state indirectly
by changing the pion profile $F(r)$.
Note also that unless stabilized by vector mesons the soliton shrinks to
zero size, $F'(0)\rightarrow -\infty$.
}
$\omega$-meson contributions are comparable
to that of the pion.
Of course this is just a rough estimate but seems to represent a
typical situation. It is clear that the vector mesons and
the higher-order derivative terms are expected
to play a nontrivial role.

\subsection{Finite Mass Corrections}
So far, we have limited ourselves to the heavy-quark limit. Thus
heavy-baryon properties have been computed to leading order in $1/m_Q$,
that is to $O (1)$.
In the heavy-quark limit, $\Sigma_Q$ and $\Sigma^*_Q$
are in a degenerate multiplet with isospin one and spin 1/2 and 3/2,
respectively. The (hyperfine) mass splitting between these states is
an $1/m_Q$ effect.

The $\Sigma^*_Q$-$\Sigma_Q$ mass difference due to the leading heavy-quark
symmetry breaking was first computed in Ref.~\citenum{JM92}.
The leading order Lagrangian in derivative expansion that breaks the
heavy quark symmetry is\cite{Wise}
\beq
{\cal L}_1 = \frac{\lambda_2}{m_Q}
\Tr \sigma^{\mu\nu} H \sigma_{\mu\nu} \bar{H}.
\eeqlb{L1}
This gives a $P_Q^*$-$P_Q^{}$ mass difference of $-8\lambda_2/m_Q$.
One can easily check that it is the next-to-leading order term
ignored in obtaining Eq.~(\ref{Lv}) by substituting Eq.~(\ref{HQF})
into Lagrangian (\ref{Yan}): {\it viz.},
\begin{eqnarray*}
{\cal L}_1 &=& \frac{1}{2\m_P}\left\{ (\m^2_P - m^2_P) P_v^{} P^\dagger_v
 - (\m^2_P - m^2_{P^*}) P^{*\mu}_v P^{*\dagger}_{v\mu} \right\}  \\
&=&  \ts\frac14 (m_{P^*} - m^{}_P) \left\{ 3 P^{}_v P_v^\dagger
+ P^{*\mu}_v P^{*\dagger}_{v\mu} \right\}.
\end{eqnarray*}

Here we shall follow Ref.~\citenum{JM92} and use the alternative
bound-state approach described in Sec.~5.3 to calculate the
$\Sigma^*_Q$-$\Sigma_Q$.
The soliton-meson states $|1,1;N\rangle$ and $|1,1;\Delta\rangle$
appearing in Eq.~(\ref{ES11}) can be written explicitly in terms of
the soliton states and the spin of the light degrees of freedom of $H$,
\beqa{1.5}{l}
|1,1,1;N\rangle = |\frac12\rangle_N |\uparrow\rangle_\ell, \\
|1,1,0;N\rangle = \sqrt{\frac12} |\frac12\rangle_N |\downarrow\rangle_\ell
   + \sqrt{\frac12} |-\frac12\rangle_N |\uparrow\rangle_\ell, \\
|0,0,0;N\rangle = \sqrt{\frac12} |\frac12\rangle_N |\downarrow\rangle_\ell
   - \sqrt{\frac12} |-\frac12\rangle_N |\uparrow\rangle_\ell, \\
|1,1,1;\Delta\rangle
= \sqrt{\frac34} |\frac32\rangle_\Delta |\downarrow\rangle_\ell
 - \sqrt{\frac14} |\frac12\rangle_\Delta |\uparrow\rangle_\ell, \\
|1,1,0;\Delta\rangle
= \sqrt{\frac12} |\frac12\rangle_\Delta |\downarrow\rangle_\ell
 - \sqrt{\frac14} |-\frac12\rangle_\Delta |\uparrow\rangle_\ell,
\eeqa
where $|\ \rangle_\ell$ is the spin state of the light degrees of
freedom in the heavy meson, $|m\rangle_{N,\Delta}$ is the soliton state
in the $N$ or $\Delta$ sector with $s_{\ell 3} = m$,
and $|i,j_\ell,m;N,\Delta\rangle$ is the bound state
$|i,j_\ell;N,\Delta\rangle$ with $j_{\ell3}=m$. Furthermore,
the tensor product of the light degrees of freedom and heavy quark
in $H$ can be re-expressed in terms of $P$ and $P^*$ mesons,
\beqa{1.5}{l}
|\uparrow\rangle_\ell |\uparrow\rangle_Q = |P^*_Q,1\rangle, \\
|\downarrow\rangle_\ell |\downarrow\rangle_Q = |P^*_Q,-1\rangle, \\
|\uparrow\rangle_\ell |\downarrow\rangle_Q
= \sqrt{\frac12} |P_Q\rangle + \sqrt{\frac12}|P^*_Q,0\rangle , \\
|\downarrow\rangle_\ell |\uparrow\rangle_Q
= \sqrt{\frac12} |P_Q\rangle - \sqrt{\frac12}|P^*_Q,0\rangle,
\eeqa
where $|P_Q^*,m\rangle$ is the $P_Q^*$ meson with $s_3=m$.
Finally, the $\Sigma_Q^*$, $\Sigma_Q$ and $\Lambda_Q$ states are
written explicitly as
\beqa{1.7}{l}
|\Sigma_Q^*, \frac32\rangle
= -\sqrt{\frac38} b |\frac32\rangle_\Delta |P_Q\rangle
 - \frac12 b |\frac12\rangle_\Delta |P^*_Q,1\rangle
 + a |\frac12\rangle_N |P^*_Q,1\rangle
 + \sqrt{\frac38} b |\frac32\rangle_\Delta |P^*_Q,0\rangle, \\
|\Sigma_Q,\frac12\rangle
= \sqrt{\frac34} a |\frac12\rangle_N |P_Q\rangle
 + \sqrt{\frac12} b |\frac32\rangle_\Delta |P^*_Q,-1\rangle
 + \sqrt{\frac16} b |-\frac12\rangle_\Delta |P^*_Q,1\rangle \\
\hskip 2cm
- \sqrt{\frac16} a |-\frac12\rangle_N |P^*_Q,1\rangle
- \sqrt{\frac13} b |\frac12\rangle_\Delta |P^*_Q,0\rangle
+ \sqrt{\frac1{12}} a |\frac12\rangle_N |P^*_Q,0\rangle, \\
|\Lambda_Q,\frac12\rangle = \frac12 |\frac12\rangle_N |P_Q^*,0\rangle
- \frac12 |\frac12\rangle_N |P_Q\rangle
-\sqrt{\frac12} |-\frac12\rangle_N |P^*_Q,1\rangle.
\eeqalb{SSLstates}
Now, the $\Sigma_Q^*$, $\Sigma_Q$ and $\Lambda_Q$ masses including
the next-to-leading order Lagrangian  can be read off directly from the
expressions for the states (\ref{SSLstates}) :
\beqa{1.5}{l}
m_{\Sigma_Q^*} = - \frac32\gF + a^2 m_N + b^2 m_\Delta
  + \frac38 b^2 m_{P} + (a^2 + \frac58 b^2) m_{P^*}, \\
m_{\Sigma_Q} = -\frac32\gF + a^2 m_N + b^2 m_\Delta
  + \frac34 a^2 m_{P} + (\frac14 a^2 + b^2) m_{P^*}, \\
m_{\Lambda_Q} = -\frac32 \gF + m_N + \frac14 m_{P} + \frac34 m_{P^*},
\eeqalb{MFJM}
where the coefficient $a^2$ is the probability that $\Sigma^*_Q$
contains a nucleon, $3b^2/8$ is the probability that $\Sigma^*_Q$
contains a $P_Q$, etc. The $\Sigma_Q^*$-$\Sigma_Q$ mass difference
is obtained as
\beq
m_{\Sigma^*_Q} - m_{\Sigma_Q} = \ts\frac38 (2a^2-b^2) (m_{P^*} - m_P)
= \dps \frac{(m_\Delta - m_N)(m_{P^*} - m_P)}{4\gF}.
\eeq
Note that the mass splittings have the dependence on
$m_Q$ and $N_c$ that agrees with the constituent quark model.
The $P^*$-$P$ mass difference is of order $1/m_Q$ and
the $\Delta$-$N$ mass difference is of order $1/N_c$.
This implies that the $\Sigma^*_Q$-$\Sigma$ mass difference is of
order $1/(m_QN_c)$. Substituting $\gF = 419$ MeV, we obtain
\beq
m_{\Sigma_c^*} - m_{\Sigma_c} = 25 \mbox{ MeV}  \and
m_{\Sigma_b^*} - m_{\Sigma_b} = 8 \mbox { MeV}.
\eeq
The experimentally measured $\Sigma_c^*$-$\Sigma_c$
mass difference $\sim 77$ MeV is three times as big as this
Skyrme model prediction and, compared to the quark model prediction,
they are reduced by a factor two. Note however, they are not
the {\it whole story} of the large $N_c$ prediction.
For example, the Lagrangian
\beq
{\cal L}_2 = \frac{\lambda_\pi}{m_Q}
\Tr (H A_\mu \bar{H} \gamma^\mu\gamma_5),
\eeq
also breaks the heavy quark symmetry, and upsets the relation
between the $P^* P^* \pi$ and $P^* P \pi$ coupling constants.

For an illustration of the equivalence of the two approaches discussed
in this paper, we now do the same calculation in the
CK bound-state approach described in Sec.~5.1 and Sec.~5.2.
With the heavy quark symmetry breaking Lagrangian ${\cal L}_1$,
the equation of motion for the heavy meson gets an additional term
\beq
i\partial_0 h(\vr,t) = \ve h(\vr,t)
= h(\vr,t) [ g \vA \cdot \vs ]
   + \frac{2\lambda_2}{m_Q} \vs \cdot h(\vr,t) \vs.
\eeq
We shall now take the last term as a perturbation and compute its effect
on the $k=1/2$ bound state obtained in Sec.~5.1.
The last term breaks only the heavy quark spin symmetry. The grand
spin is still a good symmetry of the equation of motion so that the
eigenstates can be classified by the corresponding quantum number.
Assuming the same radial functions peaked strongly at the
origin as in Sec.~5.1,
the problem reduces to finding the eigenfunction of the equation
\beq
\ve \K_{\frac12 k_3} = \ts\frac12\gF \K_{\frac12 k_3} (\tdr) [\sdt] (\tdr)
 + \dps\frac{2\lambda_2}{m_Q} \vs \cdot \K_{\frac12 k_3} \vs.
\eeq
The eigenstates $\K_{\frac12 k_3}$ can be expanded in terms of
the three possible basis states $\K^{(i)}_{\frac12 k_3}$
as in Eq.~(5.20) with the expansion coefficients
given by the solution of the secular equation
\beq
\sum_{j=1}^3\left( \M_{ij}^0 + \M_{ij}^{1} \right) c_j = - \ve c_i,
\eeqlb{M0M1}
with the matrix elements $M^{0}_{ij}$ and $M^1_{ij}$ defined by
$$ M^{0}_{ij} = \int\! d\Omega \Tr \{ \K^{(i)}_{\frac12 k_3} (\tdr)
[\ts\frac12\gF(\sdt)](\tdr) \bar{\K}^{(j)}_{\frac12 k_3} \},
\seqno{M0M1}{a}$$
$$ M^{1}_{ij} = \frac{2\lambda_2}{m_Q} \int\! d\Omega \Tr \{
\vs\cdot \K^{(i)}_{\frac12 k_3} \vs \bar{\K}^{(j)}_{\frac12 k_3} \}.
\seqno{M0M1}{b}$$
The matrix elements $M^{0}_{ij}$ were evaluated in Sec.~5.1. For
$M^1_{ij}$, we exploit the fact that $\frac12\vs$
acting on the right hand side of $\K$ is the heavy quark spin operator
while $-\frac12\vs$ acting on its left hand side is the light quark spin
operator of the heavy mesons. Actually, $K^{(i)}_{\frac12 k_3}(i=1,2,3)$
are the eigenstates of the operator:
\beq
\vs \cdot \K^{(i)}_{\frac12 k_3} \vs
= \{ -4\vec{S}_Q\cdot\vec{S}_\ell \} \K^{(i)}_{\frac12 k_3}
= - 2 (s(s+1)-3/2) \K^{(i)}_{\frac12 k_3}
= \left\{ \begin{array}{ll}
3 \K^{(i)}_{\frac12 k_3} & i=1 \\
- \K^{(i)}_{\frac12 k_3} & i=2,3
\end{array}\right. .
\eeq
The result is
\beq
M^{1} = -\frac{2\lambda_2}{m_Q} \left( \begin{array}{ccc}
3 & 0 & 0 \\ 0 & -1 & 0 \\ 0 & 0 & -1 \end{array} \right).
\eeq
Up to first order in perturbation, the bound state energy remains
unchanged: $\ve=-\frac32\gF$. On the other hand, the corresponding eigenstate
$\K^{bs}_{\frac12 k_3}$ is perturbed to
\beq
\K^{bs}_{\frac12 k_3} =
\ts\frac{1}{2} (1 + 3 \epsilon) \K^{(1)}_{\frac12 k_3}
- \frac{\sqrt3}{2} (1 - \epsilon) \K^{(2)}_{\frac12 k_3},
\eeqlb{BSptbd}
with
\beq
\epsilon = - \frac{\lambda_2}{m_Q} \frac{1}{\gF}.
\eeq
With $\epsilon=0$, the eigenstate returns to that in the heavy quark limit.

The heavy baryons can be obtained by quantizing the soliton-heavy meson
bound state in the same way as explained in Sec.~5.2. It leads to the
heavy baryon states of Eq.~(\ref{HBS}) with $|m\rangle_{bs}$ replaced
by the perturbed state of Eq.~(\ref{BSptbd}).
Due to the perturbation, the expectation value of $\vec{\Phi}(\infty)$
defined by Eq.~(\ref{Hcol}b) with respect to the bound states {\em does not
vanish}. According to Wigner-Eckart theorem, the expectation value
can be expressed as
\beq
{^{}_{bs} \langle m' |\vec{\Phi}(\infty)| m \rangle_{bs}^{}}
= - c \langle \ts\frac12 m'|\vec{K} | \frac12 m \rangle ,
\eeqlb{WETh}
where the ``hyperfine constant" is given by
\beq
c = 2\epsilon = - \frac{2\lambda_2}{m_Q} \frac{1}{\gF}.
\eeq
With the help of Eq.~(\ref{WETh}), we can compute the expectation value of
the collective Hamiltonian (\ref{Hcol})
\beq
E_{i,j} = M_{sol} + \ve_{bs} + \frac{1}{2\I}
\{ (1-c)i(i+1) + cj(j+1) - ck(k+1) + \ts\frac34 \}.
\eeq
Explicitly, we have the heavy baryon masses
\beqa{1.5}{l}
m_{\Lambda_Q}^{}
   = M_{sol} + \m_{P}^{} - \frac32\gF + {3}/{8\I}, \\
m_{\Sigma_Q}^{}
   = M_{sol} + \m_{P}^{} - \frac32\gF + (11-8c)/{8\I}, \\
m_{\Sigma_Q^*}^{}
   = M_{sol} + \m_{P}^{} - \frac32\gF + (11+4c)/{8\I}.
\eeqalb{MFLSS*}
We see that these
are identical to Eq.~(\ref{MFJM}).

One more important aspect of the bound state approach
with finite mass heavy mesons has to do with the turning-on of the
kinetic motions.
In the heavy quark limit, the heavy mesons are approximated to
sit at the origin of the soliton and the kinetic motion of the
heavy mesons is neglected. As the heavy meson masses become finite,
the effects of the kinetic motion increase. For example,
the binding energy of the heavy mesons would be reduced compared with
that in the infinite mass limit.
Such $1/m_Q$ corrections have been studied in Ref.~\citenum{OPM1}.

Let us now return to the Lagrangian (\ref{Yan}) written in terms of the
component
heavy-meson fields $P$ and $P^*_\mu$. The equations of motion are
\beqa{1.5}{c}
( D_\mu D^\mu + m_P^2 ) P = f_{_Q}  P^*_\mu A^\mu, \\
D_\mu P^{*\mu\nu} + m^2_{P^*} P^{*\nu}
= - f_{_Q}  P A^\nu + g_{_Q} \varepsilon^{\mu\nu\lambda\rho}
P^*_{\mu\rho} A_\lambda.
\eeqalb{eq_PP*}
The momenta conjugate to the meson fields $P$ and $P^*_\mu$, respectively,
are
\beqa{1.5}{rcl}
\Pi &=& \dps\frac{\partial {\cal L}}{\partial \dot{P}} = D_0 P^\dagger, \\
\Pi^{*i} &=& \dps\frac{\partial {\cal L}}{\partial \dot{P}_i^*} =
(P^{*i0})^\dagger - g_{_Q} \epsilon^{ijk} A_j P^{*\dagger}_k ,
\eeqa
and similarly for $\Pi^\dagger$ and $\Pi^{*i\dagger}$.
Since $\Pi^*_0$ vanishes identically,
$P^*_0$ is not an independent dynamical variable; it
can be eliminated by using Eq.~(\ref{eq_PP*})
\beq
P^{*0} = - \frac{1}{m_{P^*}^2} ( D_i \Pi^{*i\dagger}
 + \textstyle \frac12 g_{_Q} \epsilon^{ijk} P^*_{ij} A_k) ,
\eeqlb{time-comp}
which results in a set of coupled equations
\beqa{1.5}{c}
\dps
\dot{\vec{P}}^* = - {\vec \Pi}^{*\dagger}
+ g_{_Q} \vec{P}^* \times \vec{A} + \frac{1}{m^2_{P^*}}
 \vec D ( \vec D \cdot {\vec \Pi}^{*\dagger} )
 + \frac{g_{_Q}}{m^2_{P^*}}
\vec D( ( \vec D \times \vec{P}^* ) \cdot \vec A ) , \\
\displaystyle
\dot {\vec \Pi}{}^{*\dagger} = \vec D \times ( \vec D \times \vec{P}^* )
+ m^2_{P^*} \vec{P}^* + f_{_Q} P \vec{A}
+  g_{_Q} {\vec \Pi}^{*\dagger} \times \vec{A}
- g_{_Q}^2 (\vec{P}^* \times \vec{A}) \times \vec{A} \\
\displaystyle
 - \frac{2g_{_Q}}{m_{P^*}^2} \vec D \left\{
\vec{D} \cdot \vec{\Pi}^{*\dagger} +
g_{_Q} ( \vec{D} \times \vec{P}^* ) \cdot \vec{A} \right\} \times \vec{A}.
\eeqalb{Pi-dot}
where $\vec{D} P \equiv \vec{\nabla} P - P \vec{V}^\dagger $.

In order to express the equations
of motion only in terms of $P$ and $\vec{P}^*$,
we use the fact that
$P^*_0$ field is at most of order $1/m_Q$; {\it viz.\/},
\beq
P^{*0} \sim \frac{1}{m^2_{P^*}} D_i \dot{P}^{*i}
= O(1/m_{P^*}).
\eeq
Keeping this leading order term,
we can express the equations of motion as
\beq
\ddot{\vec{P}}{}^* = + 2 g_{_Q} \dot{\vec{P}}{}^* \times \vec A
- \vec D\times ( \vec D \times {\vec{P}}^*)
- M^2_{P^*} {\vec{P}}^*
- f_{_Q} P \vec A + \vec D ( \vec D \cdot {\vec P}^*).
\eeqlb{14}

The eigenstates are classified by the grand spin quantum numbers, $k$, $k_3$
and the parity $\pi$. The wavefunctions of the $k^\pi= \frac12^+$ state
of our interest are expanded as
\beqa{1.5}{l}
\displaystyle
P_{\frac12^+ k_3}(\vec{r},t) = e^{-i \omega t} \varphi(r)
{\cal Y}_{\frac12^+ k_3}(\hat{r}) , \\
\displaystyle
\vec{P}^*_{\frac12^+ k_3}(\vec{r},t) = e^{-i \omega t} \sum_{\kappa}
\varphi^{*}_{\kappa}(r) \vec{\cal Y}_{\frac12^+ k_3}^{(\kappa)}(\hat{r}),
\eeqa
where ${\cal Y}_{k^\pi k_3}$ and $\vec{\cal Y}_{k^\pi k_3}$
are the generalized spherical spinors and vector harmonics, respectively,
and $\kappa$ is an index to label the possible vector spherical harmonics
with the same $k$, $k_3$ and parity $\pi$. These spherical harmonics correspond
to $\K^{(i)}_{kk_3}$ in Sec.~5 with $\omega$ equal to $\m_P + \ve$.

Since the spin of the heavy mesons is represented in a different way
from the $4\times 4$ matrix representation in the heavy quark limit,
the explicit forms of ${\cal Y}_{kk_3}$ and $\vec{\cal Y}_{kk_3}$
are different from those of ${\cal K}^{(i)}_{kk_3}$.
As for the pseudoscalar meson, we have only one
spherical spinor harmonics with $k^\pi={\frac12}^+$:
\beq
{\cal Y}_{\frac12^+ \pm1/2}(\hat{r})
 = \frac{1}{\sqrt{4\pi}} \tp_{\pm} \vec\tau \cdot \hat{r}.
\eeq
Here, $\tp_\pm$ is the same isospin basis for the heavy meson anti-doublets
that we have used in Sec.~5.
For vector mesons with spin 1, we can construct two different
$k^P={\frac12}^+$ vector spherical harmonics
\footnote{Here, 
we combine first the spin  and orbital angular momentum bases
to the total spin ($\vec{J}=\vec{S}+\vec{L}$) basis
and then combine the spin and isospin.
Then, $\vec{\cal Y}_{\frac12^+ \pm\frac12}^{(1)}(\hat{r})$ and
$\vec{\cal Y}_{\frac12^+ \pm\frac12}^{(2)}(\hat{r})$ correspond
to $J=0$ and $J=1$ states, respectively. This procedure
enables us to proceed with
the simple combinations such as $\tdr$ and $\vec{\tau}\times \hat{r}$
in the forthcoming calculations.
One may obtain the vector
spherical harmonics in different ways; for example, by combining
first the isospin and orbital angular momentum
to the $\vec{\Lambda}(=\vec{I}+\vec{L})$ basis and then to the spin basis,
leading to
$$\begin{array}{l}
\displaystyle \vec{\cal Y}_{\frac12^+ \pm\frac12}^{\prime(1)}(\hat{r})
= \frac{1}{\sqrt{12\pi}} \vec\tau \cdot \hat r \vec{\tau} \chi_{\pm} =
\textstyle \sqrt{\frac13}\vec{\cal Y}_{\frac12^+ \pm\frac12}^{(1)}
+\sqrt{\frac23}\vec{\cal Y}_{\frac12^+ \pm\frac12}^{(2)},  \\
\displaystyle \vec{\cal Y}_{\frac12^+ \pm\frac12}^{\prime(2)}(\hat{r})
= \frac{1}{\sqrt{24\pi}} (\vec\tau \cdot \hat r\hat{r}-3\hat{r}) \chi_{\pm} =
\textstyle -\sqrt{\frac23}\vec{\cal Y}_{\frac12^+ \pm\frac12}^{(1)}
+\sqrt{\frac13}\vec{\cal Y}_{\frac12^+,\pm\frac12}^{(2)}.
\end{array}$$
They correspond exactly to $\K^{(2)}_{\frac12^+ k_3}$ and
$\K^{(3)}_{\frac12^+ k_3}$ of Sec.~5.1, respectively.
}
: {\it viz.\/},
\beqa{1.5}{l}
\displaystyle
\vec{\cal Y}_{\frac12^+ \pm\frac12}^{(1)}(\hat{r})
= \frac{1}{\sqrt{4\pi}} \tp_{\pm} \hat{r}, \\
\displaystyle
\vec{\cal Y}_{\frac12^+ \pm\frac12}^{(2)}(\hat{r})
= i\frac{1}{\sqrt{8\pi}} \tp_{\pm} (\vec{\tau}\times\hat{r}) .
\eeqa
Putting
\beqa{1.5}{l}
P(\vec{r},t)
= e^{-i\omega t} \varphi(r) {\cal Y}_{\frac12^+ \pm\frac12}(\hat{r}), \\
\vec{\Phi}^*(\vec{r},t)
= e^{-i\omega t}\left\{
\varphi^*_1(r) \vec{\cal Y}^{(1)}_{\frac12^+ \pm\frac12}(\hat{r})
 + \varphi^*_2(r) \vec{\cal Y}^{(2)}_{\frac12^+ \pm\frac12}(\hat{r}) \right\},
\eeqa
into the equations of motion (\ref{eq_PP*}) and (\ref{14}),
we obtain three coupled differential equations for the radial functions:
\beqa{1.5}{l}
\displaystyle
\varphi'' + \frac{2}{r} \varphi'
+ (\omega^2 - m^2_P - \frac{2}{r^2}) \varphi   =
2 \upsilon(\upsilon - \frac{2}{r} ) \varphi
 + \frac{f_{_Q}}{2} (a_1+a_2) \varphi^*_1 - \frac{1}{\sqrt2}{f_{_Q}}
a_1 \varphi^*_2 , \\
\displaystyle
\varphi^{*\prime\prime}_1 + \frac{2}{r} \varphi^{*\prime}_1
+ (\omega^2 - m^2_{P^*} - \frac{2}{r^2}) \varphi^*_1
=  \frac{f_{_Q}}{2} ( a_1 + a_2 ) \varphi + 2  \upsilon^2 \varphi^*_1 \\
\displaystyle \hskip 65mm
 + \sqrt2 ( g_{_Q} a_1 \omega - \frac{1}{r} \upsilon +
\upsilon' ) \varphi^*_2 , \\
\displaystyle
 \varphi^{*\prime\prime}_2 + \frac{2}{r} \varphi^{*\prime}_2
 + (\omega^2 - m^2_{P^*} - \frac{2}{r^2} ) \varphi^*_2
=  - \frac{f_{_Q}}{\sqrt2} a_1 \varphi + \sqrt2
( \omega g_{_Q} a_1 - \frac{1}{r} \upsilon + \upsilon') \varphi^*_1   \\
\displaystyle \hskip 65mm
+ ( - \omega g_{_Q} (a_1+a_2)
- \frac{4}{r} \upsilon + 4 \upsilon^2 ) \varphi^*_2. \\
\eeqalb{EoMRF}
The wavefunctions are normalized such that each mode carries one
corresponding heavy flavor number:
\beq
1 = \int^\infty_0\!\! r^2dr \left\{
2\omega \left[|\varphi|^2 + |\varphi^*_1|^2 + |\varphi^*_2|^2 \right]
+ g_{_Q} \left[ (a_1+a_2)|\varphi^*_2|^2 - \sqrt2
a_1 ( \varphi_1^{*\dagger} \varphi^*_2
  + \varphi_2^{*\dagger} \varphi^*_1 ) \right]
\right\},
\eeqlb{norm}
where we have kept terms up to next-to-leading order in $1/m_Q$.

Near the origin, the equations of motion become, asymptotically,
\beqa{1.5}{c}
\displaystyle
   \varphi'' + \frac{2}{r}\varphi' = 0 , \\
\displaystyle
   \varphi^{*\prime\prime}_1 + \frac{2}{r}\varphi^{*\prime}_1
     - \frac{4}{r^2} \varphi^*_1
   = -\frac{2\sqrt2}{r^2} \varphi^*_2 , \\
\displaystyle
   \varphi^{*\prime\prime}_2 + \frac{2}{r}\varphi^{*\prime}_2
- \frac{2}{r^2} \varphi^*_2
   = -\frac{2\sqrt2}{r^2} \varphi^*_1.
\eeqa
They imply that we have three independent sets of solutions:
\beqa{1.5}{l}
\mbox{(i)} \hskip 4mm \left( \begin{array}{l}
\varphi(r) = \varphi(0) + O(r^2) , \\
\varphi^*_i(r) = O(r^2) , \hskip 2mm(i=1,2)
\end{array} \right. \\
\mbox{(ii)} \hskip 4mm \left(  \begin{array}{l}
\varphi(r) = O(r^2) , \\
\varphi^*_i(r) = \varphi^*_{bi}(0) + O(r^2), \hskip 2mm(i=1,2)
\end{array} \right. \\
\mbox{(iii)} \hskip 4mm \left(  \begin{array}{l}
\varphi(r) = O(r^4) , \\
\varphi^*_i(r) =  \frac12\varphi^{*\prime\prime}_{ci}(0) r^2 + O(r^4),
\hskip 2mm(i=1,2) \end{array} \right. \\
\eeqa
with $\sqrt2\varphi^*_{1}(0)=\varphi^*_{2}(0)$ for the solution set (ii)
and $\varphi^{*\prime\prime}_{1}(0)=-\sqrt2\varphi^{*\prime\prime}_{2}(0)$
for the set (iii).
For sufficiently large $r (\gg 1/m_P)$,
the three equations decouple from each other: for example,
\beq
\varphi'' + \frac{2}{r} \varphi'
  + (\omega^2 - m_{P^*}^2) \varphi = 0.
\eeq
Thus, the bound-state solutions ($\omega < m_P^{}$) are
\beqa{1.5}{l}
\varphi(r) = \alpha \dps\frac{e^{-r\sqrt{m_P^2-\omega^2}}}{r}, \\
\varphi^*_1(r) =
\alpha_1 \dps\frac{e^{-r\sqrt{m_{P^*}^2-\omega^2}}}{r}, \\
\varphi^*_2(r) =
\alpha_2\frac{e^{-r\sqrt{m_{P^*}^2-\omega^2}}}{r},
\eeqa
with three constants $\alpha$, $\alpha_1$ and $\alpha_2$.

The lowest-energy bound states are found by solving numerically
the equations of motion (\ref{EoMRF}).
The results are shown in Fig.~6.1 and Table 6.1.
Figure~6.1 shows the radial functions
$\varphi(r)$ and $\varphi^*_1(r)$
for the $D$ and $D^*$ mesons (solid curve) and the $B$ and $B^*$ mesons
(dashed curves).
The radial function  $\varphi^*_2(r)$ -- which is hardly distinguishable
from $\sqrt2\varphi^*_1(r)$ -- is not shown there.
By comparing the two cases,
one can easily check that as the meson mass increases,
(1) the radial function becomes more sharply peaked at the origin
and (2) the role of the vector mesons becomes as important
as that of the pseudoscalar mesons. Note that
the radial function $\varphi^*_1(r)$ becomes comparable to $\varphi(r)$
[see also the ratio $\varphi^*_1(0)/\varphi(0)$].
This can be understood as follows:
due to their heavy masses, the heavy mesons are
localized in the region
$r$\raisebox{-0.6ex}{$\stackrel{\textstyle <}{\sim}$}$1/m_P$,
where
\beqa{1.5}{l}
[a_1(r)+a_2(r)] \sim [- a_1(r)] \sim F'(0)+O(r^2), \\
\upsilon(r) \sim \dps \frac{1}{r} - \ts\frac14 F^{\prime2}(0)r+\cdots,
\eeqalb{AppPot}
so that the equation of motion for $(\varphi^*_1-\frac{1}{\sqrt2}\varphi^*_2)$
is completely decoupled from those for $\varphi$ and
$(\varphi^*_1+\sqrt2\varphi^*_2)$.
Note also that in Sec.~5.1 the $\K^{(3)}$ component
is completely decoupled from the rest;
\ie, from the $\K^{(1)}$ and $\K^{(2)}$ components.
\begin{figure}
\beginpicture
\setcoordinatesystem units <1cm,1cm> point at 3 2
\setplotarea x from 0 to 15, y from 0 to 8
\put {{\bf Figure 6.1 :} $\varphi(r)$ and $\varphi_1^*(r)$ for
   $Q$=$c$ (solid) and $b$ (dashed).}  at 7.5 0.5
\put{$\varphi_2^*(r)$ is nearly equal to $\sqrt2\varphi_1^*(r)$ for both
cases.} at 7.75 0
\setcoordinatesystem units <9cm,0.40cm> point at 0 0
\setplotarea x from 0 to 1.0, y from 0 to 15.0
\linethickness=1pt
\thicklines
\axis bottom ticks in
   numbered from 0 to 1.0 by 0.5
   unlabeled short from 0 to 1.00 by 0.1 /
\axis left ticks in
   numbered  from 0 to 15 by 5
   unlabeled short from 1 to 14 by 1 /
\axis top ticks in
   unlabeled short from 0.1 to 0.9 by 0.1 /
\axis right ticks in
   unlabeled short from 1 to 14 by 1 /
\put {$r$ (fm)} at 0.75 -1
\setquadratic
\inboundscheckon
\setplotsymbol ({.})
\plot
    .00120 8.6307   .04918 8.4075    .05517 8.3504   .07317 8.2363
    .09116 8.1247   .10916 7.9766    .12715 7.7979   .14514 7.5931
    .16314 7.3658   .18113 7.1195    .19912 6.8574   .21711 6.5827
    .23511 6.2982   .27109 5.7115    .30708 5.1181   .34306 4.5357
    .37905 3.9783   .41503 3.4561    .45102 2.9760   .48701 2.5418
    .52299 2.1548   .52899 2.0949    .56497 1.7619   .60096 1.4727
    .63695 1.2241   .67293 1.0122    .70892 .83314   .74490 .68289
    .78089 .55763   .87685 .31985    1.0088 .14495 /
\put {$\varphi^c_{}$} at 0.4 4.5
\plot
    .00120 7.3999   .04918 7.2171   .05517 7.1595   .07317 7.0490
    .09116 6.9466   .10916 6.8115   .12715 6.6494   .14514 6.4641
    .16314 6.2590   .18113 6.0373   .19912 5.8021   .21711 5.5560
    .23511 5.3020   .27109 4.7805   .30708 4.2565   .34306 3.7456
    .37905 3.2603   .41503 2.8092   .45102 2.3979   .48701 2.0291
    .52299 1.7033   .52899 1.6531   .56497 1.3759   .60096 1.1374
    .63695 .93444   .67293 .76331   .70892 .62024   .74490 .50155
    .78089 .40378   .87685 .22224   1.0088 .09438 /
\put {$-\varphi_1^{*c}$} at 0.25 3.5
\setdashes
\plot
    .00120 13.786   .04318 13.363   .04918 13.186
    .06717 12.949   .08516 12.646   .10316 12.253    .12115 11.788
    .13914 11.263   .15714 10.692   .17513 10.086    .19312 9.4557
    .24110 7.7340   .27709 6.4795   .31308 5.3152    .34906 4.2743
    .38505 3.3738   .42103 2.6170   .45702 1.9973    .49300 1.5014
    .52899 1.1129   .56497 .81418   .60096 .58849    .63695 .42062
    .67293 .29753   .76889 .11301   .83487 .056235   .90084 .02738
    .96681 .01309  1.0328 .00617 /
\put {$\varphi^b_{}$} at 0.18 12
\plot
    .00120 12.923   .04318 12.465   .04918 12.359    .06717 12.131
    .08517 11.842   .10316 11.469   .12115 11.027    .13914 10.530
    .15714 9.9887   .17513 9.4148   .19312 8.8187    .24110 7.1935
    .27709 6.0125   .31308 4.9192   .34906 3.9446    .38505 3.1040
    .42103 2.3998   .45702 1.8250   .49300 1.3668    .52899 1.0091
    .56497 .73519   .60096 .52905   .63695 .37636    .67293 .26490
    .76889 .09914   .83487 .04877   .90084 .02344    .96681 .01105
    1.0328 .00513 /
\put {$-\varphi^{*b}_1$} at 0.10 9.5
\endpicture
\end{figure}

The wavefunctions of Sec.~5.1 obtained
in the heavy mass limit ($m_P^{},m_{P^*}^{}\rightarrow\infty$),
can be expressed in the same convention as
\beqa{1.5}{l}
\displaystyle P \sim \frac12 \frac{1}{\sqrt{2 \m_P}}
f(r) {\cal Y}_{1/2^+ \pm1/2}, \\
\displaystyle \vec{P}^* \sim -\frac12\frac{1}{\sqrt{2 \m_P}}
f(r) (\vec{\cal Y}^{(1)}_{1/2^+ \pm1/2}
 + \sqrt2 \vec{\cal Y}^{(2)}_{1/2^+ \pm 1/2}),
\eeqalb{IMlimit}
where  the radial function $f(r)$, normalized to
$\int\!\!r^2dr |f|^2 = 1$,  is strongly peaked at the
origin.
It implies that
$$\ts \varphi(r) = -\varphi^*_1(r) = -\frac{1}{\sqrt2} \varphi^*_2(r)
\sim \frac12\frac{1}{\sqrt{2M_{\Phi^*}}} f(r).
\seqno{IMlimit}{a} $$
These radial functions satisfy the normalization condition of Eq.~(\ref{norm})
in the leading order in $1/m_Q$; {\it viz.\/},
$$  2\omega_B^{} \int^\infty_0\!\! r^2dr
(|\varphi|^2 + |\varphi^*_1|^2 + |\varphi^*_2|^2 ) = 1.
\seqno{IMlimit}{b}$$
It is interesting to note that the pseudoscalar meson and the three vector
mesons contribute equally to the bound state.

\begin{table}
\begin{center}
{\bf Table 6.1 :} {Input parameters and the numerical results. }
\vskip 3mm
\begin{tabular}{cccrrrcrccc} \hline
$Q$  & $f_\pi^{a)}$ & $e^{c)}$ & \multicolumn{1}{c}{$m_\Phi^{a)}$}
& $m_{\Phi^*}^{a)}$ & \multicolumn{1}{c}{\ $f_{_Q}^{a)}$} & \ $g_{_Q}^{c)}$
& \multicolumn{1}{c}{$\;\omega^{a)}_{B}$}
& b.e.$^{b)}$ & $c^{c)}$ & $\varphi^*_1(0)/\varphi(0)$ \\
\hline
$c$ & 64.5 & 5.45 & 1872 & 2010 & $-$3016
 & $-$0.75 & 1481 & 494 & 0.10 & $-$0.828 \\
$b$ & 64.5 & 5.45 & 5275 & 5325 & $-$7988
 & $-$0.75 & 4722 & 590 & 0.04 & $-$0.932 \\
\hline
 & \multicolumn{10}{c}{$^{a)}$ in MeV unit, $^{b)}$ binding energy
$(\equiv {\overline m}_P^{} - \omega$) in MeV unit,} \\
 & \multicolumn{10}{c}{$^{c)}$ dimensionless quantities. }
\end{tabular}
\end{center}
\end{table} 
We have listed In Table~6.1 the numerical results on the lowest bound states
together with the input parameters.  The $SU(2)$ parameters,
$f_\pi$ and $e$, are fit to the nucleon and $\Delta$ masses and, as for
the heavy meson coupling constants, the nonrelativistic quark model
prediction, $\gQ=-0.75$,  and the heavy quark symmetric relation,
$\fQ = 2 m_P^{} \gQ$ are used.
Comparing the numerical results given in Table~6.1 with the binding energy
$\ve_{bs}=-\frac32\gF$ of Sec.~5.1. which gives $\sim 800$ MeV with the
same input parameters, one can see that
the $1/m_Q$ corrections  amount to $\sim 200$ MeV in the bottom
sector and $\sim 300$ MeV in the charm sector.

The collective quantization of the soliton-heavy-meson bound state
leads to the same heavy-baryon states of Eq.~(\ref{HBS}), so we can use the
same mass formula, Eq.~(\ref{MFLSS*}).
Here, the hyperfine constant $c$ is found to be\footnote{An
error committed in Ref.~\citenum{OPM1} is corrected here.}
\beqa{1.5}{rl}
 c = &\hskip -3mm \displaystyle \left.\int^\infty_0\!\! r^2dr \, \right\{
2\omega_B^{} \left[ \textstyle
( |\varphi|^2 - \frac13 |\varphi^*_1|^2 - \frac23 |\varphi^*_2|^2 )
- \frac43 \cos^2\!(F/2) ( |\varphi|^2 - |\varphi^*_1|^2) \right] \\
& \hskip 3mm - \frac{2f_{_Q}}{3M_{\Phi^*}} \sin F \left[
\varphi^\dagger \left( {\varphi^*_1}' + \frac{2}{r} \varphi^*_1
\right) + \left( {\varphi^*_1}^{'\dagger} + \frac{2}{r}
{\varphi^*_1}^\dagger \right) \varphi - \frac{\sqrt2}{r} \sin^2
\frac{F}{2} \left( \varphi^\dagger \varphi^*_2 + {\varphi^*_2}^\dagger
\varphi \right) \right] \\
& \hskip 3mm -\left.\frac13g_{_Q}\left[
 F' |\varphi^*_2|^2
+ \frac{\sqrt2}{r} \sin\!F ( 4\cos \frac{F}{2} - 1)
( \varphi_1^{*\dagger} \varphi^*_2
 + \varphi_2^{*\dagger} \varphi^*_1 ) \right] \right\},
\eeqa
where the terms next-to-leading order in $1/m_Q$ have been kept.
One sees that the hyperfine constant $c$
is of order $1/m_Q$. This is a consistency check. The leading order terms
proportional to $\omega_B^{}$ vanish identically when the
radial functions of Eq.~(\ref{IMlimit}) are used.

With the bound state energy $\omega_B^{}$
and the hyperfine constant $c$ given in Table~6.1,
the mass formula (\ref{MFLSS*}) predicts the heavy
baryon masses as shown in Table~6.2 (Result I).
Result II is obtained by taking the two coupling constants
as free parameters.
To fit the experimental masses of $\Lambda_c$ and
$\Sigma_c$, one should have $f_{_Q}/2M_{D^*}=-0.88$ and $g_{_Q}=-1.00$,
which indicates that the heavy quark symmetric relation $\fQ=2m_P^{} \gQ$
is broken in the charm sector and shows about 25\% difference with the
estimate of $g$ made in Sec.~2.2.

\begin{table} 
\begin{center}
{\bf Table 6.2} : {Numerical results on the heavy baryon masses.}
\vskip 3mm
\begin{tabular}{cccccclll}
\hline
$Q$ &   & $f_{_Q}/2M_{\Phi^*}$ & $g_{_Q}$ & $\omega_B^{a)}$ & $c$
& $M_{\Lambda_Q}^{a)}$ & $M_{\Sigma_Q}^{a)}$ & $M_{\Sigma^*_Q}^{a)}$ \\
\hline
    & exp. & & & & & 2285$^{b)}$ & 2453$^{b)}$ & 2530$^{c)}$  \\
$c$ &  I   & $-$0.75 & $-$0.75 & 1481 & 0.10 & 2427 & 2596 & 2625 \\
    & II   & $-$0.88 & $-$1.00 & 1339 & 0.10 & 2285 & 2454 & 2483 \\
\hline
$b$ & exp.$^{b)}$ & & & & & 5641$^{b)}$ & \multicolumn{1}{c}{---} &
\multicolumn{1}{c}{---}  \\
    &  I   & $-$0.75 & $-$0.75 & 4722 & 0.04 & 5664 & 5849 & 5860 \\
\hline
 \multicolumn{9}{c}{$^{a)}$ in MeV unit,
$^{b)}$ Particle Data Group\cite{PDG}, $^{c)}$ Ref.~\citenum{SKAT}.}
\end{tabular}
\end{center}
\end{table} 

\setcounter{equation}{0}
\section{Conclusions}

In this review, we have described how chiral symmetry of light quarks
and heavy-quark symmetry of heavy quarks can be combined in a skyrmion
description of the heavy baryons. That heavy baryons can be described
as skyrmions may appear to some readers as surprising as one would
naively think that the soliton idea would be inappropriate for a system
where one or more heavy quarks are ``bound" to the soliton which in
reality with $N_c=3$ is not so heavy compared with the heavy meson.
This would seem to give an absurd picture of the system as ``a tail
wagging a dog." So how does this work?

There are two points to this paradox. The first is that the picture
is built with both the number of colors $N_c$ and the heavy quark mass
going to infinity. In this limit, both the soliton and the meson are
infinitely heavy and so on top of each other. The second point is that
in actual fact for $N_c\neq \infty$, the heavy meson is ``wrapped" by
the soliton in contrast to the monopole-scalar field system where the
scalar field is ``pierced" by the soliton.

We have seen that the two approaches, the ``top-down" method that comes
down from heavy-quark limit and the ``bottom-up" approach which goes up
from chiral symmetry limit, give the same description. How they are
related is shown in this review but the deep reason behind this relation,
if it exists, is not known.

In the latter approach, the heavy-quark limit emerges in an intriguing
way through the vanishing of a nonabelian Berry potential, with a finite
non-vanishing Berry potential describing in an approximate but rather
accurate way the hyperfine splitting of heavy baryons. The interesting
question as to how to compute corrections to hyperfine splittings that
are not included in the Berry potentials but would be needed for
not-so-heavy baryons (such as strange and charmed hyperons) has not been
addressed in this paper and remains an open question. This may be
related to non-adiabatic corrections to Berry potentials. There is also
a problem as to at what quark mass the Wess-Zumino term vanishes in the
``bottom-up" approach. We have seen that the $\omega$-exchange term in
the CK skyrmion is attractive for $K^-$ and repulsive for $K^+$ but when
the Wess-Zumino term disappears, the role of the $\omega$ field is not
determined. So what happens to the $\omega$ field as the Wess-Zumino
term is about to disappear?

Many applications of the skyrmion description to heavy-baryon phenomenology
still remain to be worked out although some are just appearing in the
literature.
What is crucially needed however is more experiments in the field.

\section*{Acknowledgement}
\addcontentsline{toc}{section}{Acknowledgement}
We are grateful for discussions with M. A. Nowak, N. N. Scoccola and I. Zahed.
This work was supported in part by the Korea Science and Engineering
Foundation through the Center for Theoretical Physics of Seoul National
University and by the Chungnam National University Research and Scholarship
Fund. One of us (Y.O.) was supported in part by the National Science
Council of ROC under grant \#NSC84-2811-M002-036.

\addcontentsline{toc}{section}{References}


\end{document}